\documentclass[a4paper,11pt]{article}

\usepackage[utf8x]{inputenc}
\usepackage{graphicx}
\usepackage{float}
\usepackage[T1]{fontenc}
\usepackage{epsfig}
\usepackage{color}
\usepackage{amsmath,jheppub_0,mathtools}
\usepackage{subfloat}
\usepackage{amsfonts}
\usepackage{braket}
\usepackage{cleveref}
\usepackage{epstopdf}
\usepackage{caption}
\usepackage{subcaption}
\usepackage{enumitem}
\usepackage{comment}
\usepackage[titletoc,toc,title]{appendix}
\usepackage{hyperref}
\hypersetup{
	colorlinks=true,
	linkcolor=blue,
	filecolor=red,      
	urlcolor=blue,
	citecolor=red
} 

\title{Reflected entropy in BCFTs on a black hole background}

\author[]{Debarshi Basu,}
\author[]{Himanshu Chourasiya,}
\author[]{Vinayak Raj and}
\author[]{Gautam Sengupta}

\affiliation[]{
	Department of Physics,\\Indian Institute of Technology Kanpur,\\208016, India}
\emailAdd{debarshi@iitk.ac.in}
\emailAdd{chim@iitk.ac.in}
\emailAdd{vraj@iitk.ac.in}
\emailAdd{sengupta@iitk.ac.in}
	\date{}

\abstract{
We obtain the reflected entropy for bipartite mixed state configurations involving two disjoint and adjacent subsystems in two dimensional boundary conformal field theories (BCFT$_2$s) in a black hole background. The bulk dual is described by an AdS$_3$ black string geometry truncated by a Karch-Randall brane. The
entanglement wedge cross section computed for this geometry matches with the
reflected entropy obtained for the BCFT$_2$ verifying the holographic duality.
In this context, we also obtain the analogues of the Page curves for the reflected entropy and investigate the behaviour of the Markov gap.
}
	
\begin{document}

\maketitle

\section{Introduction}
The black hole information loss problem \cite{Hawking:1975vcx, Hawking:1976ra} has greatly aided the understanding of many aspects of the quantum theory of gravity. The paradox arises after the Page time when the fine grained entropy of the Hawking radiation from an evaporating black hole exceeds the coarse-grained entropy which leads to a violation of unitarity. Recently the authors \cite{Penington:2019npb, Almheiri:2019psf, Almheiri:2019hni, Almheiri:2019yqk, Penington:2019kki, Almheiri:2020cfm} proposed a novel \textit{island} formula motivated by the quantum extremal surface (QES) prescription\footnote{The QES prescription is a quantum corrected version of the Ryu-Takayanagi (RT) proposal \cite{Ryu:2006bv, Ryu:2006ef, Hubeny:2007xt}.} \cite{Faulkner:2013ana, Engelhardt:2014gca} for the fine grained entropy of subsystems in quantum field theories coupled to semi-classical theories of gravity which restores the unitarity and leads to the Page curve \cite{Page:1993wv, Page:1993df, Page:2013dx}. This involves certain disconnected regions termed islands arising at late times in the bulk entanglement wedge for subsystems in radiation baths. The authors in \cite{Penington:2019kki, Almheiri:2019qdq, Dong:2020uxp, Kawabata:2021vyo} have substantiated the island formula from a gravitational path integral for the R\'enyi entanglement entropy by considering certain replica wormhole saddles which are dominant at late times.

The \textit{doubly holographic} description \cite{Almheiri:2019hni, Rozali:2019day, Chen:2020uac, Chen:2020hmv, Deng:2020ent, Suzuki:2022xwv, Grimaldi:2022suv, Geng:2020qvw, Geng:2020fxl, Geng:2021iyq, Geng:2021mic, Geng:2021hlu} provides a more intuitive understanding of the island formula, where the radiation baths are described by holographic conformal field theories (CFTs). In the double holographic scenario (bulk perspective) a $d$-dimensional CFT coupled to the semi-classical gravity has been described as dual to a $(d+ 1)$-dimensional gravitational theory. The island formula may then be obtained through the RT prescription \cite{Ryu:2006bv, Ryu:2006ef, Hubeny:2007xt} in the corresponding higher-dimensional bulk perspective. Furthermore the authors in \cite{Suzuki:2022xwv} observed an equivalence between a CFT coupled to the semi-classical gravity and a boundary conformal field theory (BCFT) through a combination of the AdS/BCFT correspondence \cite{Takayanagi:2011zk, Fujita:2011fp} and the braneworld holography \cite{Karch:2000ct, Karch:2000gx}. The bulk dual of this BCFT is then described by an AdS$_{d+1}$ spacetime truncated by a codimension one end-of-the-world (EOW) brane, which may be identified as the bulk perspective described earlier.

In the context of the Karch-Randall (KR) braneworld with a bulk black hole geometry, a lower dimensional black hole may be induced on the EOW brane if it intersects the horizon. In the lower dimensional effective description, the CFT on the half line serves as a radiation bath for such black hole on the KR brane. Usually the BCFT is defined on a flat background\footnote{For a detailed review see \cite{Raju:2020smc} and references therein.} leading to a non gravitating radiation bath in the lower dimensional effective description implying the existence of massive gravitons \cite{Geng:2020qvw}. Interestingly in this connection the authors in \cite{Geng:2021mic, Geng:2022dua} considered a specific KR braneworld model in which the holographic BCFT$_d$ is defined on an eternal AdS$_d$ black hole background. The bulk dual geometry is then described by an AdS$_{d+1}$ black string truncated by a KR brane. In the bulk geometry each AdS$_d$ foliation of the AdS$_{d+1}$ black string may support an eternal AdS$_d$ black hole and the asymptotic boundary where the BCFT$_d$ lives is one such eternal AdS$_d$ slice. The authors in \cite{Geng:2021mic, Geng:2022dua} computed the EE for the Hawking radiation and obtained the corresponding Page curves.

On a separate note, it is well known in quantum information theory that the EE is not suitable for the characterization of mixed state entanglement. To address this issue several computable mixed state entanglement and correlation measures such as the entanglement negativity \cite{Vidal:2002zz, Plenio:2005cwa, Calabrese:2012ew, Calabrese:2012nk, Calabrese:2014yza}, reflected entropy \cite{Dutta:2019gen, Jeong:2019xdr}, entanglement of purification \cite{Horodecki:EoP, Takayanagi:2017knl}, odd entanglement entropy \cite{Tamaoka:2018ned}, balance partial entanglement \cite{Wen:2021qgx} have been proposed in the literature. These measures have also been investigated in the context of the AdS/BCFT scenarios \cite{Li:2020ceg, Li:2021dmf, Shao:2022gpg, BasakKumar:2022stg, Basu:2022reu, Afrasiar:2022ebi, Afrasiar:2022fid, Afrasiar:2023jrj, Basu:2023wmv, Kumari:2023ops}.

In this article, we focus on one such mixed state correlation measure termed the reflected entropy which involves the canonically purification of the mixed state under consideration and is bounded from below by the mutual information \cite{Dutta:2019gen}. The holographic reflected entropy was shown to be described by the bulk entanglement wedge cross section (EWCS) \cite{Dutta:2019gen}. Following this the authors in \cite{Hayden:2021gno} provided a stricter lower bound in terms of the “Markov gap” given by the difference between the holographic reflected entropy and the mutual information. It was demonstrated that the Markov gap could be interpreted geometrically in terms of the number of non-trivial boundaries of the bulk EWCS. Recently, an island formulation was proposed for the reflected entropy in \cite{Chandrasekaran:2020qtn, Li:2020ceg}. Subsequently, the authors in \cite{Vardhan:2021mdy, Akers:2022max} have obtained the analogues of the Page curve for the reflected entropy. Although the mixed state entanglement for the Hawking radiation in non-gravitating baths have been studied extensively \cite{Li:2020ceg, Li:2021dmf, Ling:2021vxe, Chandrasekaran:2020qtn, Afrasiar:2022ebi, Afrasiar:2022fid, Afrasiar:2023jrj, Basu:2022reu, Basu:2023wmv, KumarBasak:2020ams}, the same has not received significant attention for the cases involving a gravitating radiation bath. %Hence, it is worth considering the gravitating bath scenario and exploring the reflected entropy for those cases in order to investigate the mixed state entanglement structure.

In the above context, the investigation of the mixed state entanglement for Hawking radiation in gravitating baths as described in the aforementioned KR braneworld model, is a significant open issue. In the present work, we investigate the same by obtaining the reflected entropy for bipartite mixed states in a BCFT$_2$ defined on an eternal AdS$_2$ black hole background which serves as a gravitating radiation bath from the effective lower dimensional perspective.
In order to obtain the reflected entropy, first it is required to determine the EE phases for the particular mixed state configuration. Subsequently within each EE phase, we investigate various phases of the reflected entropy upon utilizing the replica technique in the large central charge limit. We also obtain the EWCS in the dual bulk black string geometry, which exactly reproduces the field theory results. Furthermore, we plot the Page curve for the EE and, within each EE phase, investigate the behaviour of the various phases of the reflected entropy with time. We then plot the holographic mutual information for bipartite mixed state configurations in the BCFT$_2$ and compare the results with the reflected entropy to investigate Markov gap \cite{Hayden:2021gno, Lu:2022cgq}.

This article is organized as follows. In \cref{Review}, we review some earlier works relevant to our computations. We start with the review of holographic BCFT$_2$ located on an eternal AdS$_2$ black hole background described in \cite{Geng:2022dua} and briefly explain the computation of the EE in this braneworld model. Subsequently, we brief recapitulate the definition of the reflected entropy, Markov gap and their holographic characterization. Following this in \cref{Two disjoint subsystems} and \cref{Two adjacent subsystems}, we describe the computation of the different possible phases for reflected entropy of bipartite mixed state configurations involving disjoint and adjacent subsystems and their corresponding bulk EWCS for different phases. Furthermore we obtain the Page curve for the reflected entropy and investigate the Markov gap. Finally in \cref{Summary} we summarize our results and discuss future directions.

\section{Review}\label{Review}

In this section we briefly review the salient features of a BCFT$_2$ on a AdS$_2$ black hole background described in \cite{Geng:2021mic, Geng:2022dua}. We also review the computation of the subregion EE for various two-sided bipartition in this setup. Subsequently we provide the definition of the reflected entropy and the replica technique for its computation in $CFT_2s$ and  then describe the holographic reflected entropy in terms of the bulk minimal EWCS \cite{Takayanagi:2017knl} as described in \cite{Dutta:2019gen}. Furthermore, we also review the issue of the Markov gap \cite{Hayden:2021gno} described as the difference between the reflected entropy and the holographic mutual information.

\subsection{Holographic BCFT$_2$ in a black hole background}\label{BCFT in BH}
\begin{figure}[h!]
	\centering
	\includegraphics[scale=1]{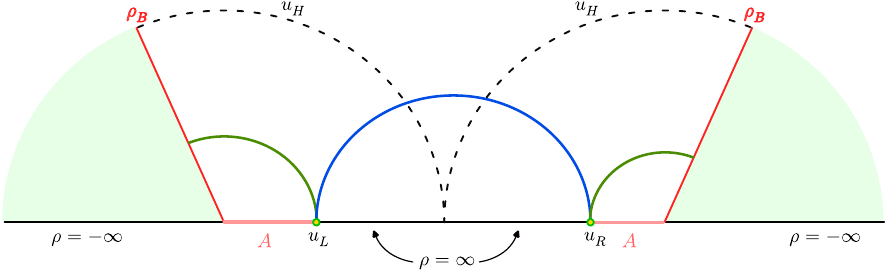}
	\caption{\it A pictorial representation of the black string geometry. A KR brane is inserted at constant angular coordinate $\rho=\rho_B$ shown in red colour. The black string horizon forms the dashed arc at the coordinate $u=u_h$. Here we have identified the horizon from both sides of the TFD. The blue curve is a Hartman-Maldacena surface and the green curves are island surfaces. Figure modified from \cite{Geng:2022dua}}
	\label{black string geometry}
\end{figure}

The authors in \cite{Geng:2021mic, Geng:2022dua} considered a holographic BCFT$_2$ on a AdS$_2$ black hole background. The bulk dual is an AdS$_3$  black string geometry truncated by a Karch-Randall brane, which is described by the metric 
\begin{align}\label{black string metric}
	ds^2= \cosh^2\rho \left[-\frac{\left(1-\frac{u}{u_h}\right)}{u^2}dt^2+ \frac{du^2}{u^2\left(1-\frac{u}{u_h}\right)}\right]+d\rho^2,
\end{align} 
where $\rho \in [-\infty, \infty]$ and $\rho= -\infty \cup \infty$ is the asymptotic boundary and the KR brane is embedded at a constant $\rho= \rho^{}_B$ slice. The accessible bulk region then extends from $\rho=\rho^{}_B$ to $\rho= \infty$ as depicted in \cref{black string geometry}. The geometry on each constant $\rho$ slice of the bulk black string is an eternal AdS$_2$ black hole which has two asymptotic boundaries. From the AdS$_3$/BCFT$_2$ correspondence, the dual field theory is a BCFT$_2$ on an AdS$_2$ black hole background with conformal boundary conditions at $u=0$. The bulk geometry may be embedded as a codimension one submanifold in $\mathbb{R}^{2,2}$, 
\begin{align}
	ds^2=\eta_{AB} dX^{A} dX^{B}, \quad \quad \eta_{AB}=\text{diag}(-1,-1,1,1),
\end{align}
with the following embedding equation
\begin{align}
	X^{}_A X^{A}=-1.
\end{align}
The metric described by \cref{black string metric} may be obtained by utilizing the following parametrization 
\begin{align}\label{embedding coordinate}
	&X_0=\frac{2 u_h-u}{u}\cosh\rho , ~~~~~~~~~~~~~~~~~~~~~~~~~~ X_1= \frac{2 \sqrt{u_h^{2}-u u_h}}{u}\sinh\frac{2\pi t}{\beta}\cosh\rho ,
	\notag\\
	&X_2= \frac{2 \sqrt{u_h^{2}-u u_h}}{u}\cosh\frac{2\pi t}{\beta}\cosh\rho ,~~~~~~~ X_3= \sinh\rho,
\end{align}
where $\beta=4\pi u_h$ is the inverse Hawking temperature. These embedding coordinates may be utilized to compute the holographic EE for the two sided bipartition in this model \cite{Geng:2022dua}.

\subsection{Entanglement entropy} 
In this subsection, we will briefly review the computation of the EE between a subsystem $A= (0^{}_L,u^{}_{L}) \cup (0^{}_R,u^{}_{R})$ and its complement in the above setup both from the field theory and the bulk perspective.

\subsubsection{Field theory computation}  Utilizing the replica trick, the computation of the EE between the subsystem $A$ and it's complement is equivalent to computing a two point function of the twist fields $\Phi_n$ inserted at the two bipartition points $u^{}_{L}$ and $u^{}_{R}$ as
\begin{align}\label{def EE}
	S_{A}=\lim_{n\rightarrow1}\frac{1}{1-n}\log\langle\Phi_{n}(u^{}_{L},t^{}_{L})\Phi_{n}(u^{}_{R},t^{}_{R})\rangle.
\end{align}
Here for the left bipartition, time coordinate is $t^{}_{L}=t$ while for the right bipartition  the time coordinate is given as $\tilde{t}=t^{}_{R} \to -t+\frac{i \beta}{2}$. The metric at the asymptotic boundary of the bulk black string is given as
\begin{align}
	ds^2= -\frac{1}{u^2}\left(1-\frac{u}{u_h}\right)dt^2+ \frac{du^2}{u^2\left(1-\frac{u}{u_h}\right)} \, ,
\end{align}
which is an AdS$_2$ planar black hole and we have two copies of such geometry corresponding to the two asymptotic boundaries of \cref{black string metric}. The BCFT$_2$ is then located in the AdS$_2$ black hole background and is not conformally flat, hence the computation of the twist field correlator in \cref{def EE} is not straightforward. It is necessary to map this field theory on the curved geometry to that on a flat background by using the following series of conformal transformations
\begin{align}\label{w-z map}
	\omega= u_h e^{\frac{u_\star+i \tau}{2 u_h}}, \quad u_\star = -u_h \log\left(1-\frac{u}{u_h}\right).
\end{align}
The metric in these new coordinates becomes conformally flat,
\begin{align}\label{bulk geometry}
	ds^2= 4 \Omega(u_\star)^2 e^{\frac{u_\star}{u_h}} d\omega d\bar{\omega},
\end{align}	
where $\Omega(u_\star(u))= \frac{1}{u}\left(1-\frac{u}{u_h}\right)$ is the conformal factor. The conformal boundary is then mapped to the circle $\omega \bar{\omega}=u^{2}_h$. It is possible to further conformally map this geometry to the upper-half-plane (UHP) through the transformations
\begin{equation}\label{pltoUHP}
	\omega=\frac{u_h}{v-\frac{i}{2}}-i u_h,
\end{equation}
where the conformal boundary is mapped to the real axis $v-\bar{v}=0$ and the metric transforms to
\begin{equation}\label{boundary geometry}
	ds^{2}=4\Omega(u_\star)^{2}u^{2}_h e^{-\frac{u_\star}{u_{H}}}\Big(e^{\frac{u_\star+i \tau}{2u_{H}}}+i\Big)^{2}\Big(e^{\frac{{u_\star-i \tau}}{2u_{H}}}-i\Big)^{2}dvd\bar{v}.
\end{equation}
Now, by utilizing the metrics and conformal factor described in \cref{bulk geometry,boundary geometry} it is possible to obtain the EE of a subsystem $A$ in the boundary and the bulk channels as \cite{Geng:2022dua}
\begin{equation}\label{Entropy}
	\begin{aligned}
		\text{$S_{A} =\left\{
			\begin{array}{ll}
				\frac{c}{3}\log\left(\frac{2}{\epsilon}\right) + 2\log g_{b}~& \text{Boundary channel},\\
				\frac{c}{6}\log\Bigg[\frac{u_h}{u^{}_Lu^{}_R}\left(\Delta_L+\Delta_R+2\sqrt{\Delta_L\Delta_R}\cosh\frac{t}{u_h}\right)\Bigg]+\frac{c}{3}\log\left(\frac{2}{\epsilon}\right)~&\text{Bulk channel}. 
			\end{array}\right.$}
	\end{aligned}
\end{equation}
where $\log g_b\equiv S_{\text{bdy}}$ is the usual boundary entropy in AdS$_3$/BCFT$_2$ and $\Delta_{ L}=u_h-u^{}_{L}$, $\Delta_{ R}=u_h-u^{}_{R}$ and $\epsilon$ is UV cut-off.

\subsubsection{The holographic calculation of EE} 
On the gravity side, the calculation of the EE for a boundary subsystem $A$ is done by using the Ryu-Takayanagi (RT) formula \cite{Ryu:2006bv}.  In this model, there are two types of RT surfaces: the Island surface, which connects the boundaries of the subsystems to nearby branes, and the Hartman-Maldacena (HM) surface, which passes through the interior of the black string to connect the boundaries of the subsystems. The island and HM surfaces are shown as solid green and blue curves in \cref{black string geometry} respectively. 
%\begin{align}$(X_0,X_1,X_2,X_3)$ and $(X^{'}_{0},X^{'}_{1},X^{'}_{2},X^{'}_{3})$\label{HEE}
%S^{}_{A}= \frac{1}{4 G_3} \min(L_{HM}, L_{Is}).\cosh^{-1}(X_0 X^{'}_{0}+X_1 X^{'}_{1}-X_2 X^{'}_{2}-X_3 X^{'}_{3})
%\end{align}
In the embedding space formalism, the geodesic length between the points $X^{A}_{1}=(t_1,u_1,\rho_1)$ and $X^{A}_{2}=(t_2,u_2,\rho_2)$ is given as
\begin{align}\label{Geodesic length}
	L&=  \cosh ^{-1}\Bigg[\left(\frac{(u_h+\Delta_1) (u_h+\Delta_2)-4 u_h \sqrt{\Delta_1 \Delta_2} \cosh \left(\frac{2 \pi  (t_1-t_2)}{\beta}\right)}{u^{}_1 u^{}_2}\right)\cosh\rho_1 \cosh\rho_2\notag\\
	& \quad \quad \quad \quad \quad \quad \quad\quad\quad\quad\quad\quad \quad\quad\quad\quad\quad\quad ~~~~~~~~~~~~~~~~~~~~~~~~~~~~~~~~~
	 -\sinh \rho_1 \sinh \rho_2\Bigg].
\end{align}
%\begin{align}
%	L&=  \cosh ^{-1}\Bigg[\left(\frac{(u_h+\Delta_1) (u_h+\Delta_2)-4 u_h \sqrt{\Delta_1 \Delta_2} \cosh \left(\frac{2 \pi  (t_1-t_2)}{\beta}\right)}{u^{}_1 u^{}_2}\right)cosh\rho_1 \cosh\rho_2\notag\\
%	& ~~~~~~~~~~~~~~~
%	-\sinh \rho_1 \sinh \rho_2\Bigg].
%\end{align}
where $\Delta_{ 1}=u_h-u^{}_{1}$ and $\Delta_{ 2}=u_h-u^{}_{2}$.

\subsubsection*{Length of island surface:} The island surface connects the bipartition points to the nearby branes, shown as solid green curves in \cref{black string geometry}. In the left copy of thermofield double (TFD) the coordinates of the end points of the island surface are $(u^{}_L,\rho_\epsilon,t)$ and $(u^{}_B,\rho^{}_B,t)$ where  $\rho_{\epsilon}$ describes the regularized asymptotic boundary and $u^{}_B$ is a dynamical point on the brane. Now by utilizing these coordinates in \cref{Geodesic length}, the length of the island surface may be obtained as
\begin{align}
	L_\text{Is}= \log \left[\left(\frac{2u_h(\sqrt{\Delta_{L}}-\sqrt{\Delta_{B}})^2 +u^{}_{L} u^{}_{B}}{u^{}_{L} u^{}_{B}}\right)\cosh \rho^{}_B-\sinh \rho^{}_B\right]+\rho_\epsilon.
\end{align}
The extremal length of the island surface may be obtained by extremizing the above expression over $u^{}_B$ as follows
\begin{align}\label{length of island}
	L_\text{Is}= \rho_\epsilon-\rho^{}_B.
\end{align}
Finally adding the contribution from the right TFD copy, the EE may be obtained by using RT formula.

\subsubsection*{Length of the HM surface:} 
Similar to the previous case, the length of the HM surface may be obtained by utilizing the embedding coordinates of the left and right bipartition in \cref{Geodesic length} as 
\begin{align}\label{length of HM}
	L_\text{HM}= \log\left[\frac{e^{2 \rho_\epsilon}u_h}{u^{}_L u^{}_R}\left(\Delta_L+\Delta_R+ 2 \sqrt{\Delta_L \Delta_R} \cosh\frac{t}{u_h}\right)\right].
\end{align}
Note that by using the following identifications,
\begin{equation}\label{bulk-bdy relation}
	\log g_{b}=-\frac{c}{6}\rho^{}_{B}\,,\quad \epsilon=2e^{-\rho_{\epsilon}},
\end{equation}
and the Brown-Henneaux formula \cite{Brown:1986nw}, it can be shown that the field theory results in \cref{Entropy} exactly match with the holographic EE obtained by utilizing \cref{length of island} and \cref{length of HM}. Here the first expression in \cref{bulk-bdy relation} describes the relation between the boundary entropy $S_\text{bdy}$ in AdS$_3$/BCFT$_2$ \cite{Fujita:2011fp} and the brane tension $\rho^{}_B$ and the second relation is a matching between the UV and IR cutoffs in the BCFT and the bulk.

\subsection{Reflected entropy in CFT$_2$}
In this subsection we briefly review the reflected entropy and its computation in CFT$_2$ as described in \cite{Dutta:2019gen}. Let us consider a bipartite quantum system $A \cup B$ in a mixed state $\rho_{AB}$. The canonical purification of this state involves the doubling of its Hilbert space to define a pure state $\ket{\sqrt{\rho_{AB}}}_{A B A^* B^*}$. 
\begin{comment}
\begin{equation} 
	\rho^{}_{AB} = \text{Tr}_{A^*B^*} \ket{\sqrt{\rho^{}_{AB}}} \bra{\sqrt{\rho^{}_{AB}}}.	
\end{equation}
\end{comment}
The reflected entropy $S_{R}(A:B)$ for the bipartite mixed state is defined as the von Neumann entropy of the reduced density matrix $\rho_{AA^{*}}$ as follows 
\begin{equation}\label{def.}
	S_{R}(A:B) = S_{vN} (\rho^{}_{AA^{*}})_{\sqrt{\rho^{}_{AB}}} \,\,\,,
\end{equation}
where $\rho_{AA^{*}}$ may be obtained by tracing out the degree of freedom of $B$ and $B^*$ from the density matrix $\ket{\sqrt{\rho^{}_{AB}}}\bra{\sqrt{\rho^{}_{AB}}}$.
The authors in \cite{Dutta:2019gen} developed a novel replica technique to compute the reflected entropy between two disjoint subsystems $A \equiv[z_1, z_2]$ and $B \equiv [z_3, z_4]$ in a CFT$_2$. The reflected entropy may then be obtained in terms of a four-point twist field correlator as follows
\begin{equation}\label{defination}
	S_R(A:B)=\lim_{n,m \to1}S_{n}(AA^{*})_{\psi_{m}}=\lim_{n,m \to1}\frac{1}{1-n} \log \frac{\left<\sigma_{g_{A}}(z_{1})\sigma_{g_{A}^{-1}}(z_{2})\sigma_{g_{B}}(z_{3})\sigma_{g_{B}^{-1}}(z_{4})\right>_{\text{CFT}^{\otimes mn}}}{\left<\sigma_{g_{m}}(z_{1})\sigma_{g_{m}^{-1}}(z_{2})\sigma_{g_{m}}(z_{3})\sigma_{g_{m}^{-1}}(z_{4})\right>^{n}_{\text{CFT}^{\otimes m}}} \, ,
\end{equation}
where $m$, $n$ are the replica indices\footnote{As discussed in \cite{Kusuki:2019evw, Akers:2021pvd, Akers:2022max}, the two replica limits $n \to 1$ and $m \to 1$ are non-commuting. In this article, we compute the reflected entropy by first taking $n \to 1$ and subsequently $m \to 1$ as suggested in \cite{Kusuki:2019evw, Akers:2021pvd}.} and twist operators $\sigma_{g_A}$ and $\sigma_{g_B}$ are inserted at the end points of the subsystems. The conformal dimensions of the operators $\sigma_{g^{}_A}, \sigma_{g_B}$ and $\sigma_{g_{m}}$ are given as \cite{Dutta:2019gen}
\begin{equation}\label{conformal weights}
	h\equiv h_{A}=h_{B}=\frac{n c}{24} \left(m-\frac{1}{m}\right), \quad  h_{m}= \frac{c}{24}\left(m-\frac{1}{m}\right), \quad  h_{AB}= \frac{2c}{24}\left(n-\frac{1}{n}\right).
\end{equation} 
It was proved in \cite{Dutta:2019gen} that the reflected entropy for a bipartite state in a CFT$_{d}$ in the large central charge limit is dual to twice the minimal EWCS for the bulk static AdS$_{d+1}$ geometry. In the next subsection we review the EWCS which describes bulk dual of the reflected entropy.

\subsection{Entanglement wedge cross section }
The bulk dual of the density matrix $\rho_{AB}$ is described by the entanglement wedge $M_{AB}$ \cite{Czech:2012bh} which is the region enclosed by the subsystems $A \cup B$ and the codimension two bulk minimal surface $\gamma_{AB}$ homologous to the subsystem $A \cup B$. The entanglement wedge cross section $E_W$ is then defined as the minimum cross sectional area of the entanglement wedge \cite{Takayanagi:2017knl}.
\begin{figure}[H]
	\centering
	\includegraphics[scale=.57]{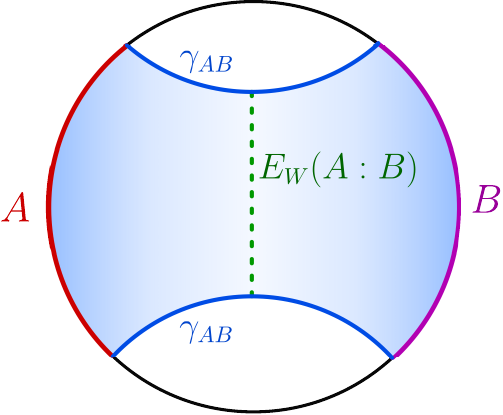}
	\caption{\it The light green region represents the entanglement wedge and dashed green line is the EWCS of subsystem $A\cup B$.}
	\label{EW}
\end{figure}
In the context of AdS$_3$/CFT$_2$, the EWCS for two disjoint subsystems $A=[X_1,X_2]$ and $B=[X_3,X_4]$ in terms of the embedding coordinates may be obtained as \cite{Kusuki:2019evw}
\begin{align}\label{Dis.EWCS}
	E_W=
	\frac{1}{4G_N} \cosh^{-1}\left(\frac{1+\sqrt{u}}{\sqrt{v}}\right),
\end{align}
where $u$ and $v$ are defined in terms of $\xi_{ij}=-X_i\cdot X_j$ as,
\begin{align}
	u=\frac{\xi_{12}\xi_{34}}{\xi_{13}\xi_{24}},\; v=\frac{\xi_{14}\xi_{23}}{\xi_{13}\xi_{24}}.
\end{align}
The length between a point $X_2$ and a spacelike geodesic connecting points $X_1$, $X_3$ may be utilized to obtained the EWCS for two adjacent subsystems as follows 
\begin{align}\label{EWCS three pt. formula}
	E_W= \frac{1}{4 G_N}\cosh ^{-1}\left(\sqrt{\frac{2 \xi_{12}\xi_{23}}{\xi_{13}}}\right).
\end{align}
The detailed derivation of this formula is described in appendix \ref{Appendix}.
%Now for two disjoint subsystems, the EWCS is given by geodesic length between two spacelike geodesics $(X_{14}(\lambda), X_{23}(\lambda^\prime))$ anchored on the boundary points. This length may be obtained from \cref{geodesic length}. Now by extremizing this, we may obtain the corresponding EWCS as. Now computing the geodesic length from \cref{geodesic length} and extremizing it 

%This holographic duality is given as follows 
%\begin{align}\label{duality}
%	S_R(A:B) = 2 E_W(A:B).
%\end{align}

\subsection{Markov gap}
In \cite{Hayden:2021gno} authors have shown that the difference between the holographic reflected entropy and holographic mutual information, termed the Markov gap, may be understood geometrically  in terms of the number of non-trivial boundaries of the EWCS. In the context of $AdS_3/CFT_2$, it was shown that 
\begin{align}\label{Markov gap}
	S_{R}(A: B)-I(A: B) \geq \frac{\log (2) \ell_{\text {AdS }}}{2 G_{N}} \times(\# \text { of boundaries of EWCS})+\mathcal{O}\left(\frac{1}{G_{N}}\right).
\end{align}
%The difference between the reflected entropy and mutual information is equal to conditional mutual information. 
It has been demonstrated that the Markov gap is bounded by the fidelity of a Markov recovery process related to the purification of the mixed state under consideration. For a perfect Markov recovery process, the Markov gap vanishes.

In the following sections, we first compute the reflected entropy for various bipartite states involving two disjoint and  adjacent subsystems in BCFT$_2$ located on a black hole background. We also explain the computation of the bulk EWCS in the context of the black string geometry which exactly reproduce the field theory results.

\section{Holographic reflected entropy: Disjoint subsystems}\label{Two disjoint subsystems}
In this section, we compute the reflected entropy and the bulk EWCS for two disjoint subsystems $A\equiv(u^{}_{L_1},u^{}_{L_2})\cup (u^{}_{R_1},u^{}_{R_2})$ and $B\equiv(u^{}_{L_3},u^{}_{L_4})\cup (u^{}_{R_3},u^{}_{R_4})$ in the AdS$_3$/BCFT$_2$ setup described in \cref{BCFT in BH} where the BCFT$_2$ is defined on an AdS$_2$ black hole background. Here we considered $u_L$ and $u_R$ to be asymmetric. The R\'enyi reflected entropy in this scenario may then be obtained in terms of the twist field correlators as follows
\begin{align}\label{Ref-dis-def}
	&S_n(AA^*)_{\psi_{m}}\notag\\
	&=\log \frac{\langle \sigma_{g^{}_A}(u^{}_{L_{1}})\sigma_{g_A^{-1}}(u^{}_{L_{2}})\sigma_{g^{}_B}(u^{}_{L_{3}})\sigma_{g^{-1}_B}(u^{}_{L_{4}})\sigma_{g^{}_B}(u^{}_{R_{4}})\sigma_{g_B^{-1}}(u^{}_{R_{3}})\sigma_{g^{}_A}(u^{}_{R_{2}})\sigma_{g_A^{-1}}(u^{}_{R_{1}}) \rangle_{\mathrm{BCFT}^{\bigotimes mn}}}{\langle \sigma_{g_m}(u^{}_{L_{1}})\sigma_{g_m^{-1}}(u^{}_{L_{2}})\sigma_{g_m}(u^{}_{L_{3}})\sigma_{g^{-1}_m}(u^{}_{L_{4}})\sigma_{g_m}(u^{}_{R_{4}})\sigma_{g_m^{-1}}(u^{}_{R_{3}})\sigma_{g_m}(u^{}_{R_{2}})\sigma_{g_m^{-1}}(u^{}_{R_{1}}) \rangle^{n}_{\mathrm{BCFT}^{\bigotimes m}}}.
\end{align}
To compute the reflected entropy and the bulk EWCS, it is first required to determine the EE phases for the two disjoint subsystems under consideration. In the following we demonstrate four possible phases of the EE depending on the subsystem size and its location. In what follows we describe the computation of the reflected entropy and the bulk EWCS for these EE phases and show that they match verifying the holographic duality mentioned earlier.

\subsection{Entanglement entropy phase 1}\label{Dis.EE phase1}
\begin{figure}[h!]
	\centering
	\includegraphics[scale=0.8]{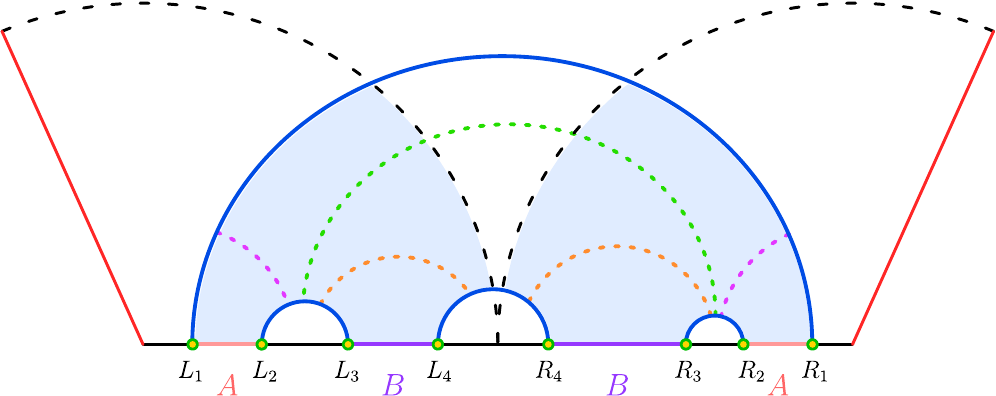}
	\caption{\it Schematic depicting the different phases of the EWCS (represented by various colored dashed curves) between subsystems $A$ and $B$ when the RT surface for $A \cup B$ is shown by the solid blue curves. } 
	\label{fig_dis.EE phase1}
\end{figure} 
In the first EE phase the subsystems $A\cup B$ are considered to be large and far away from the boundary. Hence the EE for the two disjoint subsystems $A$ and $B$ is proportional to the sum of the lengths of the two HM surfaces corresponding to the points $L_1(u^{}_{L_1},\rho_{\epsilon},t)$, $R_1(u^{}_{R_1},\rho_{\epsilon},\tilde{t})$ and $L_4(u^{}_{L_4},\rho_{\epsilon},t)$, $R_4(u^{}_{R_4},\rho_{\epsilon},\tilde{t})$ and two dome-type RT surfaces, shown as solid blue curves in \cref{fig_dis.EE phase1}. Note that $\tilde{t}$ is the time coordinate for the right TFD copy. Now by utilizing \cref{Geodesic length}, the geodesic length for the dome-type RT surface connecting points $L_2(u^{}_{L_2},\rho_{\epsilon},t)$, $L_3(u^{}_{L_3},\rho_{\epsilon},t)$, may be written as
\begin{align}\label{dome length}
	L= \log\left(\frac{u_h}{u^{}_{L_{2}}u^{}_{L_{3}}}(\sqrt{\Delta_{L_2}}-\sqrt{\Delta_{L_3}})^2
	\right)+ 2 \rho_\epsilon.
\end{align}
Utilizing \cref{length of HM} and \cref{dome length} and adding the contribution for the right TFD copy, the EE for this configuration is given as
\begin{align}
	S_1=&\frac{1}{4 G_N} \log \left[\frac{u^{2}_h \left(\Delta_{L_1}+\Delta_{R_1}+2 \sqrt{\Delta_{L_1}\Delta_{R_1}} \cosh \frac{t}{u_h}\right)\left(\Delta_{L_4}+\Delta_{R_4}+2 \sqrt{\Delta_{L_4}\Delta_{R_4}} \cosh \frac{t}{u_h}\right)}{u^{}_{L_1}u^{}_{R_1}u^{}_{L_4} u^{}_{R_4}}\right]\notag\\
	&+\frac{1}{4 G_N} \log \left[\frac{u^{2}_h}{u^{}_{L_2} u^{}_{L_3} u^{}_{R_2} u^{}_{R_3}}(\sqrt{\Delta_{L_2}}-\sqrt{\Delta_{L_3}})^2 (\sqrt{\Delta_{R_2}}-\sqrt{\Delta_{R_3}})^2\right]+\frac{2\rho_\epsilon }{G_N}.
\end{align}
For this EE phase, we observe three distinct phases of the reflected entropy and the corresponding bulk EWCS, depicted as dashed curves in \cref{fig_dis.EE phase1}. 
Here we consider both the subsystems are far away from the boundary, hence
the OPE channel for the BCFT$_2$ correlator is favoured. In this channel, the BCFT$_2$ twist field correlators may be expressed as CFT$_2$ twist field correlators  \cite{Rozali:2019day,Li:2021dmf,Shao:2022gpg}. In the following, we describe the computation of the reflected entropy and the bulk EWCS for this EE phase.

\subsection*{Phase-I}\label{dis SR phase1(I)}
\paragraph{Reflected entropy:}
In this reflected entropy phase the subsystems are considered to be large and far away from the boundary, hence the numerator of \cref{Ref-dis-def} may be factorized into two two-point twist field correlators and one four-point twist field correlator as
\begin{align}\label{fator1-dis(1(I))}
	&\langle \sigma_{g^{}_A}(u^{}_{L_{1}})\sigma_{g_A^{-1}}(u^{}_{L_{2}})\sigma_{g^{}_B}(u^{}_{L_{3}})\sigma_{g^{-1}_B}(u^{}_{L_{4}})\sigma_{g^{}_B}(u^{}_{R_{4}})\sigma_{g_B^{-1}}(u^{}_{R_{3}})\sigma_{g^{}_A}(u^{}_{R_{2}})\sigma_{g_A^{-1}}(u^{}_{R_{1}}) \rangle_{\mathrm{CFT}^{\bigotimes mn}}\notag\\
	&\quad\quad\quad =\langle\sigma_{g^{}_A}(u^{}_{L_{1}})\sigma_{g_A^{-1}}(u^{}_{R_{1}})\rangle_{\mathrm{CFT}^{\bigotimes mn}}	\langle\sigma_{g^{-1}_B}(u^{}_{L_{4}})\sigma_{g^{}_B}(u^{}_{R_{4}})\rangle_{\mathrm{CFT}^{\bigotimes mn}}\notag\\	  &\quad\quad\quad\quad\quad\quad\quad\quad\quad \times \langle\sigma_{g_A^{-1}}(u^{}_{L_{2}})\sigma_{g^{}_B}(u^{}_{L_{3}})\sigma_{g_B^{-1}}(u^{}_{R_{3}})\sigma_{g^{}_A}(u^{}_{R_{2}})\rangle_{\mathrm{CFT}^{\bigotimes mn}}.
\end{align}
The denominator of \cref{Ref-dis-def} may also be factorized similarly. Now by utilizing \cref{Ref-dis-def,fator1-dis(1(I))}, the reflected entropy in this scenario may be given as  
\begin{align}
	S_R(A:B)= \lim_{{m,n} \to 1}\frac{1}{1-n}\log \frac{\langle\sigma_{g_A^{-1}}(u^{}_{L_{2}})\sigma_{g^{}_B}(u^{}_{L_{3}})\sigma_{g_B^{-1}}(u^{}_{R_{3}})\sigma_{g^{}_A}(u^{}_{R_{2}})\rangle_{\mathrm{CFT}^{\bigotimes mn}}}{\langle\sigma_{g_m^{-1}}(u^{}_{L_{2}})\sigma_{g_m}(u^{}_{L_{3}})\sigma_{g_m^{-1}}(u^{}_{R_{3}})\sigma_{g_m}(u^{}_{R_{2}})\rangle^{n}_{\mathrm{CFT}^{\bigotimes m}}}.
\end{align}
Note that the two point twist field correlator in the above expression cancels from the numerator and the denominator. Since the field theory is described on a AdS$_2$ black hole background, it is necessary to transform the above four-point twist correlator to the flat plane twist field correlator. This transformation is obtained through the conformal map given in \cref{w-z map}. Now by utilizing this map and the form of the four point function in the large central charge limit given in \cite{Dutta:2019gen,Fitzpatrick:2014vua}, the final expression for the reflected entropy of the two disjoint subsystems may be obtained as
\begin{align}\label{SR dis.Phase1(i)}
	S_R(A:B)=\frac{c}{6}\log\left[\frac{(1+\sqrt{\eta})(1+\sqrt{\bar{\eta}})}{(1-\sqrt{\eta})(1-\sqrt{\bar{\eta}})}\right],
\end{align}
where the cross ratios $\eta$ and $\bar{\eta}$ are defined as
\begin{align}\label{HM SR}
	&\eta= \frac{\left(\sqrt{\Delta_{L_2}}+\sqrt{\Delta_{R_2}} e^{t/u_h}\right) \left(\sqrt{\Delta_{L_3}}+\sqrt{\Delta_{R_3}} e^{t/u_h}\right)}{\left(\sqrt{\Delta_{L_2}}+\sqrt{\Delta_{R_3}} e^{t/u_h}\right) \left(\sqrt{\Delta_{L_3}}+\sqrt{\Delta_{R_2}} e^{t/u_h}\right)},\notag\\
	&\bar{\eta}=\frac{\left(\sqrt{\Delta_{R_2}}+\sqrt{\Delta_{L_2}} e^{t/^{}u_h}\right) \left(\sqrt{\Delta_{R_3}}+\sqrt{\Delta_{L_3}} e^{t/^{}u_h}\right)}{\left(\sqrt{\Delta_{R_3}}+\sqrt{\Delta_{L_2}} e^{t/^{}u_h}\right) \left(\sqrt{\Delta_{R_2}}+\sqrt{\Delta_{L_3}} e^{t/^{}u_h}\right)}.
\end{align}

\paragraph{EWCS:}The bulk EWCS for this phase is obtained by the length of the geodesic between the two dome-type RT surfaces which is depicted as the dashed green curve in \cref{fig_dis.EE phase1}. By employing the embedding coordinates provided in \cref{embedding coordinate} for points $L_2$, $L_3$ and $R_2$, $R_3$, we may calculate the bulk EWCS using \cref{Dis.EWCS}, with $u$ and $v$ given as   
\begin{align}\label{HM EWCS}
	&u=\frac{\left(\Delta _{L_2}+\Delta _{R_2}+2 \sqrt{\Delta _{L_2} \Delta _{R_2}} \cosh \frac{t}{u_h}\right) \left(\Delta _{L_3}+\Delta _{R_3}+2 \sqrt{\Delta _{L_3} \Delta _{R_3}} \cosh \frac{t}{u_h}\right)}{\left(\Delta _{L_2}+\Delta _{R_3}+2 \sqrt{\Delta _{L_2} \Delta _{R_3}} \cosh \frac{t}{u_h}\right) \left(\Delta _{L_3}+\Delta _{R_2}+2 \sqrt{\Delta _{L_3} \Delta _{R_2}} \cosh \frac{t}{u_h}\right)},\notag\\
	&v= \frac{\left(\sqrt{\Delta _{L_2}}-\sqrt{\Delta _{L_3}}\right){}^2 \left(\sqrt{\Delta _{R_2}}-\sqrt{\Delta _{R_3}}\right){}^2}{\left(\Delta _{L_2}+\Delta _{R_3}+2 \sqrt{\Delta _{L_2} \Delta _{R_3}} \cosh \frac{t}{u_h}\right) \left(\Delta _{L_3}+\Delta _{R_2}+2 \sqrt{\Delta _{L_3} \Delta _{R_2}} \cosh \frac{t}{u_h}\right)}.
\end{align}
Note that the reflected entropy computed in \cref{SR dis.Phase1(i)} exactly matches with twice the bulk EWCS upon utilizing the Brown-Henneaux relation. 
%This serves as a consistency check for the duality mentioned in \cref{duality}.

\subsection*{Phase-II}
\paragraph{Reflected entropy:} In this reflected entropy phase, we consider the subsystem $A$ to be smaller than $B$, so the numerator in \cref{Ref-dis-def} may be factorized into a two-point and a six-point twist correlator as
\begin{align}\label{factor dis phase1(ii)}
	&\langle \sigma_{g^{}_A}(u^{}_{L_{1}})\sigma_{g_A^{-1}}(u^{}_{L_{2}})\sigma_{g^{}_B}(u^{}_{L_{3}})\sigma_{g^{-1}_B}(u^{}_{L_{4}})\sigma_{g^{}_B}(u^{}_{R_{4}})\sigma_{g_B^{-1}}(u^{}_{R_{3}})\sigma_{g^{}_A}(u^{}_{R_{2}})\sigma_{g_A^{-1}}(u^{}_{R_{1}}) \rangle_{\mathrm{CFT}^{\bigotimes mn}}\notag\\
	&=\langle\sigma_{g^{-1}_B}(u^{}_{L_{4}})\sigma_{g^{}_B}(u^{}_{R_{4}})\rangle_{\mathrm{CFT}^{\bigotimes mn}} \langle\sigma_{g_A}(u^{}_{L_{1}})\sigma_{g_A^{-1}}(u^{}_{L_{2}})\sigma_{g_B}(u^{}_{L_{3}})\sigma_{g_B^{-1}}(u^{}_{R_{3}})\sigma_{g_A}(u^{}_{R_{2}})\sigma_{g^{-1}_A}(u^{}_{R_{1}})\rangle_{\mathrm{CFT}^{\bigotimes mn}}.
\end{align}
Here the six-point function may be expanded in terms of the conformal block $\mathcal{F}_6$ which factorizes into a product of two four-point conformal blocks $\mathcal{F}_4$ in the OPE channel (which is termed the $\Omega_1$ channel in \cite{Banerjee:2016qca}) as
\begin{align}\label{block factorization}
	\mathcal{F}_{6}(u^{}_{L_1},u^{}_{L_2},u^{}_{L_3},u^{}_{R_3},u^{}_{R_2},u^{}_{R_1};h,h_{AB})=&\mathcal{F}_{4}(u^{}_{L_1},u^{}_{L_2},u^{}_{L_3},u^{}_{R_1};h,h_{AB})\notag\\
	&\times \mathcal{F}_{4}(u^{}_{R_1},u^{}_{R_2},u^{}_{R_3},u^{}_{L_1};h,h_{AB}).
\end{align}
The dominant contribution in the above four point conformal block arises from the primary operator $\sigma_{g^{}_A g_B^{-1}}$ with conformal dimension $h_{AB}$, in the large central charge limit. Now by utilizing \cref{factor dis phase1(ii),block factorization,w-z map} and the form of the four point function conformal block in \cref{Ref-dis-def}, we may obtain the reflected entropy in this phase as
\begin{align}\label{SR dis.Phase1(ii)}
	S_R(A:B)=\frac{c}{6}\log\left[\frac{(1+\sqrt{\eta})(1+\sqrt{\bar{\eta}})}{(1-\sqrt{\eta})(1-\sqrt{\bar{\eta}})}\right]
	+\frac{c}{6}\log\left[\frac{(1+\sqrt{\xi})(1+\sqrt{\bar{\xi}})}{(1-\sqrt{\xi})(1-\sqrt{\bar{\xi}})}\right],
\end{align}
where $\eta$, $\bar{\eta}$ and $\xi$, $\bar{\xi}$ are cross ratios. The first two cross ratios $(\eta, \bar{\eta})$ are given as
\begin{align}
	\eta= \frac{\left(\sqrt{\Delta_{L_1}}-\sqrt{\Delta_{L_2}}\right) \left(\sqrt{\Delta_{L_3}}+\sqrt{\Delta_{R_1}} e^{t/u_h}\right)}{\left(\sqrt{\Delta_{L_1}}-\sqrt{\Delta_{L_3}}\right) \left(\sqrt{\Delta_{L_2}}+\sqrt{\Delta_{R_1}} e^{t/u_h}\right)},\,\, \bar{\eta}=\frac{\left(\sqrt{\Delta_{L_1}}-\sqrt{\Delta_{L_2}}\right) \left(\sqrt{\Delta_{R_1}}+\sqrt{\Delta_{L_3}} e^{t/u_h}\right)}{\left(\sqrt{\Delta_{L_1}}-\sqrt{\Delta_{L_3}}\right) \left(\sqrt{\Delta_{R_1}}+\sqrt{\Delta_{L_2}} e^{t/u_h}\right)}.
\end{align}
The other cross ratios $\xi$ and $\bar{\xi}$ may be obtained by replacing $\Delta_{ L_i} \longleftrightarrow \Delta_{ R_i}$ in the expression of cross ratio $\eta$ and $\bar{\eta}$ respectively.

\paragraph{EWCS:} The bulk EWCS may be expressed as the sum of the lengths of the two geodesics which connect a dome-type RT surface to the HM surface on both the TFD copies and depicted as the dashed magenta curves in \cref{fig_dis.EE phase1}. The length of one of the geodesic may be obtained by employing the embedding coordinates for points $L_1$, $L_2$, $L_3$, and $R_1$ in \cref{Dis.EWCS}, with $u$ and $v$ given as
\begin{align}\label{EW dis phase1(ii)}
	&u= \left(\frac{\sqrt{\Delta_ {L_1}}-\sqrt{\Delta_{L_2}}}{\sqrt{\Delta_{L_1}}-\sqrt{\Delta_{ L_3}}}\right)^2 \left(\frac{\Delta_{L_3}+\Delta_{R_1}+2 \sqrt{\Delta_{L_3}\Delta_{R_1}}\cosh\frac{t}{u_h}}{\Delta_{L_2}+\Delta_{R_1}+2 \sqrt{\Delta_{L_2}\Delta_{R_1}}\cosh\frac{t}{u_h}}\right),\notag\\
	&v=\left(\frac{\sqrt{\Delta_{L_2}}-\sqrt{\Delta_{L_3}}}{\sqrt{\Delta_{ L_1}}-\sqrt{\Delta_{L_3}}}\right)^2 \left(\frac{\Delta_{L_1}+\Delta_{R_1}+2 \sqrt{\Delta_{L_1}\Delta_{R_1}}\cosh\frac{t}{u_h}}{\Delta_{L_2}+\Delta_{R_1}+2 \sqrt{\Delta_{L_2}\Delta_{R_1}}\cosh\frac{t}{u_h}}\right).
\end{align}  
For the right TFD copy the length of the other geodesic may be obtained by using the embedding coordinates for points $R_1$, $R_2$, $R_3$ and $L_1$ in \cref{Dis.EWCS}. The corresponding ratios $u$ and $v$ may be obtained from \cref{EW dis phase1(ii)} by replacing $\Delta_{ L_i} \longleftrightarrow \Delta_{ R_i}$. The final expression for the corresponding bulk EWCS is given by the sum of the lengths of the two geodesics. Upon utilization of the Brown-Henneaux relation, we find that the reflected entropy computed in \cref{SR dis.Phase1(ii)} exactly matches with twice the bulk EWCS. 

%This provides a consistency check for the duality mentioned earlier.

\subsection*{Phase-III}
\paragraph{Reflected entropy:} In this reflected entropy phase, we assume that the subsystem $B$ is smaller than $A$, hence the numerator of \cref{Ref-dis-def} may be factored into a two-point twist correlator and a six-point twist correlator as follows
\begin{align}\label{factor dis phase1(iii)}
	&\langle \sigma_{g^{}_A}(u^{}_{L_{1}})\sigma_{g_A^{-1}}(u^{}_{L_{2}})\sigma_{g^{}_B}(u^{}_{L_{3}})\sigma_{g^{-1}_B}(u^{}_{L_{4}})\sigma_{g^{}_B}(u^{}_{R_{4}})\sigma_{g_B^{-1}}(u^{}_{R_{3}})\sigma_{g^{}_A}(u^{}_{R_{2}})\sigma_{g_A^{-1}}(u^{}_{R_{1}}) \rangle_{\mathrm{CFT}^{\bigotimes mn}}\notag\\
	&=\langle\sigma_{g^{}_A}(u^{}_{L_{1}})\sigma_{g^{-1}_A}(u^{}_{R_{1}})\rangle_{\mathrm{CFT}^{\bigotimes mn}} \langle\sigma_{g^{-1}_A}(u^{}_{L_{2}})\sigma_{g^{}_B}(u^{}_{L_{3}})\sigma_{g^{-1}_B}(u^{}_{L_{4}})\sigma_{g^{}_B}(u^{}_{R_{4}})\sigma_{g_B^{-1}}(u^{}_{R_{3}})\sigma_{g^{}_A}(u^{}_{R_{2}})\rangle_{\mathrm{CFT}^{\bigotimes mn}}
\end{align}
Here, we may expand the six-point function in terms of the conformal block $\mathcal{F}_6$, which, in OPE channel, may be factorized as the product of two four-point conformal blocks $\mathcal{F}_4$ as detailed in \cite{Banerjee:2016qca}
\begin{align}\label{block factorization2}
	\mathcal{F}^{}_{6}(u^{}_{L_2},u^{}_{L_3},u^{}_{L_4},u^{}_{R_4},u^{}_{R_3},u^{}_{R_2};h,h_{AB})=&\mathcal{F}_{4}(u^{}_{L_2},u^{}_{L_3},u^{}_{L_4},u^{}_{R_4};h,h_{AB})\notag\\
	&\times \mathcal{F}_{4}(u^{}_{R_2},u^{}_{R_3},u^{}_{R_4},u^{}_{L_4};h,h_{AB}).
\end{align}
Now by utilizing \cref{factor dis phase1(iii),block factorization2,w-z map} and the conformal block in \cref{Ref-dis-def}, we may obtain the expression for the reflected entropy identical to \cref{SR dis.Phase1(ii)} with the cross ratios $\eta, \bar{\eta}$ defined as follows
\begin{comment}
	\begin{align}\label{SR dis.Phase1(iii)}
		S_R(A:B)=\frac{c}{6}\log\left[\frac{(1+\sqrt{\eta})(1+\sqrt{\bar{\eta}})}{(1-\sqrt{\eta})(1-\sqrt{\bar{\eta}})}\right]
		+\frac{c}{6}\log\left[\frac{(1+\sqrt{\xi})(1+\sqrt{\bar{\xi}})}{(1-\sqrt{\xi})(1-\sqrt{\bar{\xi}})}\right],
	\end{align}
\end{comment}
\begin{align}
	\eta= \frac{\left(\sqrt{\Delta_{L_2}}-\sqrt{\Delta_{L_4}}\right) \left(\sqrt{\Delta_{L_3}}+\sqrt{\Delta_{R_4}} e^{t/u_h}\right)}{\left(\sqrt{\Delta_{L_3}}-\sqrt{\Delta_{L_4}}\right) \left(\sqrt{\Delta_{L_2}}+\sqrt{\Delta_{R_4}} e^{t/u_h}\right)},~~ \bar{\eta}=\frac{\left(\sqrt{\Delta_{L_2}}-\sqrt{\Delta_{L_4}}\right) \left(\sqrt{\Delta_{R_4}}+\sqrt{\Delta_{L_3}} e^{t/u_h}\right)}{\left(\sqrt{\Delta_{L_3}}-\sqrt{\Delta_{L_4}}\right) \left(\sqrt{\Delta_{R_4}}+\sqrt{\Delta_{L_2}} e^{t/u_h}\right)}.
\end{align}
The other cross ratios $\xi$ and $\bar{\xi}$ may be obtained by replacing $\Delta_{ L_i} \longleftrightarrow \Delta_{ R_i}$ in the expression of cross ratio $\eta$ and $\bar{\eta}$ respectively.

\paragraph{EWCS:} The bulk EWCS for this phase is obtained by the sum of the lengths of two geodesics starting from a dome-shaped RT surface and ending at the HM surface, shown as dashed orange curves in the \cref{fig_dis.EE phase1}. The EWCS in this phase may be obtained by utilizing the embedding coordinates of points $L_2$, $L_3$, $L_4$ and $R_4$ in \cref{Dis.EWCS} as
\begin{align}\label{dis EW(3)}
	E_W(A:B)=& \frac{1}{4G_N}\cosh^{-1}\left[\frac{1+\left(\frac{\sqrt{\Delta_{L_2}}-\sqrt{\Delta_{L_4}}}{\sqrt{\Delta_{L_3}}-\sqrt{\Delta_{L_4}}}\right)\sqrt{\frac{\Delta_{L_3}+\Delta_{R_4}+2 \sqrt{\Delta_{L_3}\Delta_{R_4}}\cosh\frac{t}{u_h}}{\Delta_{L_2}+\Delta_{R_4}+2 \sqrt{\Delta_{L_2}\Delta_{R_4}}\cosh\frac{t}{u_h}}}}{\left(\frac{\sqrt{\Delta_{L_2}}-\sqrt{\Delta_{L_3}}}{\sqrt{\Delta_{L_3}}-\sqrt{\Delta_{L_4}}}\right)\sqrt{\frac{\Delta_{L_4}+\Delta_{R_4}+2 \sqrt{\Delta_{L_4}\Delta_{R_4}}\cosh\frac{t}{u_h}}{\Delta_{L_2}+\Delta_{R_4}+2 \sqrt{\Delta_{L_2}\Delta_{R_4}}\cosh\frac{t}{u_h}}} }\right]\nonumber\\
	& \qquad \qquad \qquad \qquad \qquad \qquad \qquad \qquad \qquad \qquad \quad +(\Delta_{ L_i}\longleftrightarrow \Delta_{ R_i}),
\end{align}
where second term in the preceding equation represents the right TFD contribution in the bulk EWCS. It should be noted that when the Brown-Henneaux relation is used, the above expression of the bulk EWCS matches with half of the reflected entropy computed earlier in this phase.

\subsection{Entanglement entropy phase 2}\label{Dis.EE phase2}
\begin{figure}[h!]
	\centering
	\includegraphics[scale=0.8]{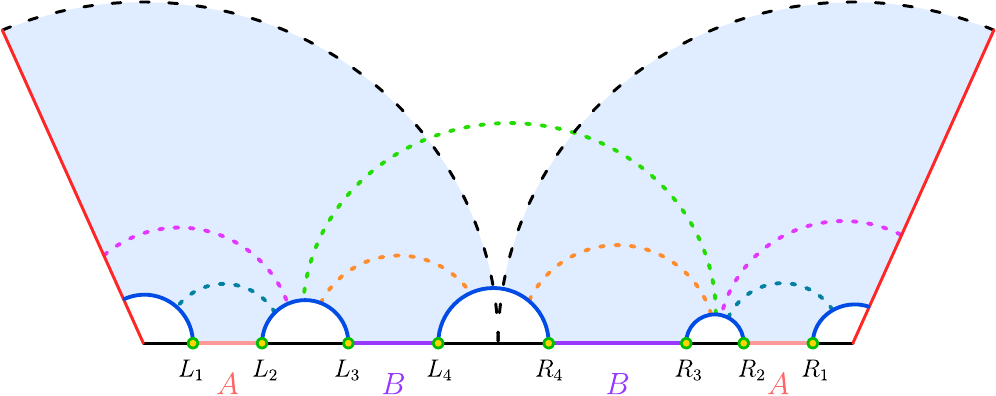}
	\caption{\it Schematic illustrating the various phases of the EWCS between subsystems A and B (represented by various colored dashed curves), while the RT surface for $A \cup B$ is represented by the solid blue curves. } 
	\label{fig_dis.EE phase2}
\end{figure}
For this phase, we consider that the subsystem $A$ is close to the boundary while $B$ is far away. So in this case the EE corresponds to the sum of the lengths of the HM surface, two dome-type RT surfaces and two island surfaces shown as blue curves in \cref{fig_dis.EE phase2}. Now by using \cref{length of HM,length of island,dome length}, the EE for this phase may be written as
\begin{align}
	S_2=&\frac{1}{4 G_N} \log \left[\frac{u^{2}_h}{u^{}_{L_2} u^{}_{L_3}u^{}_{R_2} u^{}_{R_3}}(\sqrt{\Delta_{L_2}}-\sqrt{\Delta_{L_3}})^2 (\sqrt{\Delta_{R_2}}-\sqrt{\Delta_{R_3}})^2\right]\notag\\
	&+\frac{1}{4 G_N} \log \left[\frac{u_h}{u^{}_{L_{4}}u^{}_{R_{4}}}\left(\Delta_{L_4}+\Delta_{R_4}+2 \sqrt{\Delta_{L_4}\Delta_{R_4}} \cosh \frac{t}{u_h}\right)\right]+\frac{1}{2G_N}\left(4\rho_\epsilon-\rho_B\right).
\end{align}
As seen in \cref{fig_dis.EE phase2}, this EE phase has four distinctive phases of the reflected entropy or the  bulk EWCS. The computation of the reflected entropy and the bulk EWCS for each phase is described in the following subsection.

\subsection*{Phase-I}
The reflected entropy or the bulk EWCS in this phase is similar to the first case of the previous EE phase, shown as dashed green curve in \cref{fig_dis.EE phase2}. Therefore, the reflected entropy is given by \cref{SR dis.Phase1(i)}.

\subsection*{Phase-II}
\paragraph{Reflected entropy:} In this reflected entropy phase, we consider that the subsystem $A$ is large enough. Therefore the eight-point twist field correlator in the numerator of \cref{Ref-dis-def} factorizes into two one-point and three two-point twist field correlator in a BCFT$_2$ as follows
\begin{align}
	&\langle \sigma_{g^{}_A}(u^{}_{L_{1}})\sigma_{g_A^{-1}}(u^{}_{L_{2}})\sigma_{g^{}_B}(u^{}_{L_{3}})\sigma_{g^{-1}_B}(u^{}_{L_{4}})\sigma_{g^{}_B}(u^{}_{R_{4}})\sigma_{g_B^{-1}}(u^{}_{R_{3}})\sigma_{g^{}_A}(u^{}_{R_{2}})\sigma_{g_A^{-1}}(u^{}_{R_{1}}) \rangle_{\mathrm{BCFT}^{\bigotimes mn}}\notag\\
	&=\langle\sigma_{g^{}_A}(u^{}_{L_{1}})\rangle_{\mathrm{BCFT}^{\bigotimes mn}}\langle\sigma_{g^{-1}_A}(u^{}_{R_{1}})\rangle_{\mathrm{BCFT}^{\bigotimes mn}}\langle\sigma_{g^{-1}_B}(u^{}_{L_{4}})\sigma_{g^{}_B}(u^{}_{R_{4}})\rangle_{\mathrm{BCFT}^{\bigotimes mn}}\notag\\
	&~~~~~\times
	\langle\sigma_{g_A^{-1}}(u^{}_{L_{2}})\sigma_{g^{}_B}(u^{}_{L_{3}})\rangle_{\mathrm{BCFT}^{\bigotimes mn}}\langle\sigma_{g^{}_A}(u^{}_{R_{2}})\sigma_{g^{-1}_B}(u^{}_{R_{3}})\rangle_{\mathrm{BCFT}^{\bigotimes mn}}.
\end{align}
The first three twist field correlators in the right end side of the above equation cancels with a similar factorization in the denominator of \cref{Ref-dis-def}.  In order to compute  $\langle\sigma_{g_A^{-1}}(u^{}_{L_{2}})\sigma_{g^{}_B}(u^{}_{L_{3}})\rangle_{\mathrm{BCFT}^{\bigotimes mn}}$ and  $\langle\sigma_{g^{}_A}(u^{}_{R_{2}})\sigma_{g^{-1}_B}(u^{}_{R_{3}})\rangle_{\mathrm{BCFT}^{\bigotimes mn}}$ it is necessary to transform these to the twist field correlators defined on the conformally flat cylindrical background described by the coordinates $(u_\star,\tau)$ defined in \cref{w-z map}. Note that the conformal boundary in these coordinates is located at $u_\star=0$.
We may now  utilize the doubling trick \cite{Cardy:2004hm} to map these BCFT$_2$ twist field correlator to the chiral twist field correlator in a CFT$_2$ defined on the full complex plane, leading to an expression for the reflected entropy as \cite{Li:2021dmf}

\begin{align}\label{SR-dis2(II)}
	S_R(A:B)=& \lim_{{m,n} \to 1}\frac{1}{1-n}\log\frac{\langle\sigma_{g^{}_A}(-u_{\star}({L_{2}}))\sigma_{g_A^{-1}}(u_{\star}({L_{2}}))\sigma_{g^{}_B}(u_{\star}(u^{}_{L_{3}}))\sigma_{g^{-1}_B}(-u_{\star}({L_{3}}))\rangle_{\mathrm{CFT}^{\bigotimes mn}}}{\langle\sigma_{g_m}(-u_{\star}({L_{2}}))\sigma_{g_m^{-1}}(u_{\star}({L_{2}}))\sigma_{g_m}(u_{\star}({L_{3}}))\sigma_{g^{-1}_m}(-u_{\star}({L_{3}}))\rangle^n_{\mathrm{CFT}^{\bigotimes m}}}\notag\\
	+& \lim_{{m,n} \to 1}\frac{1}{1-n}\log\frac{\langle\sigma_{g^{-1}_A}(-u_{\star}({R_{2}}))\sigma_{g^{}_A}(u_{\star}({R_{2}}))\sigma_{g^{-1}_B}(u_{\star}({R_{3}}))\sigma_{g^{}_B}(-u_{\star}({R_{3}}))\rangle_{\mathrm{CFT}^{\bigotimes mn}}}{\langle\sigma_{g^{-1}_m}(-u_{\star}({R_{2}}))\sigma_{g_m}(u_{\star}({R_{2}}))\sigma_{g^{-1}_m}(u_{\star}({R_{3}}))\sigma_{g_m}(-u_{\star}({R_{3}}))\rangle^n_{\mathrm{CFT}^{\bigotimes m}}},
\end{align}
where $-u_{\star}$ corresponds to the mirror image of $u_\star$.
Note that to compute the reflected entropy, it is required to further transform the above four-point twist correlators to the flat plane ($\omega$-plane) twist field correlators.
%** (modify) Note that as the field theory is located on a AdS$_2$ black hole background, it is necessary to transform the above four-point twist correlators to the flat plane ($\omega$-plane) twist field correlators. **
To proceed, we recall that the reflected entropy between two disjoint subsystems $A = [\omega_1, \omega_2]$ and $B = [\omega_3, \omega_4]$ corresponding to the above boundary channel in a BCFT$_2$ may be obtained in the large central charge limit as \cite{Dutta:2019gen, Fitzpatrick:2014vua,BasakKumar:2022stg}
\begin{align}\label{SR-Dutta}
	S_R(A:B)= \frac{2c}{3}\log\left(\frac{1+\sqrt{1-{\eta}}}{\sqrt{{\eta}}}\right)+ 2S_\text{bdy},
\end{align}
where ${\eta}= \frac{\omega^{}_{23} \omega^{}_{14}}{\omega^{}_{12}\omega^{}_{24}}$ is the cross ratio and the OPE coefficient in the four-point twist field correlator involves the contribution from the boundary entropy as well as the usual OPE coefficient given as \cite{Dutta:2019gen, BasakKumar:2022stg} 
\begin{align}\label{OPE coefficient}
	C_{n,m} = e^{2 (1-n)S_\text{bdy}} (2m)^{-4h}.
\end{align}
Now by utilizing \cref{SR-Dutta,OPE coefficient,w-z map} in \cref{SR-dis2(II)} and accounting for the second term in \cref{SR-dis2(II)}, the reflected entropy in this phase may be obtained as
\begin{align}\label{SR dis.Phase2(ii)}
	S_R(A:B)=\frac{2c}{3}\log\left(\frac{1+\sqrt{1-\eta}}{\sqrt{\eta}}\right)+\frac{2c}{3}\log\left(\frac{1+\sqrt{1-\xi}}{\sqrt{\xi}}\right)+4S_{\text{bdy}}\,,
\end{align}
where $\eta$ and $\xi$ are the cross ratios on the black hole background defined as
\begin{align}
	\eta=\frac{u_h \left(\sqrt{\Delta_{L_2}}-\sqrt{\Delta_{L_3}}\right)^2}{\left(u_h-\sqrt{\Delta_{L_2}} \sqrt{\Delta_{L_3}}\right)^2}, \quad\quad \xi= \frac{u_h \left(\sqrt{\Delta_{R_2}}-\sqrt{\Delta_{R_3}}\right)^2}{\left(u_h-\sqrt{\Delta_{R_2}} \sqrt{\Delta_{R_3}}\right)^2}.
\end{align}
\vspace{.05mm}
\paragraph{EWCS:} The bulk EWCS for this phase is equivalent to the sum of two geodesic lengths which start from the dome-shaped RT surface and end at the EOW brane on both copies of the TFD. These geodesics are shown as dashed magenta curves in the \cref{fig_dis.EE phase2}. We may now obtain an expression for the length of geodesic which ends on the brane at an arbitrary point $(u^{}_{B},\rho^{}_B,t)$, by using the embedding coordinates of the points $L_2$, $L_3$ and $(u^{}_{B},\rho^{}_B,t)$ in \cref{EWCS three pt. formula} as follows
\begin{align}\label{geodesic length CZ}
	L=\cosh ^{-1}\Bigg[\frac{1}{u^{}_B \sqrt{u_h} (\sqrt{\Delta_{ L_2}}-\sqrt{\Delta_{ L_3}})}&\sqrt{2 u_h (\sqrt{\Delta_{L_B}}-\sqrt{\Delta_{L_2}})^2 \cosh \rho_B-u^{}_{L_2} u^{}_B e^{\rho^{}_B}}\notag\\
	&\times\sqrt{2 u_h (\sqrt{\Delta_{L_B}}-\sqrt{\Delta_{L_3}})^2 \cosh \rho_B-u^{}_{L_3} u^{}_B e^{\rho^{}_B}}\Bigg].
\end{align}
The bulk EWCS may be computed by extremizing this length over the brane coordinate $u^{}_B$. The process of extremizing is complicated in the present scenario, which could be simplified by using a variable change, $u^{}_i=u_h (1-x_{i}^2)$. The extremal value of $x^{}_B$ is then given as 
\begin{align}
	x^{}_B= \frac{1+x^{}_{L_2}x^{}_{L_3} -\sqrt{\left(x^{2}_{L_2}-1\right) \left(x^{2}_{L_3}-1\right)}}{x^{}_{L_2}+x^{}_{L_3}}.
\end{align}
Now restoring the $u$ coordinate and using this extremal value in \cref{geodesic length CZ} and adding the contribution from the right TFD copy, the corresponding bulk EWCS in this phase may be obtained as
\begin{align}\label{Brane EWCS}
	E_W(A:B)=\frac{1}{4 G_N}\Bigg[&\cosh ^{-1}\left(\frac{u_h-\sqrt{\Delta _{L_2} \Delta _{L_3}}}{\sqrt{u_h} \left(\sqrt{\Delta _{L_2}}-\sqrt{\Delta _{L_3}}\right)}\right)\notag\\
	&+\cosh ^{-1}\left(\frac{u_h-\sqrt{\Delta _{R_2} \Delta _{R_3}}}{\sqrt{u_h} \left(\sqrt{\Delta _{R_2}}-\sqrt{\Delta _{R_3}}\right)}\right)
	-2 \rho _B\Bigg].
\end{align}
Here also using the Brown-Henneaux relation, we find that the reflected entropy computed in \cref{SR dis.Phase2(ii)} exactly matches with twice the bulk EWCS. 
%This provides a consistency check for the duality mentioned earlier. 

\subsection*{Phase-III}
\paragraph{Reflected entropy:} In this reflected entropy phase, we consider that the subsystem $A$ is smaller than $B$, hence the numerator of \cref{Ref-dis-def} may be factorized into a two-point and two three-point twist field correlators in the BCFT$_2$ as follows
\begin{align}
	&\langle \sigma_{g^{}_A}(u_{L_{1}})\sigma_{g_A^{-1}}(u_{L_{2}})\sigma_{g^{}_B}(u_{L_{3}})\sigma_{g^{-1}_B}(u_{L_{4}})\sigma_{g^{}_B}(u_{R_{4}})\sigma_{g_B^{-1}}(u_{R_{3}})\sigma_{g^{}_A}(u_{R_{2}})\sigma_{g_A^{-1}}(u_{R_{1}}) \rangle_{\mathrm{BCFT}^{\bigotimes mn}}\notag\\
	&=\langle\sigma_{g^{-1}_B}(u_{L_{4}})\sigma_{g^{}_B}(u_{R_{4}})\rangle_{\mathrm{BCFT}^{\bigotimes mn}}\langle\sigma_{g^{}_A}(u_{L_{1}})\sigma_{g_A^{-1}}(u_{L_{2}})\sigma_{g^{}_B}(u_{L_{3}})\rangle_{\mathrm{BCFT}^{\bigotimes mn}}\notag\\
	&~~~~~\times
	\langle\sigma_{g^{-1}_A}(u_{R_{1}})\sigma_{g^{}_A}(u_{R_{2}})\sigma_{g^{-1}_B}(u_{R_{3}})\rangle_{\mathrm{BCFT}^{\bigotimes mn}}.
\end{align}
The first twist field correlator of the above equation cancels with a similar factorization in the denominator of \cref{Ref-dis-def}.  To compute  $\langle\sigma_{g^{}_A}(u_{L_{1}})\sigma_{g_A^{-1}}(u_{L_{2}})\sigma_{g^{}_B}(u_{L_{3}})\rangle_{\mathrm{BCFT}^{\bigotimes mn}}$ and  $\langle\sigma_{g^{-1}_A}(u_{R_{1}})\sigma_{g^{}_A}(u_{R_{2}})\sigma_{g^{-1}_B}(u_{R_{3}})\rangle_{\mathrm{BCFT}^{\bigotimes mn}}$ it is required to transform these to the twist field correlators defined on the conformally flat cylindrical background.
Now by using the doubling trick \cite{Cardy:2004hm} and the similar factorization in the denominator of \cref{Ref-dis-def}, we have the following expression of the reflected entropy for two disjoint subsystems as \cite{Li:2021dmf}
\begin{align}
	S_R(A:B)=& \lim_{{m,n} \to 1}\frac{1}{1-n}\log\frac{\langle\sigma_{g^{-1}_A}(-u_{\star}({L_{1}}))\sigma_{g_A^{}}(u_{\star}({L_{1}}))\sigma_{g^{-1}_A}(u_{\star}({L_{2}}))\sigma_{g^{}_B}(u_{\star}({L_{3}}))\rangle_{\mathrm{CFT}^{\bigotimes mn}}}{\langle\sigma_{g^{-1}_m}(-u_{\star}({L_{1}}))\sigma_{g_m^{}}(u_{\star}({L_{1}}))\sigma_{g^{-1}_m}(u_{\star}({L_{2}}))\sigma_{g^{}_m}(u_{\star}({L_{3}}))\rangle^n_{\mathrm{CFT}^{\bigotimes m}}}\notag\\
+& \lim_{{m,n} \to 1}\frac{1}{1-n}\log\frac{\langle\sigma_{g^{}_A}(-u_{\star}({R_{1}}))\sigma_{g^{-1}_A}(u_{\star}({R_{1}}))\sigma_{g^{}_A}(u_{\star}({R_{2}}))\sigma_{g^{-1}_B}(u_{\star}({R_{3}}))\rangle_{\mathrm{CFT}^{\bigotimes mn}}}{\langle\sigma_{g^{}_m}(-u_{\star}({R_{1}}))\sigma_{g^{-1}_m}(u_{\star}({R_{1}}))\sigma_{g^{}_m}(u_{\star}({R_{2}}))\sigma_{g^{-1}_m}(u_{\star}({R_{3}}))\rangle^n_{\mathrm{CFT}^{\bigotimes m}}}.
\end{align}
%\begin{align}
%	S_R(A:B)=& \lim_{{m,n} \to 1}\frac{1}{1-n}\log\frac{\langle\sigma_{g^{-1}_A}(\bar{u}^{}_{L_{1}})\sigma_{g_A}(u_{L_{1}})\sigma_{g_A^{-1}}(u_{L_{2}})\sigma_{g_B}(u_{L_{3}})\rangle_{\mathrm{CFT}^{\bigotimes mn}}}{\langle\sigma_{g^{-1}_m}(\bar{u}^{}_{L_{1}})\sigma_{g_m}(u_{L_{1}})\sigma_{g_m^{-1}}(u_{L_{2}})\sigma_{g_m}(u_{L_{3}})\rangle_{\mathrm{CFT}^{\bigotimes m}}}\notag\\
%	& +\lim_{{m,n} \to 1}\frac{1}{1-n}\log\frac{\langle\sigma_{g_A}(\bar{u}^{}_{R_{1}})\sigma_{g^{-1}_A}(u_{R_{1}})\sigma_{g_A}(u_{R_{2}})\sigma_{g^{-1}_B}(u_{R_{3}})\rangle_{\mathrm{CFT}^{\bigotimes mn}}}{\langle\sigma_{g_m}(\bar{u}^{}_{R_{1}})\sigma_{g^{-1}_m}(u_{R_{1}})\sigma_{g_m}(u_{R_{2}})\sigma_{g^{-1}_m}(u_{R_{3}})\rangle_{\mathrm{CFT}^{\bigotimes m}}}.
%\end{align}
Now by utilizing \cref{w-z map} which map these twist field correlators to the flat plane ($\omega$-plane) twist field correlators and the form of the four point conformal block in the large central charge limit \cite{Dutta:2019gen,Fitzpatrick:2014vua}, the reflected entropy in this phase may be obtained as 
\begin{align}\label{SR dis phase2(iii)}
	S_R(A:B)=\frac{c}{3}\log\left[\frac{1+\sqrt{\eta}}{1-\sqrt{\eta}}\right]
	+\frac{c}{3}\log\left[\frac{1+\sqrt{{\xi}}}{1-\sqrt{{\xi}}}\right],
\end{align}
where $\eta$ and $\xi$ are the cross ratios defined as
\begin{align}
	\eta= \frac{\left(\sqrt{\Delta_{L_1}}-\sqrt{\Delta_{L_2}}\right) \left(u_h-\sqrt{\Delta_{L_1}} \sqrt{\Delta_{L_3}}\right)}{\left(\sqrt{\Delta_{L_1}}-\sqrt{\Delta_{L_3}}\right) \left(u_h-\sqrt{\Delta_{L_1}} \sqrt{\Delta_{L_2}}\right)}, ~ \xi= \frac{\left(\sqrt{\Delta_{R_1}}-\sqrt{\Delta_{R_2}}\right) \left(u_h-\sqrt{\Delta_{R_1}} \sqrt{\Delta_{R_3}}\right)}{\left(\sqrt{\Delta_{R_1}}-\sqrt{\Delta_{R_3}}\right) \left(u_h-\sqrt{\Delta_{R_1}} \sqrt{\Delta_{R_2}}\right)}.
\end{align}

\paragraph{EWCS:} The bulk EWCS for this phase is equivalent to the sum of two geodesic lengths depicted as dashed blue curves in \cref{fig_dis.EE phase2}. These geodesics connect dome-type RT surface to the island surface on both copies of the TFD. The EWCS may now be calculated by using the embedding coordinates of the three boundary points $L_1$, $L_2$, $L_3$ and one bulk point $(u^{}_{L_1},\rho^{}_B,t)$ in \cref{bulk point EWCS formula} and adding the contribution from the right TFD copy as
\begin{align}\label{Island EWCS}
	E_W(A:B)=\frac{1}{4 G_N}\cosh ^{-1}\Bigg[&\frac{\left(\sqrt{\Delta^{}_{L_1}}-\sqrt{\Delta^{} _{L_3}}\right) \left(u_h-\sqrt{\Delta _{L_1} \Delta^{} _{L_2}}\right)}{u^{}_{L_{1}}\left(\sqrt{\Delta^{} _{L_2}}-\sqrt{\Delta^{}_{L_3}}\right)}\notag\\
	&+\frac{\left(\sqrt{\Delta _{L_1}}-\sqrt{\Delta^{} _{L_2}}\right) \left(u_h-\sqrt{\Delta _{L_1} \Delta^{} _{L_3}}\right)}{u^{}_{L_{1}}\left(\sqrt{\Delta^{} _{L_2}}-\sqrt{\Delta^{}_{L_3}}\right)}\Bigg]+ ~ (\Delta_{ L_i} \longrightarrow \Delta_{R_i})
\end{align}
Note that when the Brown-Henneaux relation is used, the preceding expression of the bulk EWCS equals half of the reflected entropy calculated in \cref{SR dis phase2(iii)}.
 
%This ensures that the duality discussed in \cref{duality} is consistent.

\subsection*{Phase-IV}
The reflected entropy or the bulk EWCS in this phase is similar to the third case of the previous EE phase, shown as dashed orange curves in \cref{fig_dis.EE phase2}. Therefore, the bulk EWCS is given by \cref{dis EW(3)}.

\subsection{Entanglement entropy phase 3}\label{Dis.EE phase3}
\begin{figure}[h!]
	\centering
	\includegraphics[scale=0.8]{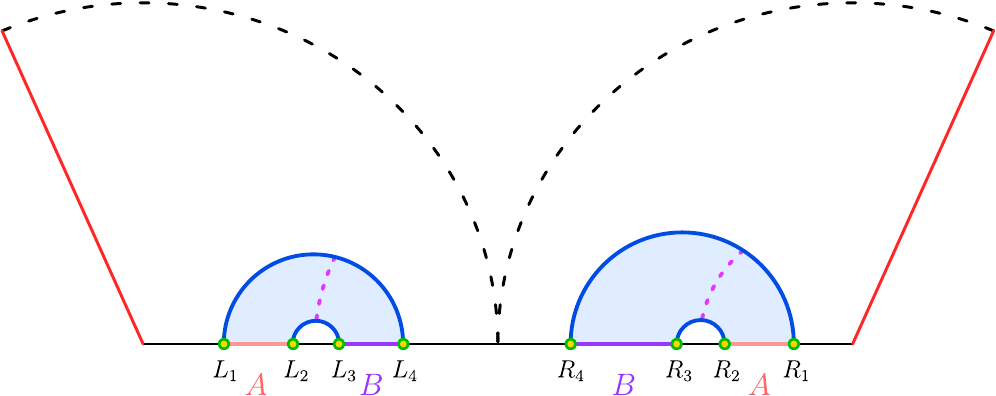}
	\caption{\it Diagrammatic illustration of the EWCS between subsystems $A$ and $B$ (depicted as dashed curves), where the solid blue curves represent the RT surface for $A \cup B$.} 
	\label{fig_dis.EE phase3}
\end{figure}
In this EE phase both subsystems are small and located close to one another away from the boundary. Hence the EE for this phase is given by the sum of the lengths of four dome-type RT surfaces, displayed as blue curves in \cref{fig_dis.EE phase3}. Now, the EE for this phase may be obtained by using \cref{dome length} as follows
\begin{align}
	S_3=&\frac{1}{4 G_N} \log \left[\frac{u^{2}_h}{u^{}_{L_2} u^{}_{L_3}u^{}_{R_2} u^{}_{R_3}}(\sqrt{\Delta_{L_2}}-\sqrt{\Delta_{L_3}})^2 (\sqrt{\Delta_{R_2}}-\sqrt{\Delta_{R_3}})^2\right]\notag\\
	&+\frac{1}{4 G_N} \log \left[\frac{u^{2}_h}{u^{}_{L_1} u^{}_{L_4}u^{}_{R_1} u^{}_{R_4}}(\sqrt{\Delta_{L_1}}-\sqrt{\Delta_{L_4}})^2 (\sqrt{\Delta_{R_1}}-\sqrt{\Delta_{R_4}})^2\right]+\frac{2 \rho_\epsilon}{G_N}.
\end{align}
For this EE phase we observe only one phase for the reflected entropy or the bulk EWCS shown as dashed magenta curves in \cref{fig_dis.EE phase3}. Note that here we assume that the subsystems are away from the boundary, hence the OPE channel for the BCFT$_2$ correlator is favoured.

\paragraph{Reflected entropy:} For the computation of the reflected entropy in this EE phase, the numerator of \cref{Ref-dis-def} may be factorized into two four-point twist field correlators as follows
\begin{align}
	&\langle \sigma_{g^{}_A}(u_{L_{1}})\sigma_{g_A^{-1}}(u_{L_{2}})\sigma_{g^{}_B}(u_{L_{3}})\sigma_{g^{-1}_B}(u_{L_{4}})\sigma_{g^{}_B}(u_{R_{4}})\sigma_{g_B^{-1}}(u_{R_{3}})\sigma_{g^{}_A}(u_{R_{2}})\sigma_{g_A^{-1}}(u_{R_{1}}) \rangle_{\mathrm{CFT}^{\bigotimes mn}}\notag\\
	&= \langle \sigma_{g_A}(u^{}_{L_{1}})\sigma_{g_A^{-1}}(u^{}_{L_{2}})\sigma_{g_B}(u^{}_{L_{3}})\sigma_{g^{-1}_B}(u^{}_{L_{4}})\rangle_{\mathrm{CFT}^{\bigotimes mn}} \langle\sigma_{g_B}(u^{}_{R_{4}})\sigma_{g_B^{-1}}(u^{}_{R_{3}})\sigma_{g_A}(u^{}_{R_{2}})\sigma_{g_A^{-1}}(u^{}_{R_{1}}) \rangle_{\mathrm{CFT}^{\bigotimes mn}}.
\end{align}
The denominator of \cref{Ref-dis-def} admits a similar factorization. Hence the reflected entropy in this phase is given as
\begin{align}
	S_R(A:B)=& \lim_{{m,n} \to 1}\frac{1}{1-n}\log\frac{\langle \sigma_{g_A}(u^{}_{L_{1}})\sigma_{g_A^{-1}}(u^{}_{L_{2}})\sigma_{g_B}(u^{}_{L_{3}})\sigma_{g^{-1}_B}(u^{}_{L_{4}})\rangle_{\mathrm{CFT}^{\bigotimes mn}}}{\langle \sigma_{g_m}(u^{}_{L_{1}})\sigma_{g_m^{-1}}(u^{}_{L_{2}})\sigma_{g_m}(u^{}_{L_{3}})\sigma_{g^{-1}_m}(u^{}_{L_{4}})\rangle^n_{\mathrm{CFT}^{\bigotimes m}}}\notag\\
	& + \lim_{{m,n} \to 1}\frac{1}{1-n}\log\frac{\langle\sigma_{g_B}(u^{}_{R_{4}})\sigma_{g_B^{-1}}(u^{}_{R_{3}})\sigma_{g_A}(u^{}_{R_{2}})\sigma_{g_A^{-1}}(u^{}_{R_{1}}) \rangle_{\mathrm{CFT}^{\bigotimes mn}}}{\langle\sigma_{g_m}(u^{}_{R_{4}})\sigma_{g_m^{-1}}(u^{}_{R_{3}})\sigma_{g_m}(u^{}_{R_{2}})\sigma_{g_m^{-1}}(u^{}_{R_{1}}) \rangle^n_{\mathrm{CFT}^{\bigotimes m}}}.
\end{align}
Now by utilizing \cref{w-z map} to map these four point twist field correlator to flat plane twist field correlator and then using the form of four point function in the large central charge limit \cite{Dutta:2019gen, Fitzpatrick:2014vua}, we may obtain the reflected entropy which is identical to \cref{SR dis phase2(iii)} where the cross ratios $\eta$ and $\xi$ are modified to
%\begin{align}\label{SR dis phase3}
%	S_R(A:B)=\frac{c}{3}\log\left[\frac{1+\sqrt{\eta}}{1-\sqrt{\eta}}\right]
%	+\frac{c}{3}\log\left[\frac{1+\sqrt{{\xi}}}{1-\sqrt{{\xi}}}\right],
%\end{align} 
\begin{align}\label{SR dis 3 cross-ratio}
	\eta= \frac{\left(\sqrt{\Delta _{L_1}}-\sqrt{\Delta _{L_2}}\right) \left(\sqrt{\Delta _{L_3}}-\sqrt{\Delta _{L_4}}\right)}{\left(\sqrt{\Delta _{L_1}}-\sqrt{\Delta _{L_3}}\right) \left(\sqrt{\Delta _{L_2}}-\sqrt{\Delta _{L_4}}\right)}, ~ \xi= \frac{\left(\sqrt{\Delta _{R_1}}-\sqrt{\Delta _{R_2}}\right) \left(\sqrt{\Delta _{R_3}}-\sqrt{\Delta _{R_4}}\right)}{\left(\sqrt{\Delta _{R_1}}-\sqrt{\Delta _{R_3}}\right) \left(\sqrt{\Delta _{R_2}}-\sqrt{\Delta _{R_4}}\right)}.
\end{align}

\paragraph{EWCS:} The corresponding EWCS for this phase is proportional to the sum of the lengths of two geodesics shown as dashed magenta curves in \cref{fig_dis.EE phase3}. Now by utilizing the coordinates of the points $L_1$, $L_2$, $L_3$ and $L_4$ in \cref{Dis.EWCS} and adding the contribution from the right TFD copy, the bulk EWCS in this case may be obtained as
\begin{align}\label{EW dis phase3}
	E_{W}(A:B)=\frac{1}{4 G_N}\log\left[\frac{1+\sqrt{\eta}}{1-\sqrt{\eta}}\right]
	+\frac{1}{4 G_N}\log\left[\frac{1+\sqrt{{\xi}}}{1-\sqrt{{\xi}}}\right],
\end{align}
where $\eta$ and $\xi$ are defined in \cref{SR dis 3 cross-ratio}.
Here also we observe that the above expression for the bulk EWCS is precisely equal to half of the reflected entropy computed earlier upon utilizing the Brown-Henneaux relation. %which once again verify the duality mentioned in \cref{duality}.

\subsection{Entanglement entropy phase 4}\label{Dis.EE phase4}
\begin{figure}[h!]
	\centering
	\includegraphics[scale=0.8]{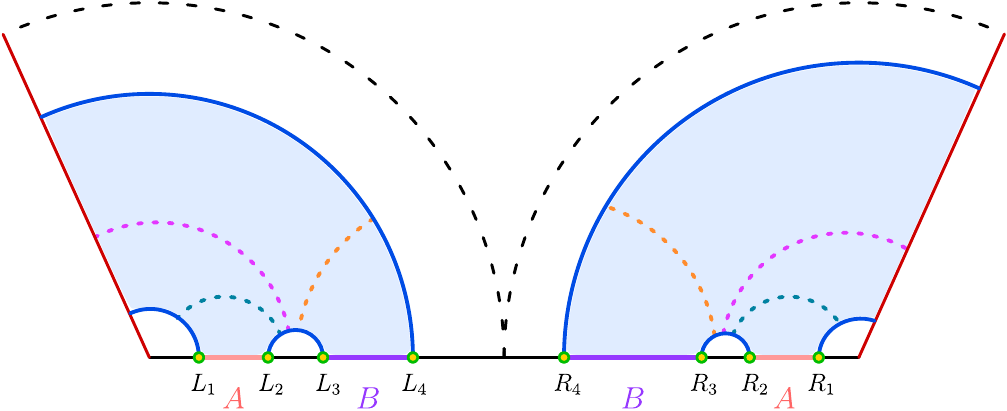}
	\caption{\it Schematic diagram of the various phases of the EWCS (represented by dashed curves) between subsystems $A$ and $B$ when the RT surface for $A\cup B$ is as depicted by the solid blue curves. } 
	\label{fig_dis.EE phase4}
\end{figure}
In the last phase we consider that both the subsystems $A$ and $B$ are very close to the boundary, so the boundary channel is dominant. Hence the EE is proportional to sum of the lengths of two dome-type RT surfaces and four island surfaces shown as blue curves in \cref{fig_dis.EE phase4}. The EE in this configuration may be obtained by using \cref{length of island,dome length} as
\begin{align}
	S_4= \frac{1}{4 G_N} \log \left[\frac{u^{2}_h}{u^{}_{L_2} u^{}_{L_3}u^{}_{R_2} u^{}_{R_3}}(\sqrt{\Delta_{L_2}}-\sqrt{\Delta_{L_3}})^2 (\sqrt{\Delta_{R_2}}-\sqrt{\Delta_{R_3}})^2\right]+\frac{1}{G_N}\left(2\rho_\epsilon-\rho_B\right).
\end{align}
 In this entropy phase, we observe three different phases for the reflected entropy or the bulk EWCS as shown in \cref{fig_dis.EE phase4}. We will explain the computation of the reflected entropy and the bulk EWCS phases in the following subsections.

\subsection*{Phase-I}
The reflected entropy or the bulk EWCS in this phase is similar to the second case of \cref{Dis.EE phase2}, shown as dashed magenta curves in \cref{fig_dis.EE phase4}. Therefore, the bulk EWCS is given by \cref{Brane EWCS}.

\subsection*{Phase-II}
\paragraph{Reflected entropy:} In this reflected entropy phase we consider the subsystem $B$ to be smaller than $A$, hence the eight-point twist field correlator in the numerator of \cref{Ref-dis-def} factorizes into two one-point and two three-point twist field correlator in BCFT$_2$ as follows
\begin{align}
	&\langle \sigma_{g^{}_A}(u_{L_{1}})\sigma_{g_A^{-1}}(u_{L_{2}})\sigma_{g^{}_B}(u_{L_{3}})\sigma_{g^{-1}_B}(u_{L_{4}})\sigma_{g^{}_B}(u_{R_{4}})\sigma_{g_B^{-1}}(u_{R_{3}})\sigma_{g^{}_A}(u_{R_{2}})\sigma_{g_A^{-1}}(u_{R_{1}}) \rangle_{\mathrm{BCFT}^{\bigotimes mn}}\notag\\
	&=\langle\sigma_{g^{}_A}(u_{L_{1}})\rangle_{\mathrm{BCFT}^{\bigotimes mn}}\langle\sigma_{g^{-1}_A}(u_{R_{1}})\rangle_{\mathrm{BCFT}^{\bigotimes mn}}\langle\sigma_{g_A^{-1}}(u_{L_{2}})\sigma_{g^{}_B}(u_{L_{3}})\sigma_{g^{-1}_B}(u_{L_{4}})\rangle_{\mathrm{BCFT}^{\bigotimes mn}}\notag\\
	&~~~~~\times
	\langle\sigma_{g^{}_A}(u_{R_{2}})\sigma_{g^{-1}_B}(u_{R_{3}})\sigma_{g^{}_B}(u_{R_{4}})\rangle_{\mathrm{BCFT}^{\bigotimes mn}}.
\end{align}
The first two twist field correlators of the preceding expression cancels with a similar factorization in the denominator of \cref{Ref-dis-def}. To compute  $\langle\sigma_{g_A^{-1}}(u_{L_{2}})\sigma_{g^{}_B}(u_{L_{3}})\sigma_{g^{-1}_B}(u_{L_{4}})\rangle_{\mathrm{BCFT}^{\bigotimes mn}}$ and  $\langle\sigma_{g^{}_A}(u_{R_{2}})\sigma_{g^{-1}_B}(u_{R_{3}})\sigma_{g^{}_B}(u_{R_{4}})\rangle_{\mathrm{BCFT}^{\bigotimes mn}}$ we need to transform these to the twist field correlators defined on the conformally flat cylindrical background.
Now using the doubling trick \cite{Cardy:2004hm} and similar factorization in the denominator of \cref{Ref-dis-def}, the final expression for the reflected entropy in this case may be written as \cite{Li:2021dmf}
\begin{align}
	S_R(A:B)=& \lim_{{m,n} \to 1}\frac{1}{1-n}\log\frac{\langle\sigma_{g^{-1}_A}(u_{\star}({L_{2}}))\sigma_{g_B^{}}(u_{\star}({L_{3}}))\sigma_{g^{-1}_B}(u_{\star}({L_{4}}))\sigma_{g^{}_B}(-u_{\star}({L_{4}}))\rangle_{\mathrm{CFT}^{\bigotimes mn}}}{\langle\sigma_{g^{-1}_m}(u_{\star}({L_{2}}))\sigma_{g_m^{}}(u_{\star}({L_{3}}))\sigma_{g^{-1}_m}(u_{\star}({L_{4}}))\sigma_{g^{}_m}(-u_{\star}({L_{4}}))\rangle^n_{\mathrm{CFT}^{\bigotimes m}}}\notag\\
	+& \lim_{{m,n} \to 1}\frac{1}{1-n}\log\frac{\langle\sigma_{g^{}_A}(u_{\star}({R_{2}}))\sigma_{g^{-1}_B}(u_{\star}({R_{3}}))\sigma_{g^{}_B}(u_{\star}({R_{4}}))\sigma_{g^{-1}_B}(-u_{\star}({R_{4}}))\rangle_{\mathrm{CFT}^{\bigotimes mn}}}{\langle\sigma_{g^{}_m}(-u_{\star}({R_{2}}))\sigma_{g^{-1}_m}(u_{\star}({R_{3}}))\sigma_{g^{}_m}(u_{\star}({R_{4}}))\sigma_{g^{-1}_m}(-u_{\star}({R_{4}}))\rangle^n_{\mathrm{CFT}^{\bigotimes m}}}.
\end{align}
%\begin{align}
%	S_R(A:B)=& \lim_{{m,n} \to 1}\frac{1}{1-n}\log\frac{\langle\sigma_{g_A^{-1}}(u_{L_{2}})\sigma_{g_B}(u_{L_{3}})\sigma_{g^{-1}_B}(u_{L_{4}})\sigma_{g_B}(\bar{u}^{}_{L_{4}})\rangle_{\mathrm{CFT}^{\bigotimes mn}}}{\langle\sigma_{g_m^{-1}}(u_{L_{2}})\sigma_{g_m}(u_{L_{3}})\sigma_{g^{-1}_m}(u_{L_{4}})\sigma_{g_m}(\bar{u}^{}_{L_{4}})\rangle^n_{\mathrm{CFT}^{\bigotimes m}}} \notag\\
%	& + \lim_{{m,n} \to 1}\frac{1}{1-n}\log\frac{\langle\sigma_{g_A}(u_{R_{2}})\sigma_{g^{-1}_B}(u_{R_{3}})\sigma_{g_B}(u_{R_{4}})\sigma_{g^{-1}_B}(\bar{u}^{}_{R_{4}})\rangle_{\mathrm{CFT}^{\bigotimes mn}}}{\langle\sigma_{g_m}(u_{R_{2}})\sigma_{g^{-1}_m}(u_{R_{3}})\sigma_{g_m}(u_{R_{4}})\sigma_{g^{-1}_m}(\bar{u}^{}_{R_{4}})\rangle^n_{\mathrm{CFT}^{\bigotimes m}}}.
%\end{align}
Now by utilizing \cref{w-z map} and the form of the four point twist field correlator in the large central charge, we may obtain the reflected entropy for two disjoint subsystems which is identical to the expression given in \cref{SR dis phase2(iii)} with the cross ratios defined as follows
%
%\begin{align}\label{SR dis phase4(ii)}
%	S_R(A:B)=\frac{c}{3}\log\left[\frac{1+\sqrt{\eta}}{1-\sqrt{\eta}}\right]
%	+\frac{c}{3}\log\left[\frac{1+\sqrt{{\xi}}}{1-\sqrt{{\xi}}}\right],
%\end{align}
\begin{align}
	\eta= \frac{\left(\sqrt{\Delta_{L_3}}-\sqrt{\Delta_{L_4}}\right) \left(u_h-\sqrt{\Delta_{L_2}\Delta_{L_4}} \right)}{\left(\sqrt{\Delta_{L_2}}-\sqrt{\Delta_{L_4}}\right) \left(u_h-\sqrt{\Delta_{L_3}\Delta_{L_4}}\right)} , ~ \xi= \frac{\left(\sqrt{\Delta_{R_3}}-\sqrt{\Delta_{R_4}}\right) \left(u_h-\sqrt{\Delta_{R_2}\Delta_{R_4}} \right)}{\left(\sqrt{\Delta_{R_2}}-\sqrt{\Delta_{R_4}}\right) \left(u_h-\sqrt{\Delta_{R_3}\Delta_{R_4}} \right)}.
\end{align}

\paragraph{EWCS:} The bulk EWCS for this phase is the sum of the lengths of two geodesics which connect dome-type RT surface to the island surface on both the TFD copies. These geodesics are shown as dashed orange curves in \cref{fig_dis.EE phase4}. The bulk EWCS may be obtained by utilizing the coordinates of the three boundary points $L_2$, $L_3$, $L_4$ and one bulk point $(u^{}_{L_{4}},\rho^{}_B, t)$ in \cref{bulk point EWCS formula} and adding the contribution from the right TFD copy as follows
\begin{align}
	E_W(A:B)= \frac{1}{4 G_N}\cosh ^{-1}\Bigg[&\frac{\left(\sqrt{\Delta _{L_3}}-\sqrt{\Delta _{L_4}}\right) \left(u_h-\sqrt{\Delta _{L_2} \Delta _{L_4}}\right)}{u^{}_{L_{4}}\left(\sqrt{\Delta _{L_2}}-\sqrt{\Delta _{L_3}}\right)}\notag\\
	&+\frac{\left(\sqrt{\Delta _{L_2}}-\sqrt{\Delta _{L_4}}\right) \left(u_h-\sqrt{\Delta _{L_3} \Delta _{L_4}}\right)}{u^{}_{L_{4}}\left(\sqrt{\Delta _{L_2}}-\sqrt{\Delta _{L_3}}\right)}\Bigg]+ ~ (\Delta_{ L_i} \longrightarrow \Delta_{R_i}).
\end{align}
Here also using the Brown-Henneaux relation, we notice that the reflected entropy computed earlier precisely matches with twice the bulk EWCS. 
%This serves as a consistency check for the duality mentioned earlier.

\subsection*{Phase-III}
The reflected entropy or the bulk EWCS in this phase is similar to the third case of \cref{Dis.EE phase2}, shown as dashed blue curves in \cref{fig_dis.EE phase4}. Therefore, the bulk EWCS is given by \cref{Island EWCS}.

\subsection{Page curve}
In this subsection, we describe the Page curves for the reflected entropy for two disjoint subsystems in a BCFT$_2$ on an AdS$_2$ black hole background. To plot the analogue of the Page curve for the reflected entropy, it is necessary to determine the phase transitions in the EE. Within each EE phase, we observe various phases for the reflected entropy depending on the subsystem sizes and their locations.

\subsubsection{Case-I}\label{dis 1HM-island}
\begin{figure}[H]
	\centering
	\begin{subfigure}[b]{0.45\textwidth}
		\centering
		\includegraphics[scale=.37]{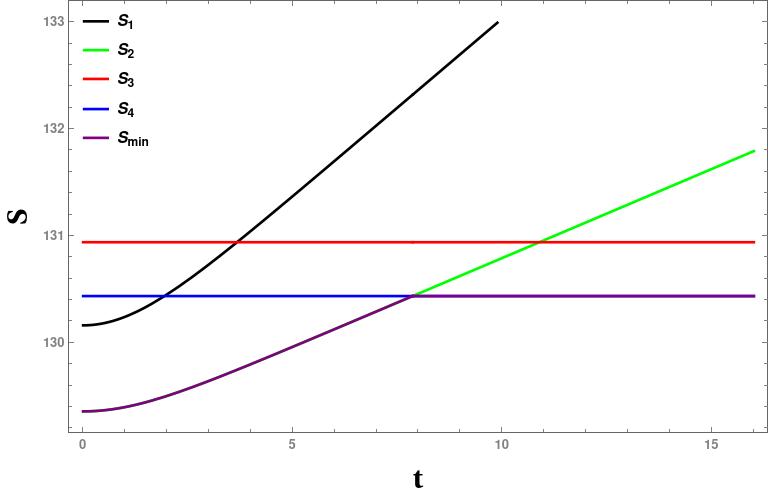} 
		\caption{}
		\label{dis plot(1HM-island)}
	\end{subfigure}
	\hfill
	\begin{subfigure}[b]{0.45\textwidth}
		\centering
		\includegraphics[scale=.347]{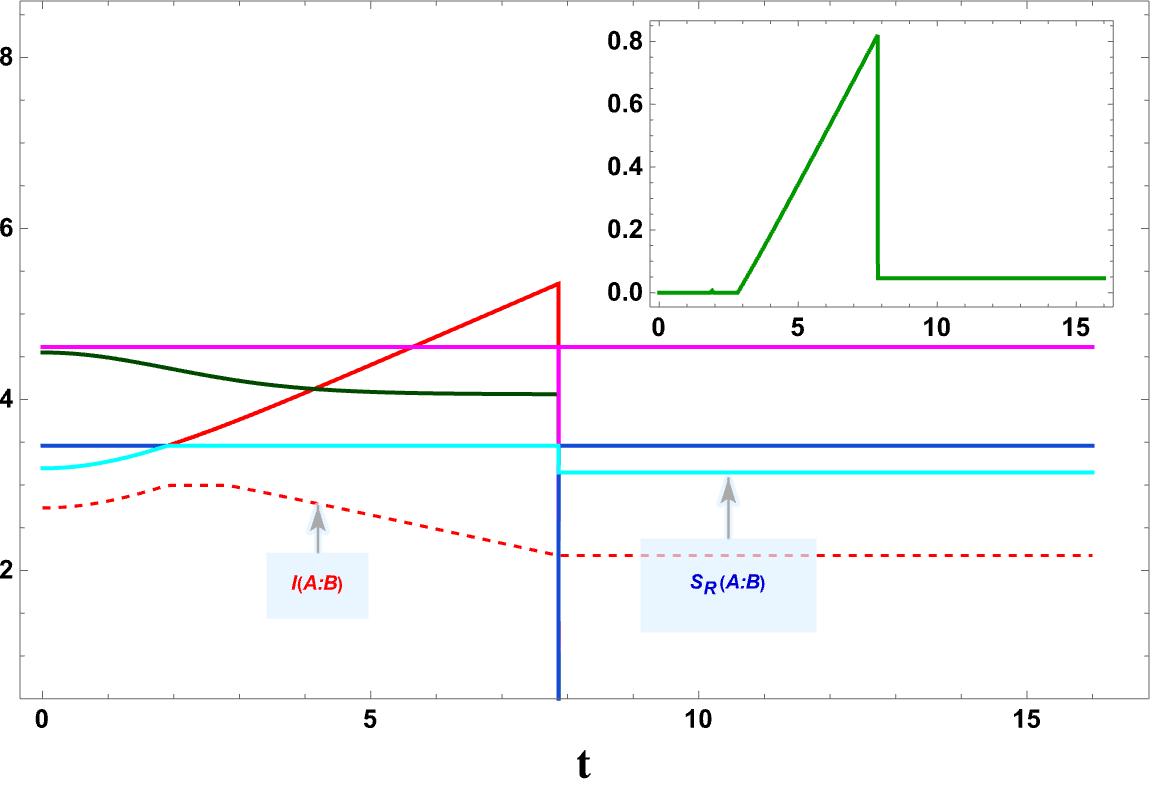}
		\caption{}
		\label{dis SR plot(1HM-island)}
	\end{subfigure}
	\caption{(a) Page curve of the EE for subsystems $A \cup B$. Here purple curve shows the minimum EE among all the phases. (b) Page curve of the reflected entropy between subsystems $A$ and $B$. Here cyan curve represents minimum $S_R$ and red dashed curve is mutual information. (Both graphs are in units of $c$). The inset plot shows the deviation from saturation of the Markov gap \cref{Markov gap}. The Page curves for the EE and $S_R$ is obtained with $u_h=1, \Delta_{L_1}=0.89, \Delta_{L_2}=0.109, \Delta_{L_3}=0.102, \Delta_{L_4}=5 \times 10^{-4}, \Delta_{R_1}=0.79, \Delta_{R_2}=0.109, \Delta_{R_3}=0.102, \Delta_{R_4}=4 \times 10^{-4}, \rho_B=-0.075,  \rho_\epsilon=100$. }
\end{figure}

The EE phase transition between \hyperref[Dis.EE phase2]{phase-2} and \hyperref[Dis.EE phase4]{phase-4} occurs for a small brane angle and the subsystem $B$ far away from the boundary as shown in \cref{dis plot(1HM-island)}. The Page time $T^{\text{disj}}_{{2\rightarrow 4}}$ for this transition is given as 
\begin{align}
	T^{\text{disj}}_{{2\rightarrow 4}}=u_h \cosh^{-1}\left(\frac{u^{2}_h-u_h\left(e^{2 \rho_B}+1\right)(\Delta_{L_4}+\Delta_{R_4})+\Delta_{L_4} \Delta_{R_4}}{2u_h e^{2 \rho_B} \sqrt{\Delta_{L_4}\Delta_{R_4}}}\right).\label{Page-time24}
\end{align}
From the Page curve of the reflected entropy we observe that in the EE \hyperref[Dis.EE phase2]{phase-2} the reflected entropy increases initially as the bulk EWCS is the HM surface which grows over time and then remains constant until the Page time as the bulk EWCS lands on the EOW brane. After the Page time \cref{Page-time24} the reflected entropy   saturates to another smaller constant value in the EE \hyperref[Dis.EE phase4]{phase-4} as depicted in \cref{dis SR plot(1HM-island)}. The Page curve for the reflected entropy indicates that before the Page time $T^{\text{disj}}_{{2\rightarrow 4}}$, the Markov gap is always greater than or equal to the anticipated lower bound $\frac{2c}{3}\log{2}$ which is in conformity with \cref{Markov gap} because the bulk EWCS phases have two non-trivial boundaries. Additionally, after the Page time $T^{\text{disj}}_{{2\rightarrow 4}}$, since there are four non-trivial boundaries of the bulk EWCS, this gap increases to a value larger than $\frac{4c}{3}\log{2}$.

\subsubsection{Case-II}\label{sec:dis 1HM-dome}
\begin{figure}[H]
	\centering
	\begin{subfigure}[b]{0.45\textwidth}
		\centering
		\includegraphics[scale=.37]{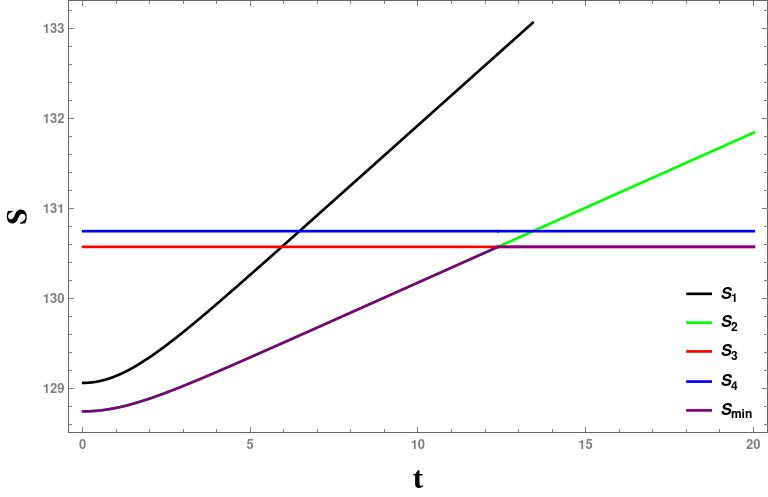}
		\caption{\it}
		\label{dis plot(1HM-dome)}
	\end{subfigure}
	\hfill
	\begin{subfigure}[b]{0.45\textwidth}
		\centering
		\includegraphics[scale=.342]{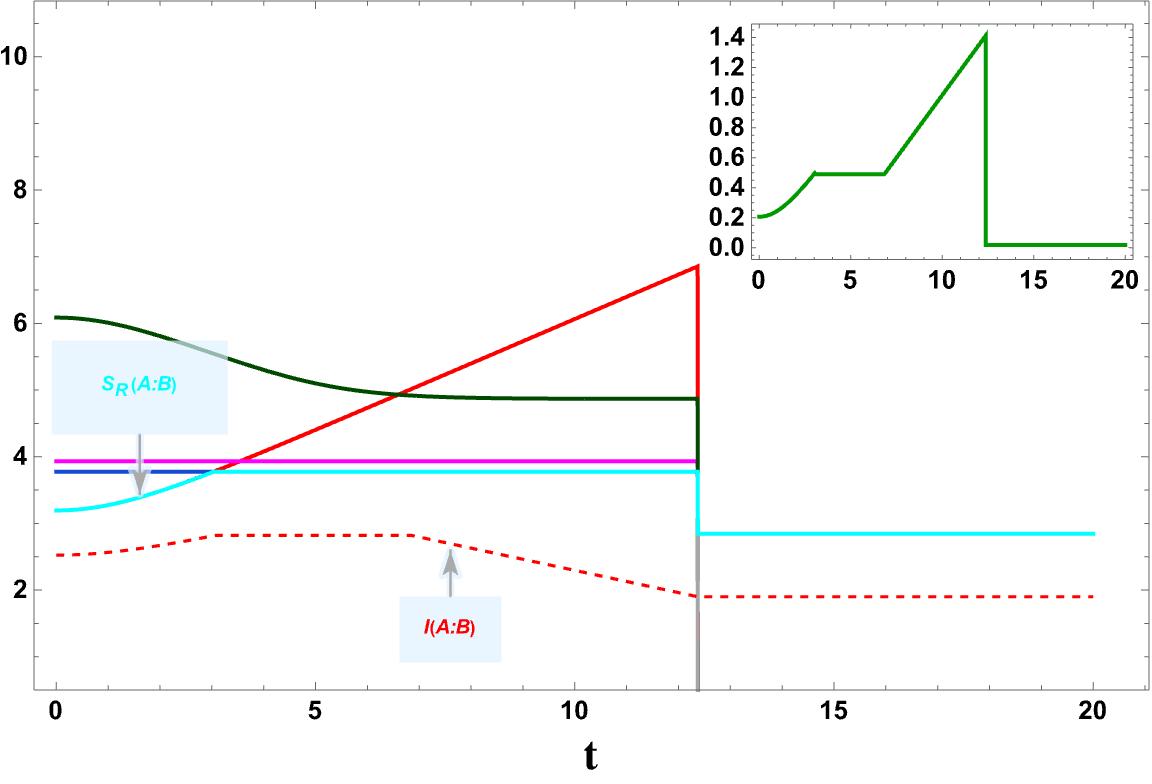}
		\caption{\it }
		\label{dis SR plot(1HM-dome)}
	\end{subfigure}
	\caption{(a) Page curve of the EE for subsystems $A \cup B$. Here purple curve shows the minimum EE among all the phases. (b) Page curve of the reflected entropy between subsystems $A$ and $B$. Here cyan curve represents minimum $S_R$ and red dashed curve is mutual information (All graphs are in units of $c$). The inset plot shows the deviation from saturation of the Markov gap \cref{Markov gap}. The Page curves for the EE and $S_R$ is obtained with $u_h=1, \Delta_{L_1}=0.69, \Delta_{L_2}=0.109, \Delta_{L_3}=0.102, \Delta_{L_4}=5 \times 10^{-6}, \Delta_{R_1}=0.59, \Delta_{R_2}=0.109, \Delta_{R_3}=0.102, \Delta_{R_4}=4 \times 10^{-6}, \rho_B=-0.55,   \rho_\epsilon=100$. }
\end{figure}

The EE transition between \hyperref[Dis.EE phase2]{phase-2} and \hyperref[Dis.EE phase3]{phase-3} may be obtained by considering the subsystem $B$ to be far away from the boundary with the brane angle relatively larger than the one described in the previous case. This EE phase transition is depicted in \cref{dis plot(1HM-dome)} and the Page time $T^{\text{disj}}_{{2\rightarrow3}}$ is given as follows
\begin{align}\label{Page time 1HM-dome}
	T^{\text{disj}}_{{2 \rightarrow 3}}=	u_h \cosh ^{-1}\left(\frac{u_h e^{2 \rho_B}  \left(\sqrt{\Delta_{L_1}}-\sqrt{\Delta_{L_4}}\right)^2 \left(\sqrt{\Delta_{R_1}}-\sqrt{\Delta_{R_4}}\right)^2}{2 u^{}_{L_1}u^{}_{R_1} \sqrt{\Delta_{L_4}\Delta_{R_4}} }-\frac{(\Delta_{L_4}+\Delta_{R_4}) }{2 \sqrt{\Delta_{L_4}\Delta_{R_4}}}\right).
\end{align}
The reflected entropy transition for these EE phases is shown in \cref{dis SR plot(1HM-dome)}. In the EE \hyperref[Dis.EE phase2]{phase-2}, the reflected entropy increases initially as the bulk EWCS is the HM surface and stays constant until $T^{\text{disj}}_{{2 \rightarrow 3}}$ as the bulk EWCS lands on the EOW brane and finally after the Page time, it saturates to another constant value in the EE \hyperref[Dis.EE phase3]{phase-3}. From the Page curve of the reflected entropy, as earlier we observe that before the Page time, the Markov gap is always greater than $\frac{2c}{3}\log{2}$, which is consistent with \cref{Markov gap} as there are two non-trivial boundaries for the bulk EWCS phases. Additionally, after the Page time, this gap saturates to a value greater than $\frac{4c}{3}\log{2}$ due to the four non-trivial boundaries of the bulk EWCS.

\subsubsection{Case-III}\label{sec:dis 2HM-1HM-dome}

\begin{figure}[H]
	\centering
	\begin{subfigure}[b]{0.45\textwidth}
		\centering
		\includegraphics[scale=.37]{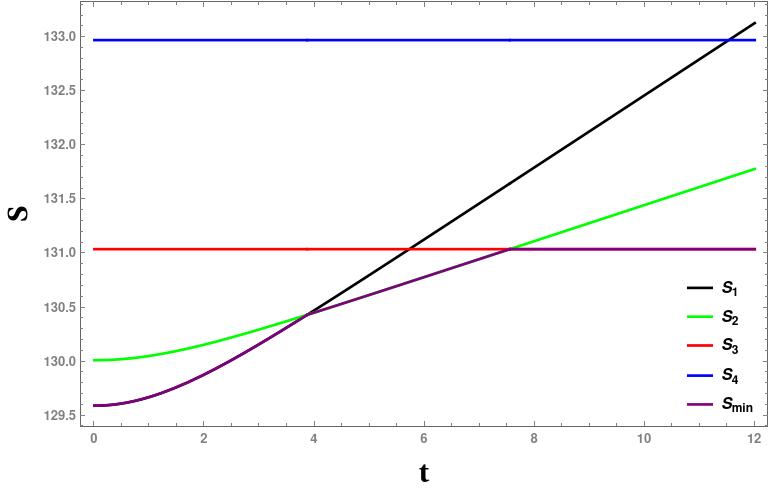}
		\caption{}
		\label{dis plot(2HM-1HM-dome)}
	\end{subfigure}
	\hfill
	\begin{subfigure}[b]{0.45\textwidth}
		\centering
		\includegraphics[scale=.342]{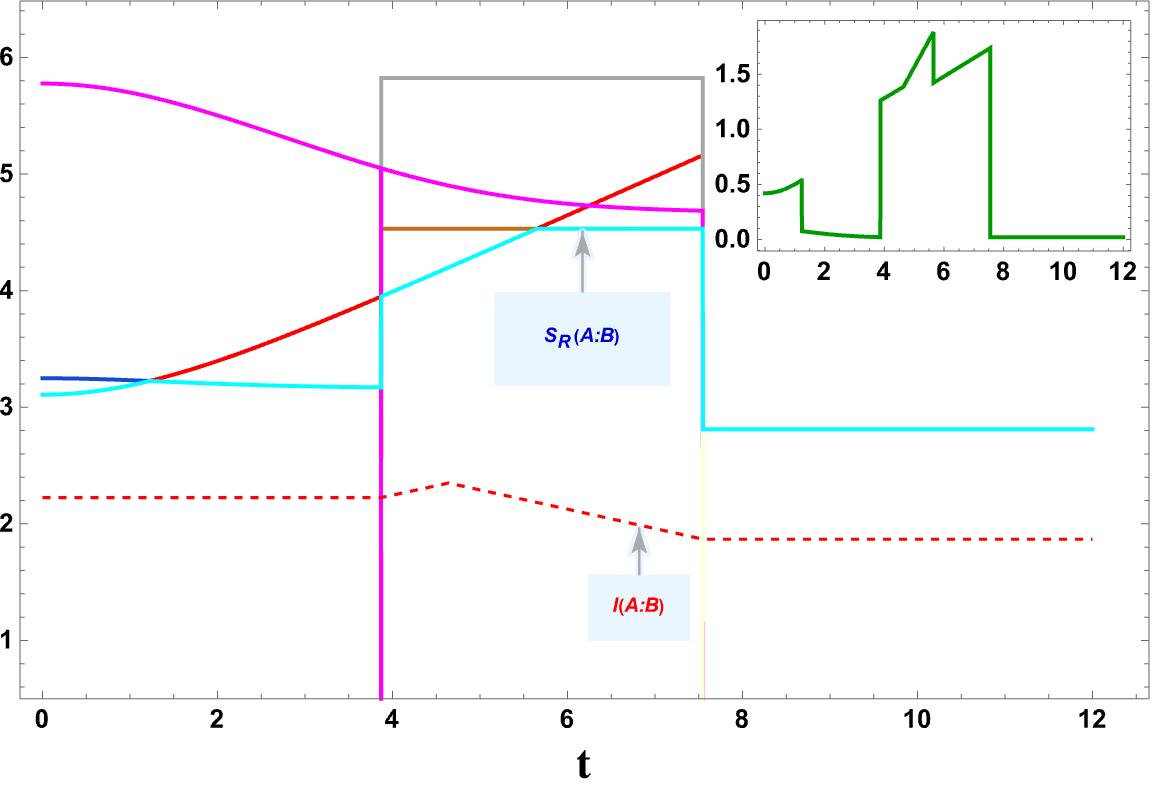}
		\caption{}
		\label{dis SR plot(2HM-1HM-dome)}
	\end{subfigure}
\caption{(a) Page curve of the EE for subsystems $A \cup B$. Here purple curve shows the minimum EE among all the phases. (b) Page curve of the reflected entropy between subsystems $A$ and $B$. Here cyan curve represents minimum $S_R$ and red dashed curve is mutual information (All graphs are in units of $c$). The inset plot shows the deviation from saturation of the Markov gap \cref{Markov gap}. The Page curves for the EE and $S_R$ is obtained with $u_h=1,\Delta_{L_1} = 0.89 ,\Delta_{L_2}=0.109, \Delta_{L_3}=0.102, \Delta_{L_4}=9\times10^{-6},\Delta_{R_1}=0.79, \Delta_{R_2}=0.109, \Delta_{R_3}=0.1,  \Delta_{R_4}=8.5\times10^{-6}, \rho_B=-3.75, \rho_\epsilon=100 $. }
\label{Dis Page curve1}
\end{figure}

By implementing a sufficiently larger brane angle than the previous two cases, it is possible to obtain the EE transition between \hyperref[Dis.EE phase1]{phase-1} and \hyperref[Dis.EE phase2]{phase-2} at time $T^{\text{disj}}_{1\rightarrow2}$ and \hyperref[Dis.EE phase2]{phase-2} and \hyperref[Dis.EE phase3]{phase-3} at time $T^{\text{disj}}_{{2\rightarrow3}}$. Here, we additionally consider that subsystem $B$ is located far away from the boundary. This entropy transition is depicted in \cref{dis plot(2HM-1HM-dome)}. The Page time $T^{\text{disj}}_{{2\rightarrow3}}$ is given in \cref{Page time 1HM-dome} and $T^{\text{disj}}_{{1\rightarrow2}}$ may be written as 
\begin{align}\label{Page time 2HM-1HM}
	T^{\text{disj}}_{{1\rightarrow2}}=u_h \cosh ^{-1}\left(\frac{u^{2}_h-u_h \left(e^{2 \rho_B}+1\right) (\Delta_{L_1}+\Delta_{R_1})+\Delta_{L_1} \Delta_{R_1}}{2 u_h \sqrt{\Delta_{L_1}\Delta_{R_1}} e^{2 \rho_B}}\right).
\end{align}
In the EE \hyperref[Dis.EE phase1]{phase-1}, the reflected entropy initially rises and then slowly falls until $T^{\text{disj}}_{{1\rightarrow2}}$ as the growth rate of the bulk EWCS which lands on the HM surface is lower than that of the HM surface. After $T^{\text{disj}}_{{1\rightarrow2}}$, in the EE \hyperref[Dis.EE phase2]{phase-2}, it increases again and then remains constant until $T^{\text{disj}}_{{2\rightarrow3}}$ as the bulk EWCS lands on the RT surface which does not cross the horizon. Finally in the EE \hyperref[Dis.EE phase3]{phase-3}, it saturates to a lesser constant value. This reflected entropy Page curve is shown in \cref{dis SR plot(2HM-1HM-dome)}. From the Page curve of the reflected entropy we observe that initially the Markov gap is greater than $\frac{2c}{3}\log{2}$ due to the two non-trivial boundaries of the bulk EWCS and after that it is always greater than $\frac{4c}{3}\log{2}$ as for all the bulk EWCS phases have four non-trivial boundaries which verify the inequality mentioned in \cref{Markov gap}.

\subsubsection{Case-IV}\label{sec:dis 2HM-dome}
\begin{figure}[H]
	\centering
	\begin{subfigure}[b]{0.45\textwidth}
		\centering
		\includegraphics[scale=.37]{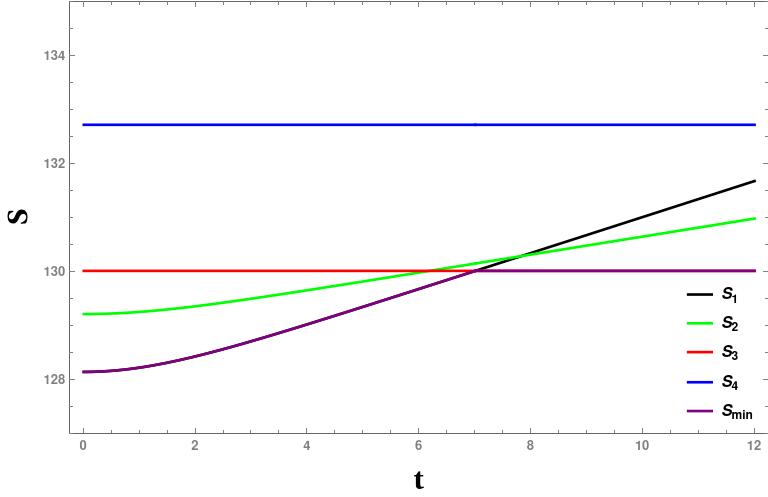}
		\caption{}
		\label{dis plot(2HM-dome)}
	\end{subfigure}
	\hfill
	\begin{subfigure}[b]{0.45\textwidth}
		\centering
		\includegraphics[scale=.342]{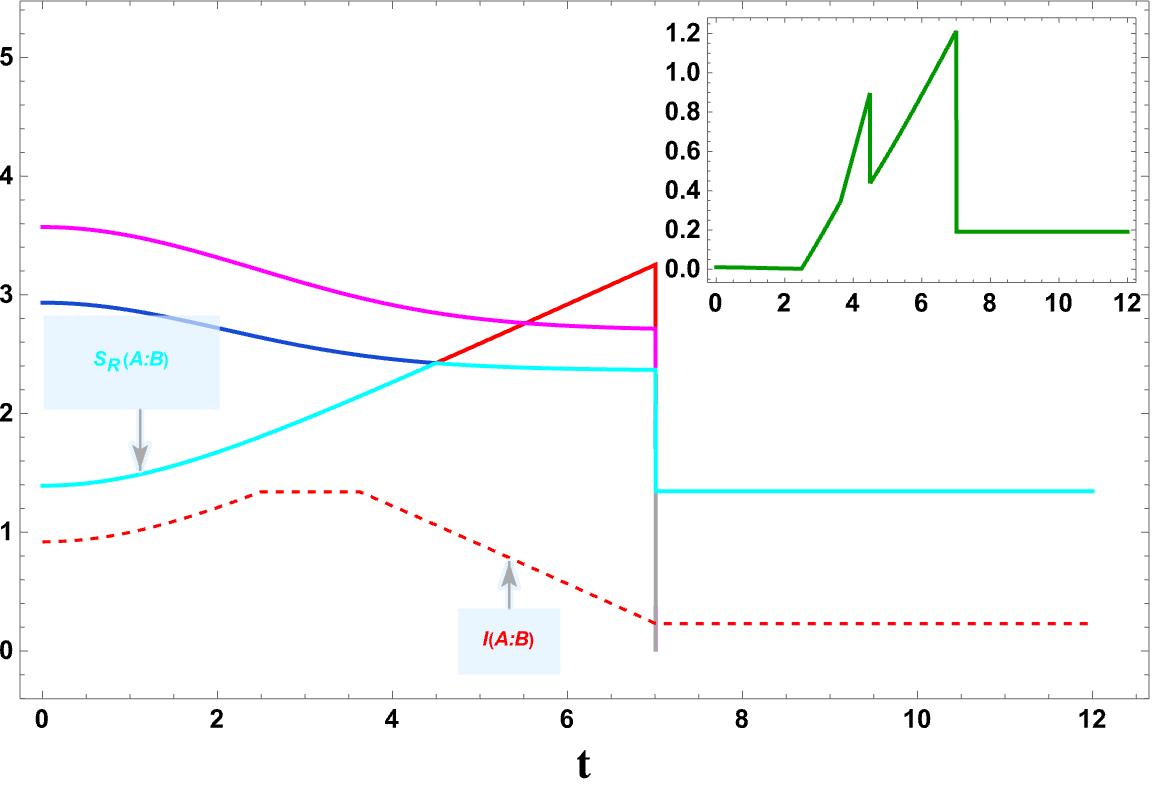}
		\caption{}
		\label{dis SR plot(2HM-dome)}
	\end{subfigure}
	\caption{(a) Page curve of the EE for subsystems $A \cup B$. Here purple curve shows the minimum EE among all the phases. (b) Page curve of the reflected entropy between subsystems $A$ and $B$. Here cyan curve represents minimum $S_R$ and red dashed curve is mutual information (All graphs are in units of $c$). The inset plot shows the deviation from saturation of the Markov gap \cref{Markov gap}. The Page curves for the EE and $S_R$ is obtained with $u_h=1, \Delta_{L_1}=0.3, \Delta_{L_2}=9.9 \times 10^{-4}, \Delta_{L_3}=4 \times 10^{-4},\Delta_{L_4}=2 \times 10^{-7}, \Delta_{R_1}=0.2, \Delta_{R_2}=9 \times 10^{-4}, \Delta_{R_3}=3 \times 10^{-4},\Delta_{R_4}=2 \times 10^{-7}, \rho_B=-3.5,   \rho_\epsilon=100$.}
\end{figure}

The EE transition between \hyperref[Dis.EE phase1]{phase-1} and \hyperref[Dis.EE phase3]{phase-3} may be obtained by using a relatively large brane angle than first two cases. Here we also take both subsystems away from the boundary. The Page time for this EE phase transition is given as
\begin{align}\label{Page time 2HM-dome(dis)}
	T^{\text{disj}}_{{1\rightarrow3}}=\cosh ^{-1}\left(\frac{\Delta_{L_1}+\Delta_{R_1}}{\sqrt{\Delta_{L_1} \Delta_{R_1}}}+\frac{\Delta_{L_4}+\Delta_{R_4}}{\sqrt{\Delta_{L_4} \Delta_{R_4}}}-A^2\right),
\end{align}
where 
\begin{align}
	A^2=& \left(\frac{\Delta_{L_1}}{\Delta_{R_1}}+\frac{\Delta_{R_1}}{\Delta_{L_1}}\right)+\left(\frac{\Delta_{L_4}}{\Delta_{R_4}}+\frac{\Delta_{R_4}}{\Delta_{L_4}}\right)-8 \Delta_{L_1} \Delta_{L_4} \Delta_{R_1} \Delta_{R_4} \left(\frac{\Delta_{L_1}+\Delta_{L_3}}{\sqrt{\Delta_{L_1} \Delta_{L_4}}}+\frac{\Delta_{R_1}+\Delta_{R_4}}{\sqrt{\Delta_{R_1} \Delta_{R_4}}}\right)\nonumber\\
	&+20+\frac{2 \Delta_{L_1} (-\Delta_{L_4}+2 \Delta_{R_1}+\Delta_{R_4})+2 \Delta_{L_4} (\Delta_{R_1}+2 \Delta_{R_4})-2 \Delta_{R_1} \Delta_{R_4}}{\sqrt{\Delta_{L_1} \Delta_{L_4} \Delta_{R_1} \Delta_{R_4}}}.
\end{align}
This EE transition and the Page curve of the reflected entropy are depicted in \cref{dis plot(2HM-dome)} and \cref{dis SR plot(2HM-dome)} respectively.  In the EE \hyperref[Dis.EE phase1]{phase-1} initially the reflected entropy increases with time as the bulk EWCS is the HM surface and then slowly falls until Page time as the growth rate of the bulk EWCS is lesser than the growth rate of the HM surface. After that in the EE \hyperref[Dis.EE phase3]{phase-3}, it saturates to a constant value. The Page curve of the reflected entropy shows that initially the Markov gap is always larger than $\frac{2c}{3}\log{2}$, which agrees with \cref{Markov gap}, since there are two non-trivial boundaries for the bulk EWCS phase. After that it is always greater than $\frac{4c}{3}\log{2}$ due to the four non-trivial boundaries of the bulk EWCS.

\section{Holographic reflected entropy: Adjacent subsystems}\label{Two adjacent subsystems}	

In this section we investigate the computation of the reflected entropy and the bulk EWCS corresponding to two adjacent subsystems $A\equiv(u^{}_{L_1},u^{}_{L_2})\cup (u^{}_{R_1},u^{}_{R_2})$ and $B\equiv(u^{}_{L_2},u^{}_{L_3})\cup (u^{}_{R_2},u^{}_{R_3})$ in the AdS$_3$/BCFT$_2$ setup described in \cref{BCFT in BH} where the BCFT$_2$ is defined on an AdS$_2$ black hole background. The R\'enyi reflected entropy in this scenario is defined in terms of the six-point twist field correlator as
\begin{align}\label{Ref-adj-def}
	S_n(AA^*)_{\psi_{m}} = \frac{1}{1-n}\log \frac{\langle \sigma_{g^{}_A}(u^{}_{L_1})\sigma_{g^{}_B g_A^{-1}}(u^{}_{L_{2}})\sigma_{g_B^{-1}}(u^{}_{L_{3}})\sigma_{g^{}_B}(u^{}_{R_{3}})\sigma_{g^{}_A g_B^{-1}}(u^{}_{R_{2}})\sigma_{g_A^{-1}}(u^{}_{R_{1}}) \rangle_{\mathrm{BCFT}^{\bigotimes mn}}}{\langle\sigma_{g^{}_m}(u^{}_{L_{1}})\sigma_{g_m^{-1}}(u^{}_{L_{3}})\sigma_{g^{}_m}(u^{}_{R_{3}})\sigma_{g_m^{-1}}(u^{}_{R_{1}}) \rangle^n_{\mathrm{BCFT}^{\bigotimes m}}}.
\end{align}
Note that for two adjacent subsystems, there are four possible phases of the EE depending on the subsystem size and its location. We will explain the computation of the various reflected entropy phases and the corresponding bulk EWCS for these EE phases in the following subsections.

\subsection{Entanglement entropy phase 1}\label{adj.EE phase1}
\begin{figure}[h!]
	\centering
	\includegraphics[scale=.8]{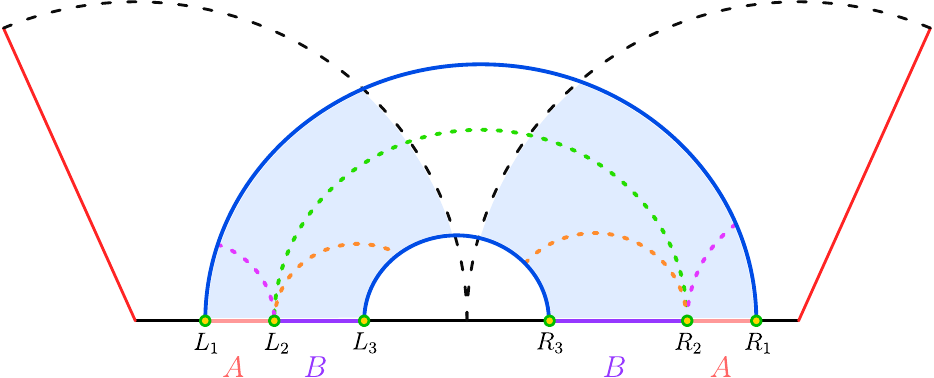}
	\caption{\it Schematic diagram depicting different phases for the EWCS (shown by different colored dashed curves) between subsystems $A$ and $B$ when the RT surface for $A \cup B$ is as shown by the solid blue curves. }
	\label{fig_adj.EE phase1}
\end{figure}
In this EE phase, both subsystems are considered to be large and far away from the boundary. So the EE in this phase is proportional to the lengths of two HM surfaces corresponding to points $L^{}_1(u^{}_{L_1},\rho^{}_\epsilon,t)$, $R^{}_1(u^{}_{R_1},\rho^{}_\epsilon,\tilde{t})$ and $L^{}_3(u^{}_{L_3},\rho^{}_\epsilon,t)$ and $R^{}_3(u^{}_{R_3},\rho_\epsilon,\tilde{t})$, depicted as solid blue curves in \cref{fig_adj.EE phase1}. Now by utilizing \cref{length of HM}, we may obtain the EE for this configuration as
\begin{align}
	S_1=& \frac{1}{4 G_N} \log \left[\frac{e^{2\rho_\epsilon} u_h}{u^{}_{L_{1}}u^{}_{R_{1}}}\left(\Delta_{L_1}+\Delta_{R_1}+2 \sqrt{\Delta_{L_1}\Delta_{R_1}} \cosh \frac{t}{u_h}\right)\right]\notag\\
	&+\frac{1}{4 G_N} \log \left[\frac{e^{2\rho_\epsilon} u_h}{u^{}_{L_{3}}u^{}_{R_{3}}}\left(\Delta_{L_3}+\Delta_{R_3}+2 \sqrt{\Delta_{L_3}\Delta_{R_3}} \cosh \frac{t}{u_h}\right)\right].
\end{align}
In this EE phase, we observe three possible phases of the reflected entropy or the bulk EWCS, shown as dashed curves in the figure \ref{fig_adj.EE phase1}. Here we assume that both the subsystems are far away from the boundary, therefore the OPE channel for the BCFT$_2$ correlator is favoured.
In the following, we explain the computation of the reflected entropy and corresponding bulk EWCS for this EE phase.

\subsection*{Phase-I} 
\paragraph{Reflected entropy:} In this phase we consider that both the subsystems are large and far away from the boundary. Hence, the six-point twist correlator in the numerator of \cref{Ref-adj-def} may be factorized into three two-point twist correlators  as follows
\begin{align}
	&\langle \sigma_{g^{}_A}(u^{}_{L_1})\sigma_{g^{}_B g_A^{-1}}(u^{}_{L_{2}})\sigma_{g_B^{-1}}(u^{}_{L_{3}})\sigma_{g^{}_B}(u^{}_{R_{3}})\sigma_{g^{}_A g_B^{-1}}(u^{}_{R_{2}})\sigma_{g_A^{-1}}(u^{}_{R_{1}}) \rangle_{\mathrm{CFT}^{\bigotimes mn}}\notag\\
	&=\langle \sigma_{g^{}_A}(u^{}_{L_{1}})\sigma_{g_A^{-1}}(u^{}_{R_{1}}) \rangle_{\mathrm{CFT}^{\bigotimes mn}} \langle\sigma_{g^{-1}_B}(u^{}_{L_{3}})\sigma_{g^{}_B}(u^{}_{R_{3}}) \rangle_{\mathrm{CFT}^{\bigotimes mn}}\notag\\
	&~~~~~~~~~\times
	\langle\sigma_{g^{}_B g_A^{-1}}(u^{}_{L_2})\sigma_{g^{}_A g_B^{-1}}(u^{}_{R_{2}})\rangle_{\mathrm{CFT}^{\bigotimes mn}}.
\end{align}
The twist field correlator in the denominator of \cref{Ref-adj-def} admits similar factorization and hence we have the following expression for the reflected entropy between the two adjacent subsystems 
\begin{align}\label{Ref-adj-1(i)}
	S_{R}(A:B)= \lim_{{m,n} \to 1}\frac{1}{1-n}\log \langle\sigma_{g^{}_B g_A^{-1}}(u^{}_{L_2})\sigma_{g^{}_A g_B^{-1}}(u^{}_{R_{2}})\rangle_{\mathrm{CFT}^{\bigotimes mn}}.
\end{align}
Since the field theory is described on a AdS$_2$ black hole background, therefore the computation of the above twist correlator is not straightforward. But we can map this field theory on the curved geometry to that on a flat background by using the conformal map given in \cref{w-z map}. Hence the above twist field correlator may be written as flat plane twist field correlator with an appropriate conformal factor as
\begin{align}\label{2p bulk-2p flat}
	\langle\sigma_{g^{}_B g_A^{-1}}(u^{}_{L_2})\sigma_{g^{}_A g_B^{-1}}(u^{}_{R_{2}})\rangle=\frac{4^{-h_{AB}}e^{\frac{u_{* }(u^{}_{L_2})h_{AB}}{2u_h}}\Omega(u^{}_{L_{2}})^{-h_{AB}}e^{\frac{u_{* }(u^{}_{R_{2}})h_{AB}}{2u_h}}\Omega(u^{}_{R_{2}})^{-h_{AB}}\,
		\epsilon^{2h_{AB}}}{({w^{}_{L_{2}}-w^{}_{R_{2}}})^{2h_{AB}}}\,.
\end{align}
Utilizing \cref{Ref-adj-1(i),2p bulk-2p flat,conformal weights}, the reflected entropy in this phase may be obtained as 
\begin{align}\label{SR-adj.phase1(i)}
	S_R(A:B)=\frac{c}{3} \log \left[\frac{4 u_h}{ u^{}_{L_{2}}u^{}_{R_{2}}\epsilon^2}\left(\Delta_{L_2}+\Delta_{R_2}+2 \sqrt{\Delta_{L_2}\Delta_{R_2}} \cosh \frac{t}{u_h}\right)\right].
\end{align}
\vspace{.05mm}
\paragraph{EWCS:} The bulk EWCS for this phase is proportional to the length of the HM surface corresponding to points $L^{}_{2}(u^{}_{L_2},\rho^{}_\epsilon,t)$ and $R^{}_2(u^{}_{R_2},\rho^{}_\epsilon,\tilde{t})$, shown as dashed green curve in \cref{fig_adj.EE phase1}. Now by utilizing \cref{length of HM}, the corresponding bulk EWCS may be obtained as 
\begin{equation}\label{EWCS-adj.phase1(i)}
	E_{W}(A:B)= \frac{1}{4 G_N} \log \left[\frac{e^{2\rho_\epsilon}  u_h }{u^{}_{L_{2}}u^{}_{R_{2}}}\left(\Delta_{L_2}+\Delta_{R_2}+2 \sqrt{\Delta_{L_2}\Delta_{R_2}} \cosh \frac{t}{u_h}\right)\right].
\end{equation}
Note that the above expression for the EWCS matches exactly with half of the reflected entropy computed in \cref{SR-adj.phase1(i)} upon utilizing \cref{bulk-bdy relation} and the standard Brown-Henneaux relation in $AdS_3/CFT_2$.
% which serves as a consistency check and also verify the holographic duality mentioned in \cref{duality}.

\subsection*{Phase-II} 
\paragraph{Reflected entropy:} For this case, consider that the subsystem $A$ is smaller than the subsystem $B$. Therefore, the six points twist correlator in the numerator of \cref{Ref-adj-def} may be factorized as one two-point twist correlators and one four-point twist correlator as
\begin{align}
	&\langle \sigma_{g^{}_A}(u^{}_{L_1})\sigma_{g^{}_B g_A^{-1}}(u^{}_{L_{2}})\sigma_{g_B^{-1}}(u^{}_{L_{3}})\sigma_{g^{}_B}(u^{}_{R_{3}})\sigma_{g^{}_A g_B^{-1}}(u^{}_{R_{2}})\sigma_{g_A^{-1}}(u^{}_{R_{1}}) \rangle_{\mathrm{CFT}^{\bigotimes mn}}\notag\\
	&= \langle \sigma_{g^{-1}_B}(u^{}_{L_{3}})\sigma_{g^{}_B}(u^{}_{R_{3}}) \rangle_{\mathrm{CFT}^{\bigotimes mn}}\langle\sigma_{g^{}_A}(u^{}_{L_{1}})\sigma_{g^{}_B g_A^{-1}}(u^{}_{L_2}) \sigma_{g^{}_A g_B^{-1}}(u^{}_{R_2})\sigma_{g_A^{-1}}(u^{}_{R_{1}})\rangle_{\mathrm{CFT}^{\bigotimes mn}}.
\end{align}
The computation of the four-point twist field correlator in the above expression is not straightforward. It is necessary that the composite twist field $\sigma_{g^{}_B g_A^{-1}}$ at $u^{}_{L_{2}}$ and $u^{}_{R_{2}}$ may be expanded in terms of twist fields $\sigma_{g^{}_B}$ and $\sigma_{g^{-1}_A}$. Using this the four-point twist field correlator may be written as
\begin{align}
	&\langle\sigma_{g^{}_A}(u^{}_{L_{1}})\sigma_{g^{}_B g_A^{-1}}(u^{}_{L_2}) \sigma_{g^{}_A g_B^{-1}}(u^{}_{R_2})\sigma_{g_A^{-1}}(u^{}_{R_{1}})\rangle_{\mathrm{CFT}^{\bigotimes mn}}\notag\\
	&=\langle\sigma_{g^{}_A}(u^{}_{L_{1}})\sigma_{g^{}_B}(u^{}_{L_2})\sigma_{g^{-1}_A}(u^ {\prime}_{L_2}) \sigma_{g^{}_A }(u^{}_{R_2})\sigma_{g^{-1}_B }(u^{\prime}_{R_2})\sigma_{g_A^{-1}}(u^{}_{R_{1}})\rangle_{\mathrm{CFT}^{\bigotimes mn}},
\end{align}
where we assume that $u^{\prime}_{L_2}$ $($$u^{\prime}_{R_2})$ are close to $u^{}_{L_2}$ ($u^{}_{R_2}$). This six-point twist field correlator may now be expanded in terms of two four-point conformal block as described in \cite{Banerjee:2016qca}. Finally, taking the OPE limit for the for the twist field located at $u^{}_{L_2}, u^{'}_{L_2}$ and $u^{}_{R_2}, u^{'}_{R_2}$, the above expression may be written as two three-point twist correlator as follows
\begin{align}
	&\langle\sigma_{g^{}_A}(u^{}_{L_{1}})\sigma_{g^{}_B g_A^{-1}}(u^{}_{L_2}) \sigma_{g^{}_A g_B^{-1}}(u^{}_{R_2})\sigma_{g_A^{-1}}(u^{}_{R_{1}})\rangle_{\mathrm{CFT}^{\bigotimes mn}} \notag \\
	&=\langle\sigma_{g^{}_A}(u^{}_{L_{1}})\sigma_{g^{}_B g_A^{-1}}(u^{}_{L_2})\sigma_{g_A^{-1}}(u^{}_{R_{1}})\rangle_{\mathrm{CFT}^{\bigotimes mn}} \langle\sigma_{g^{}_A}(u^{}_{L_{1}}) \sigma_{g^{}_A g_B^{-1}}(u^{}_{R_2})\sigma_{g_A^{-1}}(u^{}_{R_{1}})\rangle_{\mathrm{CFT}^{\bigotimes mn}}.
\end{align}
The denominator of \cref{Ref-adj-def} may be factorized into two two-point twist correlators and hence the reflected entropy between two adjacent subsystems may be written as 
\begin{align}
	S_R(A:B)= & \lim_{{m,n} \to 1}\frac{1}{1-n}\log\frac{\langle\sigma_{g^{}_A}(u^{}_{L_{1}})\sigma_{g^{}_B g_A^{-1}}(u^{}_{L_2})\sigma_{g_A^{-1}}(u^{}_{R_{1}})\rangle_{\mathrm{CFT}^{\bigotimes mn}} }{\langle\sigma_{g_m}(u_{L_1})\sigma_{g_m^{-1}}(u_{R_{1}}) \rangle^n_{\mathrm{CFT}^{\bigotimes m}}}\notag\\
	&+\lim_{{m,n} \to 1}\frac{1}{1-n}\log \langle\sigma_{g^{}_A}(u^{}_{L_{1}}) \sigma_{g^{}_A g_B^{-1}}(u^{}_{R_2})\sigma_{g_A^{-1}}(u^{}_{R_{1}})\rangle_{\mathrm{CFT}^{\bigotimes mn}}.
\end{align}
Now by utilizing \cref{w-z map} and form of the three point function and taking the replica limit, we may obtain the reflected entropy for this phase as
\begin{align}\label{SR adj.Phase1(ii)}
	S_{R}(A:B)=&\frac{c}{3} \log\left[\frac{4 u_h(\sqrt{\Delta_{L_1}}-\sqrt{\Delta_{L_2}})}{\epsilon (u_h- \Delta_{ L_2})}
	\sqrt{\frac{(\sqrt{\Delta_{L_2}}+e^\frac{t}{u_h}\sqrt{\Delta_{R_1}})(\sqrt{\Delta_{R_1}}+e^\frac{t}{u_h}\sqrt{\Delta_{L_2}})}{(\sqrt{\Delta_{L_1}}+ e^{\frac{t}{u_h}}\sqrt{\Delta_{R_1}})(\sqrt{\Delta_{R_1}}+ e^{\frac{t}{u_h}}\sqrt{\Delta_{L_1}})}}  \right]\nonumber\\
	&+(\Delta_{ L_i} \longleftrightarrow \Delta_{ R_i}).
\end{align}
\begin{comment}
\begin{align}
	S_{R}(A:B)=&\frac{c}{3} \log\left[\frac{4 u_h(\sqrt{\Delta_{L_1}}-\sqrt{\Delta_{L_2}})}{u^{}_{L_2}\epsilon}
	\sqrt{\frac{(\sqrt{\Delta_{L_2}}+e^\frac{t}{u_h}\sqrt{\Delta_{R_1}})(\sqrt{\Delta_{R_1}}+e^\frac{t}{u_h}\sqrt{\Delta_{L_2}})}{(\sqrt{\Delta_{L_1}}+ e^{\frac{t}{u_h}}\sqrt{\Delta_{R_1}})(\sqrt{\Delta_{R_1}}+ e^{\frac{t}{u_h}}\sqrt{\Delta_{L_1}})}}  \right]\nonumber\\
	&+\frac{c}{3} \log\left[\frac{4(\sqrt{\Delta_{R_1}}-\sqrt{\Delta_{R_2}})}{ u^{}_{R_2}\epsilon}
	\sqrt{\frac{(\sqrt{\Delta_{L_1}}+e^\frac{t}{u_h}\sqrt{\Delta_{R_2}})(\sqrt{\Delta_{R_2}}+e^\frac{t}{u_h}\sqrt{\Delta_{L_1}})}{(\sqrt{\Delta_{L_1}}+ e^{\frac{t}{u_h}}\sqrt{\Delta_{R_1}})(\sqrt{\Delta_{R_1}}+ e^{\frac{t}{u_h}}\sqrt{\Delta_{L_1}})}}
	\right].
\end{align}
\end{comment}
\paragraph{EWCS:} The corresponding bulk EWCS is proportional to the sum of the lengths of two geodesics, shown as dashed magenta curves in \cref{fig_adj.EE phase1}. The first geodesic connects $L_2$ to the HM surface, while the second geodesic joins $R_2$ to the HM surface. The bulk EWCS may be obtained by utilizing the embedding coordinates of point $L_1$, $L_2$ and $R_1$ in \cref{EWCS three pt. formula} and adding the contribution from the right TFD copy as
\begin{align} 
	E_{W}(A:B)=&\frac{1}{4 G_N} \log\left[\frac{2 e^{\rho_\epsilon} u_h(\sqrt{\Delta_{L_1}}-\sqrt{\Delta_{L_2}})}{(u_h - \Delta_{ L_2})} \sqrt{\frac{\Delta_{L_2}+\Delta_{R_1}+ 2 \sqrt{\Delta_{L_2}\Delta_{R_1}}\cosh\frac{t}{u_h}}{\Delta_{L_1}+\Delta_{R_1}+2 \sqrt{\Delta_{L_1} \Delta_{R_1}}\cosh\frac{t}{u_h}}}\right]\nonumber\\
	&+(\Delta_{ L_i} \longleftrightarrow \Delta_{ R_i}).
\end{align}
Again by utilizing \cref{bulk-bdy relation} and the Brown-Henneaux relation, we find that the reflected entropy computed in \cref{SR adj.Phase1(ii)} matches exactly with twice of the bulk EWCS.
% which verify the duality given in \cref{duality}. 

\subsection*{Phase-III} 
\paragraph{Reflected entropy:} In this phase, we assume that subsystem $B$ is smaller than $A$, hence the six points twist correlator in the numerator of \cref{Ref-adj-def} can be factorized into one two-point twist correlator and one four-point twist correlator as follows
\begin{align}
	&\langle \sigma_{g^{}_A}(u^{}_{L_1})\sigma_{g^{}_B g_A^{-1}}(u^{}_{L_{2}})\sigma_{g_B^{-1}}(u^{}_{L_{3}})\sigma_{g^{}_B}(u^{}_{R_{3}})\sigma_{g^{}_A g_B^{-1}}(u^{}_{R_{2}})\sigma_{g_A^{-1}}(u^{}_{R_{1}}) \rangle_{\mathrm{CFT}^{\bigotimes mn}}\notag\\
	&=\langle \sigma_{g_A}(u^{}_{L_{1}})\sigma_{g_A^{-1}}(u^{}_{R_{1}}) \rangle_{\mathrm{CFT}^{\bigotimes mn}}\langle\sigma_{g_B g_A^{-1}}(u^{}_{L_2})\sigma_{g_B^{-1}}(u^{}_{L_{3}})\sigma_{g^{}_B}(u^{}_{R_{3}}) \sigma_{g^{}_A g_B^{-1}}(u^{}_{R_2})\rangle_{\mathrm{CFT}^{\bigotimes mn}}.
\end{align}
As explained in the previous subsection, the above four-point twist correlator may be written as two three-point twist correlator as follows
\begin{align}
	&\langle\sigma_{g_B g_A^{-1}}(u^{}_{L_2})\sigma_{g_B^{-1}}(u^{}_{L_{3}})\sigma_{g^{}_B}(u^{}_{R_{3}}) \sigma_{g^{}_A g_B^{-1}}(u^{}_{R_2})\rangle_{\mathrm{CFT}^{\bigotimes mn}}\notag \\
	&=\langle\sigma_{g^{}_B}(u^{}_{R_{3}})\sigma_{g^{}_B g_A^{-1}}(u^{}_{L_2})\sigma_{g_B^{-1}}(u^{}_{L_{3}})\rangle_{\mathrm{CFT}^{\bigotimes mn}} \langle\sigma_{g^{}_B}(u^{}_{R_{3}}) \sigma_{g^{}_A g_B^{-1}}(u^{}_{R_2})\sigma_{g_B^{-1}}(u^{}_{L_{3}})\rangle_{\mathrm{CFT}^{\bigotimes mn}}.
\end{align}
The denominator of \cref{Ref-adj-def} may also be factorized into two two-point twist correlator and therefore the expression for the reflected entropy for this phase may be written as 
\begin{align}
	S_R(A:B)= & \lim_{{m,n} \to 1}\frac{1}{1-n}\log\frac{\langle\sigma_{g^{}_B}(u^{}_{R_{3}})\sigma_{g^{}_B g_A^{-1}}(u^{}_{L_2})\sigma_{g_B^{-1}}(u^{}_{L_{3}})\rangle_{\mathrm{CFT}^{\bigotimes mn}} }{\langle\sigma_{g_m^{-1}}(u^{}_{L_3})\sigma_{g_m}(u^{}_{R_{3}}) \rangle^n_{\mathrm{CFT}^{\bigotimes m}}}\notag\\
	&+\lim_{{m,n} \to 1}\frac{1}{1-n}\log \langle\sigma_{g^{}_B}(u^{}_{R_{3}}) \sigma_{g^{}_A g_B^{-1}}(u^{}_{R_2})\sigma_{g_B^{-1}}(u^{}_{L_{3}})\rangle_{\mathrm{CFT}^{\bigotimes mn}}.
\end{align}
Now by utilizing \cref{w-z map} and form of the three point function, we may obtain the reflected entropy in this phase as
\begin{align}\label{SR adj.Phase1(iii)}
	S_{R}(A:B)=&\frac{c}{3} \log\left[\frac{4 u_h(\sqrt{\Delta_{L_2}}-\sqrt{\Delta_{L_3}})}{\epsilon (u_h-\Delta_{ L_2})}
	\sqrt{\frac{(\sqrt{\Delta_{L_2}}+e^\frac{t}{u_h}\sqrt{\Delta_{R_3}})(\sqrt{\Delta_{R_3}}+e^\frac{t}{u_h}\sqrt{\Delta_{L_2}})}{(\sqrt{\Delta_{L_3}}+ e^{\frac{t}{u_h}}\sqrt{\Delta_{R_3}})(\sqrt{\Delta_{R_3}}+ e^{\frac{t}{u_h}}\sqrt{\Delta_{L_3}})}}  \right]\nonumber\\
	&+(\Delta_{ L_i} \longleftrightarrow \Delta_{ R_i}).
\end{align}

\paragraph{EWCS:} The bulk EWCS corresponds to the sum of two geodesic lengths depicted as dashed orange curves in \cref{fig_adj.EE phase1}. The first geodesic connects $L_2$ to the HM surface and the second connects $R_2$ to the HM surface. This may now be calculated by using the embedding coordinates of points $L_2$, $L_3$, and $R_3$ in the  \cref{EWCS three pt. formula} and adding the contribution from the right TFD copy as 
\begin{align}\label{EWCS adj.Phase1(iii)}
	E_{W}(A:B)=&\frac{1}{4 G_N} \log\left[\frac{2e^{\rho_\epsilon} u_h(\sqrt{\Delta_{L_2}}-\sqrt{\Delta_{L_3}})}{(u_h-\Delta_{ L_2})} \sqrt{\frac{\Delta_{L_2}+\Delta_{R_3}+ 2 \sqrt{\Delta_{L_2}\Delta_{R_3}}\cosh\frac{t}{u_h}}{\Delta_{L_3}+\Delta_{R_3}+2 \sqrt{\Delta_{L_3} \Delta_{R_3}}\cosh\frac{t}{u_h}}}\right]\nonumber\\
	&+(\Delta_{ L_i} \longleftrightarrow \Delta_{ R_i}).
\end{align}
Note that by utilizing \cref{bulk-bdy relation} and the Brown-Henneaux relation, we find that the bulk EWCS matches with half of the reflected entropy obtained in \cref{SR adj.Phase1(iii)}.
% which is again consistent with the holographic duality mentioned earlier.

\subsection{Entanglement entropy phase 2}\label{adj.EE phase2}
\begin{figure}[h!]
	\centering
	\includegraphics[scale=.8]{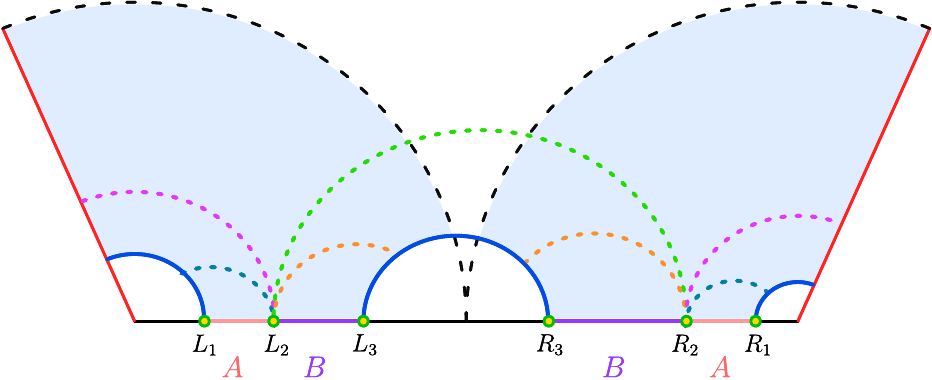}
	\caption{\it Diagrammatic representation of various phases of the EWCS (depicted as dashed curves) between subsystems $A$ and $B$ when the RT surface for $A \cup B$ is represented by the solid blue curves. }
	\label{fig_adj.EE phase2}
\end{figure}
In this EE phase, we consider the subsystem $A$ to be close to boundary while B is far away. Hence, the EE for this phase corresponds to the sum of the lengths of the HM surface between points $L_2$ and $R_3$ and two island surfaces, shown as solid blue curves in \cref{fig_adj.EE phase2}. The length of the HM surface and island surface is computed in \cref{length of HM} and \cref{length of island} respectively. So by utilizing these equations, the EE for this phase may be written as
\begin{align}
	S_2= \frac{1}{4 G_N} \log \left[\frac{u_h}{u^{}_{L_{3}}u^{}_{R_{3}}} \left(\Delta_{L_3}+\Delta_{R_3}+2 \sqrt{\Delta_{L_3}\Delta_{R_3}} \cosh \frac{t}{u_h}\right)\right]+\frac{1}{2 G_N}(2\rho_\epsilon-\rho^{}_B).
\end{align}
As illustrated in \cref{fig_adj.EE phase2}, there are four different phases of the reflected entropy or the bulk EWCS. In the following subsections, we explain the computation of the reflected entropy and bulk EWCS.

\subsection*{Phase-I}
The reflected entropy or the bulk EWCS in this phase is similar to the first case of the \cref{adj.EE phase1}, shown as dashed green curve in \cref{fig_adj.EE phase2}. Therefore, the bulk EWCS is given by \cref{EWCS-adj.phase1(i)}.

\subsection*{Phase-II}
\paragraph{Reflected entropy:} In this reflected entropy phase, we assume that the subsystem $A$ is large enough, hence the six points twist correlator in numerator of \cref{Ref-adj-def} may be factorized into a two-point and four one-point twist correlators in the BCFT$_2$ as follows
\begin{align}
	&\langle \sigma_{g^{}_A}(u^{}_{L_1})\sigma_{g^{}_B g_A^{-1}}(u^{}_{L_{2}})\sigma_{g_B^{-1}}(u^{}_{L_{3}})\sigma_{g^{}_B}(u^{}_{R_{3}})\sigma_{g^{}_A g_B^{-1}}(u^{}_{R_{2}})\sigma_{g_A^{-1}}(u^{}_{R_{1}}) \rangle_{\mathrm{BCFT}^{\bigotimes mn}}\notag\\
	&= \langle \sigma_{g^{}_A}(u^{}_{L_1})\rangle_{\mathrm{BCFT}^{\bigotimes mn}}\langle\sigma_{g_A^{-1}}(u^{}_{R_{1}}) \rangle_{\mathrm{BCFT}^{\bigotimes mn}}\langle\sigma_{g_B^{-1}}(u^{}_{L_{3}})\sigma_{g^{}_B}(u^{}_{R_{3}})\rangle_{\mathrm{CFT}^{\bigotimes mn}}\notag\\
	&~~~~~~~~~~~~\times  \langle\sigma_{g^{}_B g_A^{-1}}(u^{}_{L_{2}})\rangle_{\mathrm{BCFT}^{\bigotimes mn}}\langle\sigma_{g^{}_A g_B^{-1}}(u^{}_{R_{2}})\rangle_{\mathrm{BCFT}^{\bigotimes mn}},
\end{align}
where as the subsystem $A$ is close to the boundary, the BOE channel is favoured for the twist field correlator corresponding to the end points of $A$ while for the twist field correlator corresponding to the points ${L_{3}}$ and ${R_{3}}$ the OPE channel is favoured. The twist field correlator in the denominator of \cref{Ref-adj-def} is also factorized similarly. Now by using \cref{Ref-adj-def} the reflected entropy for this phase may be written as
\begin{align}
	S_R (A:B)=\lim_{{m,n} \to 1}\frac{1}{1-n}\log \langle\sigma_{g^{}_B g_A^{-1}}(u_{L_{2}})\rangle_{\mathrm{BCFT}^{\bigotimes mn}}\langle\sigma_{g^{}_A g_B^{-1}}(u_{R_{2}})\rangle_{\mathrm{BCFT}^{\bigotimes mn}}.
\end{align}
By utilizing \cref{w-z map} and the expression of the one point function for the BCFT$_2$ on a black hole background \cite{Geng:2022dua}, we may obtain the reflected entropy between the two adjacent subsystems as 
\begin{align}\label{SR adj.phase2(ii)}
	S_R(A:B)= 4 \log g^{}_B +\frac{2c}{3}\log\left(\frac{2}{\epsilon}\right).
\end{align}
\paragraph{EWCS:} The corresponding bulk EWCS is proportional to the sum of the lengths of two island surfaces, one beginning at point $L_2$ and ending at the EOW brane and the other starting at point $R_2$ and ending at the nearby EOW brane in the right TFD copy. These surfaces are depicted as dashed magenta curves in \cref{fig_adj.EE phase2}.  Now by utilizing the length of the island surface given in \cref{length of island}, we may obtain the corresponding EWCS as
\begin{align}\label{adj.island EWCS}
	E_{W}(A:B)= \frac{1}{2 G_N}(\rho_\epsilon-\rho^{}_B).
\end{align}
Note that the reflected entropy computed in \cref{SR adj.phase2(ii)} matches exactly with twice of the EWCS upon using \cref{bulk-bdy relation} and Brown-Henneaux relation.
% which serves as a consistency check.   

\subsection*{Phase-III} 
\paragraph{Reflected entropy:} For this phase we consider that the subsystem $A$ is smaller than $B$, hence the six-point twist correlator in the numerator of \cref{Ref-adj-def} may be factorized into three two-point twist correlators as follows
\begin{align}
	&\langle \sigma_{g^{}_A}(u^{}_{L_1})\sigma_{g^{}_B g_A^{-1}}(u^{}_{L_{2}})\sigma_{g_B^{-1}}(u^{}_{L_{3}})\sigma_{g^{}_B}(u^{}_{R_{3}})\sigma_{g^{}_A g_B^{-1}}(u^{}_{R_{2}})\sigma_{g_A^{-1}}(u^{}_{R_{1}}) \rangle_{\mathrm{BCFT}^{\bigotimes mn}}\notag\\
	&=\langle \sigma_{g^{}_A}(u^{}_{L_{1}})\sigma_{g^{}_B g_A^{-1}}(u^{}_{L_{2}})\rangle_{\mathrm{BCFT}^{\bigotimes mn}}\langle\sigma_{g^{}_A g_B^{-1}}(u^{}_{R_{2}})\sigma_{g_A^{-1}}(u^{}_{R_{1}})\rangle_{\mathrm{BCFT}^{\bigotimes mn}}\langle\sigma_{g_B^{-1}}(u^{}_{L_{3}})\sigma_{g^{}_B}(u^{}_{R_{3}}) \rangle_{\mathrm{BCFT}^{\bigotimes mn}}.
\end{align}
The last twist field correlator of the above equation cancels with a similar correlator originating from the factorization of the denominator in \cref{Ref-adj-def}. The remaining twist field correlators may be computed by transforming these to the twist field correlators defined on the conformally flat cylindrical background where the doubling trick is implemented \cite{Cardy:2004hm}. The expression for the reflected entropy for this phase may then be written as follows 
\begin{align}\label{SR 2(III)}
	S_R(A:B)= &\lim_{{m,n} \to 1}\frac{1}{1-n}\log \frac{\langle \sigma_{g^{}_A}(u_{\star}({L_{1}}))\sigma_{g^{-1}_A}(-u_{\star}({L_{1}}))\sigma_{g^{}_B g_A^{-1}}(u_{\star}({L_{2}}))\rangle_{\mathrm{CFT}^{\bigotimes mn}}}{\langle\sigma_{g^{}_m}(u_{\star}({L_{1}}))\sigma_{g^{-1}_m}(-u_{\star}({L_{1}}))\rangle^n_{\mathrm{CFT}^{\bigotimes m}}}\notag\\
	&+\lim_{{m,n} \to 1}\frac{1}{1-n}\log\frac{ \langle\sigma_{g^{}_A g_B^{-1}}(u_{\star}({R_{2}}))\sigma_{g_A^{-1}}(u_{\star}({R_{1}}))\sigma_{g^{}_A}(-u_{\star}({R_{1}}))\rangle_{\mathrm{CFT}^{\bigotimes mn}}}{\langle\sigma_{g_m^{-1}}(u_{\star}({R_{1}}))\sigma_{g^{}_m}(-u_{\star}({R_{1}})) \rangle^n_{\mathrm{CFT}^{\bigotimes m}}}.
\end{align}
%\begin{align}
%	S_R(A:B)= &\lim_{{m,n} \to 1}\frac{1}{1-n}\log \frac{\langle \sigma_{g^{}_A}(u^{}_{L_{1}})\sigma_{g^{-1}_A}(\bar{u}^{}_{L_{1}})\sigma_{g^{}_B g_A^{-1}}(u^{}_{L_{2}})\rangle_{\mathrm{CFT}^{\bigotimes mn}}}{\langle\sigma_{g^{}_m}(u^{}_{L_{1}})\sigma_{g^{-1}_m}(\bar{u}^{}_{L_{1}})\rangle^n_{\mathrm{CFT}^{\bigotimes m}}}\notag\\
%	&+\lim_{{m,n} \to 1}\frac{1}{1-n}\log\frac{ \langle\sigma_{g^{}_A g_B^{-1}}(u^{}_{R_{2}})\sigma_{g_A^{-1}}(u^{}_{R_{1}})\sigma_{g^{}_A}(\bar{u}^{}_{R_{1}})\rangle_{\mathrm{CFT}^{\bigotimes mn}}}{\langle\sigma_{g_m^{-1}}(u^{}_{R_{1}})\sigma_{g^{}_m}(\bar{u}^{}_{R_{1}}) \rangle^n_{\mathrm{CFT}^{\bigotimes m}}}.
%\end{align}
Now by utilizing \cref{w-z map} and the form of the flat plane three point twist correlator in \cref{SR 2(III)}, the reflected entropy between two adjacent subsystems may be obtained as
\begin{align}\label{SR adj.Phase2(iii)}
	S_R(A:B)= & \frac{c}{3}\log\left[\frac{4 u_h(\sqrt{\Delta_{L_1}}-\sqrt{\Delta_{L_2}}) (-u_h+\Delta_{L_1} \Delta_{L_2})}{u^{}_{L_{1}}u^{}_{L_{2}}\epsilon}\right]\nonumber\\
	&+\frac{c}{3}\log\left[\frac{4(\sqrt{\Delta_{R_1}}-\sqrt{\Delta_{R_2}}) (-u_h+\Delta_{R_1} \Delta_{R_2})}{u^{}_{R_{1}}u^{}_{R_{2}}\epsilon} \right]. 
\end{align}
\paragraph{EWCS:} The bulk EWCS for this phase is proportional to the sum of the length of two geodesics one of which connects the point $L_2$ to an arbitrary point $(u^{}_{L_1},\rho^{}_E,t)$ on the left island surface and second connects the point $R_2$ to an arbitrary point $(u^{}_{R_1},\rho^{}_F,\tilde{t})$ on the right island surface. These geodesics are shown as dark green dashed curves in \cref{fig_adj.EE phase2}. Now by using the embedding coordinates of points $L_2$ and $(u^{}_{L_1},\rho^{}_E,t)$ in \cref{Geodesic length}, the length of the first geodesic may be written as 
\begin{align}\label{length PE}
	L=\log \left[\left(\frac{2u_h(\sqrt{\Delta_{L_{1}}}-\sqrt{\Delta_{L_{2}}})^2 +u^{}_{L_{1}} u^{}_{L_{2}}}{u^{}_{L_{1}} u^{}_{L_{2}}}\right)\cosh\rho^{}_E-\sinh \rho^{}_E\right]+\rho_\epsilon.
\end{align}
The EWCS is obtained by extremizing the above expression with respect to $\rho^{}_E$. The extremum value of $\rho^{}_E$ is then given as
\begin{align}
	\rho^{}_E= \frac{1}{2}\log \left(\frac{2u_h(\sqrt{\Delta_{L_{1}}}-\sqrt{\Delta_{L_{2}}})^2 +u^{}_{L_{1}} u^{}_{L_{2}}}{u_h (\sqrt{\Delta_{L_{1}}}-\sqrt{\Delta_{L_{2}}})^2}\right).
\end{align}
Substituting this in \cref{length PE} and adding contribution from the right TFD copy, we may obtain the corresponding EWCS as
\begin{align}\small\label{island ewcs}
	E_{W}(A:B)=& \frac{1}{4 G_N}\log\left[\frac{2e^{\rho_\epsilon} u_h (\sqrt{\Delta_{L_1}}-\sqrt{\Delta_{L_2}}) (-u_h+\Delta_{L_1} \Delta_{L_2})}{u^{}_{L_{1}}u^{}_{L_{2}}}\right]\nonumber\\
	&+\frac{1}{4 G_N}\log\left[\frac{2e^{\rho_\epsilon}(\sqrt{\Delta_{R_1}}-\sqrt{\Delta_{R_2}}) (-u_h+\Delta_{R_1} \Delta_{R_2})}{u^{}_{R_{1}}u^{}_{R_{2}}} \right].
\end{align}
Here also the above expression of the EWCS is matches exactly with half of the reflected entropy obtained in \cref{SR adj.Phase2(iii)} by utilizing \cref{bulk-bdy relation} and Brown-Henneaux relation.

\subsection*{Phase-IV}
The reflected entropy or the bulk EWCS in this phase is similar to the first case of the \cref{adj.EE phase1}, shown as dashed orange curves in \cref{fig_adj.EE phase2}. Therefore, the bulk EWCS is given by \cref{EWCS adj.Phase1(iii)}.

\subsection{Entanglement entropy phase 3}\label{adj.EE phase3}
\begin{figure}[h!]
	\centering
	\includegraphics[scale=.8]{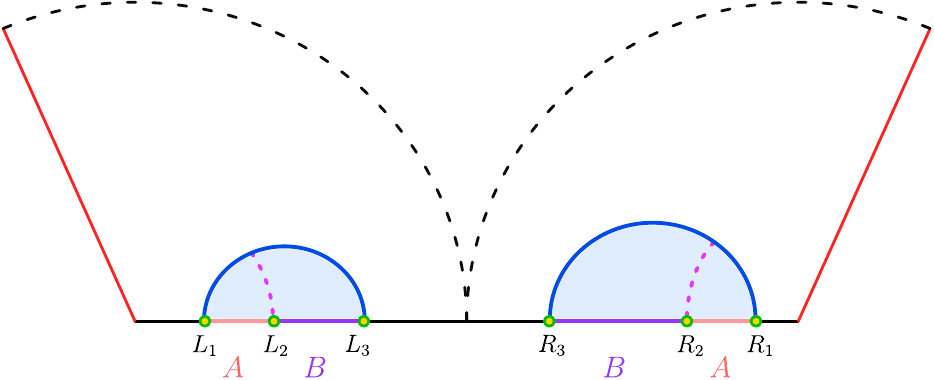}
	\caption{\it Schematic diagram of the EWCS (represented by dashed curves) between subsystems $A$ and $B$ when the RT surface for $A cup B$ is as depicted by the solid blue curves.}
	\label{fig_adj.EE phase3}
\end{figure}
For this EE phase we assume that both the subsystems are very small and close to each other away from the boundary. Hence, the EE for this phase is proportional to the length of two dome-type RT surfaces, shown as solid blue curves in \cref{fig_adj.EE phase3}. Now by utilizing the embedding coordinates of points $L_1$ and $L_3$ in \cref{Geodesic length}, the geodesic length of the dome-type RT surface may be written as
\begin{align}
	L= \log\left(\frac{u_h}{u^{}_{L_{1}}u^{}_{L_{3}}}(\sqrt{\Delta_{L_1}}-\sqrt{\Delta_{L_3}})^2
	\right)+ 2 \rho_\epsilon.
\end{align}
Adding the contribution from the right TFD copy, the EE for this phase may be obtained as
\begin{align}\label{dome EE}
	S_3=\frac{1}{4 G_N} \log \left(\frac{e^{2\rho^{}_\epsilon} u_h}{u^{}_{L_{1}}u^{}_{L_{3}}}(\sqrt{\Delta_{L_1}}-\sqrt{\Delta_{L_3}})^2
	\right)+\frac{1}{4 G_N} \log \left(\frac{e^{2\rho^{}_\epsilon} u_h}{u^{}_{R_{1}}u^{}_{R_{3}}}(\sqrt{\Delta_{R_1}}-\sqrt{\Delta_{R_3}})^2
	\right).
\end{align}
For this EE phase we observe only one phase for the reflected entropy or the bulk EWCS, shown as dashed magenta curves in \cref{fig_adj.EE phase3}. Here also the subsystems are away from the boundary, hence the OPE channel for the BCFT$_2$ correlator is favoured.

\paragraph{Reflected entropy:} For the reflected entropy computation in this phase, the six-point twist correlator in the numerator of \cref{Ref-adj-def} may be factorized into two three-point twist correlators as 
\begin{align}
	&\langle \sigma_{g^{}_A}(u^{}_{L_1})\sigma_{g^{}_B g_A^{-1}}(u^{}_{L_{2}})\sigma_{g_B^{-1}}(u^{}_{L_{3}})\sigma_{g^{}_B}(u^{}_{R_{3}})\sigma_{g^{}_A g_B^{-1}}(u^{}_{R_{2}})\sigma_{g_A^{-1}}(u^{}_{R_{1}}) \rangle_{\mathrm{CFT}^{\bigotimes mn}}\notag\\
	&= \langle \sigma_{g^{}_A}(u^{}_{L_{1}})\sigma_{g^{}_B g_A^{-1}}(u^{}_{L_{2}})\sigma_{g^{-1}_B}(u^{}_{L_{3}})\rangle_{\mathrm{CFT}^{\bigotimes mn}} \langle\sigma_{g^{-1}_A}(u^{}_{R_{1}})\sigma_{g^{}_A g_B^{-1}}(u^{}_{R_{2}})\sigma_{g^{}_B}(u^{}_{R_{3}})\rangle_{\mathrm{CFT}^{\bigotimes mn}}.
\end{align}
The denominator of \cref{Ref-adj-def} factorizes into two two-point twist correlators in the CFT$_2$. Hence the reflected entropy in this phase may be expressed as 
\begin{align}
	S_R(A:B)= &\lim_{{m,n} \to 1}\frac{1}{1-n}\log \frac{\langle \sigma_{g^{}_A}(u^{}_{L_{1}})\sigma_{g^{}_B g_A^{-1}}(u^{}_{L_{2}})\sigma_{g^{-1}_B}(u^{}_{L_{3}})\rangle_{\mathrm{CFT}^{\bigotimes mn}}}{\langle\sigma_{g^{}_m}(u^{}_{L_{1}})\sigma_{g^{-1}_m}(u^{}_{L_{3}})\rangle^n_{\mathrm{CFT}^{\bigotimes m}}}\notag\\
	&+ \lim_{{m,n} \to 1}\frac{1}{1-n}\log\frac{ \langle\sigma_{g^{-1}_A}(u^{}_{R_{1}})\sigma_{g^{}_A g_B^{-1}}(u^{}_{R_{2}})\sigma_{g^{}_B}(u^{}_{R_{3}})\rangle_{\mathrm{CFT}^{\bigotimes mn}}}{\langle\sigma_{g_m^{-1}}(u^{}_{R_{1}})\sigma_{g^{}_m}(u^{}_{R_{3}}) \rangle^n_{\mathrm{CFT}^{\bigotimes m}}}.
\end{align}
Now by utilizing \cref{w-z map} and the form of three point twist correlator in the previous expression, the reflected entropy for this phase may be obtained as
\begin{align}\label{SR adj.Phase3}
	S_R(A:B)=& \frac{c}{3}\log\left[\frac{4 u_h(\sqrt{\Delta_{L_1}}-\sqrt{\Delta_{L_2}})(\sqrt{\Delta_{L_2}}-\sqrt{\Delta_{L_3}})}{\epsilon \, u^{}_{L_2}(\sqrt{\Delta_{L_1}}-\sqrt{\Delta_{L_3}})}\right]\nonumber\\
	&+\frac{c}{3}\log\left[\frac{4(\sqrt{\Delta_{R_1}}-\sqrt{\Delta_{R_2}})(\sqrt{\Delta_{R_2}}-\sqrt{\Delta_{R_3}})}{\epsilon \, u^{}_{R_2}(\sqrt{\Delta_{R_1}}-\sqrt{\Delta_{R_3}})}\right].
\end{align}
\paragraph{EWCS:} The bulk EWCS for this phase corresponds to the sum of the length of two geodesic, depicted as dashed magenta curves in \cref{fig_adj.EE phase3}. Now, using the embedding coordinates of the points $L_1$, $L_2$ and $L_3$ in \cref{EWCS three pt. formula} and adding contribution from the right TFD copy, the bulk EWCS may be obtained as
\begin{align}
	E_W(A:B)=& \frac{1}{4 G_N}\log\left[\frac{2 e^{\rho_\epsilon} u_h(\sqrt{\Delta_{L_1}}-\sqrt{\Delta_{L_2}})(\sqrt{\Delta_{L_2}}-\sqrt{\Delta_{L_3}})}{u^{}_{L_2}(\sqrt{\Delta_{L_1}}-\sqrt{\Delta_{L_3}})}\right]\nonumber\\
	&+\frac{1}{4 G_N}\log\left[\frac{2 e^{\rho_\epsilon}(\sqrt{\Delta_{R_1}}-\sqrt{\Delta_{R_2}})(\sqrt{\Delta_{R_2}}-\sqrt{\Delta_{R_3}})}{u^{}_{R_2}(\sqrt{\Delta_{R_1}}-\sqrt{\Delta_{R_3}})}\right].
\end{align}
Note that upon utilizing \cref{bulk-bdy relation} and Brown-Henneaux relation, the above expression of the bulk EWCS matches exactly with the half of the reflected entropy computed in \cref{SR adj.Phase3}.
% This serves a consistency check for the holographic duality mentioned earlier.

\subsection{Entanglement entropy phase 4}\label{adj.EE phase4}
\begin{figure}[h!]
	\centering
	\includegraphics[scale=.8]{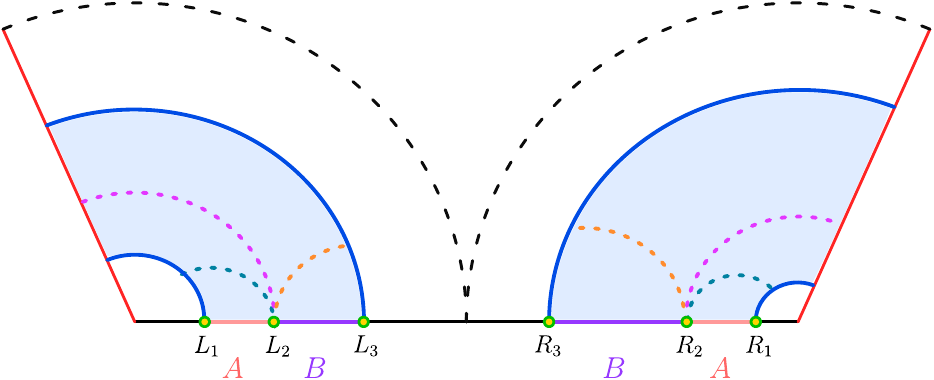}
	\caption{\it Schematic depicting the different phases of the EWCS (represented by various colored dashed curves) between subsystems $A$ and $B$ when the RT surface for $A \cup B$ is shown by the solid blue curves. }
	\label{fig_adj.EE phase4}
\end{figure}
In this EE phase, we consider that both the subsystems $A$ and $B$ are very close to the boundary. So, the EE is proportional to the sum of the lengths of four island surfaces, shown as solid blue curves in \cref{fig_adj.EE phase4}. Now by using \cref{length of island}, the EE in this phase may be obtained as
\begin{align}
	S_4= \frac{1}{G_N}(\rho_\epsilon-\rho^{}_B).
\end{align}
For this EE phase, there are three possible phases of the reflected entropy or the bulk EWCS, shown as dashed curves in \cref{fig_adj.EE phase4}. In the following subsections, we describe the computation of the reflected entropy and the bulk EWCS.

\subsection*{Phase-I}
The reflected entropy or the bulk EWCS in this phase is similar to the second case of the \cref{adj.EE phase2}, shown as dashed orange curves in \cref{fig_adj.EE phase4}. Therefore, the bulk EWCS is given by \cref{adj.island EWCS}.

\subsection*{Phase-II} 
\paragraph{Reflected entropy:} In this phase, we assume subsystem $B$ is smaller than $A$. Therefore, the six-point twist field correlator in the numerator of \cref{Ref-adj-def} may be factorized into two one-point twist field correlator and two two-point twist field correlator in BCFT$_2$ as
\begin{align}
	&\langle \sigma_{g^{}_A}(u^{}_{L_1})\sigma_{g^{}_B g_A^{-1}}(u^{}_{L_{2}})\sigma_{g_B^{-1}}(u^{}_{L_{3}})\sigma_{g^{}_B}(u^{}_{R_{3}})\sigma_{g^{}_A g_B^{-1}}(u^{}_{R_{2}})\sigma_{g_A^{-1}}(u^{}_{R_{1}}) \rangle_{\mathrm{BCFT}^{\bigotimes mn}}\notag\\
	&=\langle \sigma_{g^{}_A}(u^{}_{L_{1}})\rangle_{\mathrm{BCFT}^{\bigotimes mn}}\langle\sigma_{g^{}_B g_A^{-1}}(u^{}_{L_{2}})\sigma_{g_B^{-1}}(u^{}_{L_{3}})\rangle_{\mathrm{BCFT}^{\bigotimes mn}}\notag\\
	&\quad \quad \quad \times \langle \sigma_{g_A^{-1}}(u^{}_{R_{1}})\rangle_{\mathrm{BCFT}^{\bigotimes mn}}\langle\sigma_{g^{}_A g_B^{-1}}(u^{}_{R_{2}})\sigma_{g^{}_B}(u^{}_{R_{3}})\rangle_{\mathrm{BCFT}^{\bigotimes mn}}.
\end{align}
The two one-point twist field correlators of the above equation cancels with the denominator of \cref{Ref-adj-def}. To compute the remaining two two-point twist field correlator it is necessary to transform these to the twist field correlators defined on the conformally flat cylindrical background.
Now using the doubling trick \cite{Cardy:2004hm} and similar factorization in the denominator of \cref{Ref-adj-def}, the final expression for
the reflected entropy in this case may be written as \cite{Li:2021dmf}
\begin{align}
	S_R(A:B)=&\lim_{{m,n} \to 1}\frac{1}{1-n}\log\frac{ \langle\sigma_{g^{}_B g_A^{-1}}(u_{\star}({L_{2}}))\sigma_{g_B^{-1}}(u_{\star}({L_{3}}))\sigma_{g_B}(-u_{\star}({L_{3}}))\rangle_{\mathrm{CFT}^{\bigotimes mn}}}{\langle\sigma_{g_m^{-1}}(u_{\star}({L_{3}}))\sigma_{g_m}(-u_{\star}({L_{3}})) \rangle^n_{\mathrm{CFT}^{\bigotimes m}}}\notag\\
	&+\lim_{{m,n} \to 1} \frac{1}{1-n}\log\frac{ \langle\sigma_{g^{}_A g_B^{-1}}(u_{\star}({R_{2}}))\sigma_{g_B^{-1}}(u_{\star}({R_{3}}))\sigma_{g^{}_B}(-u_{\star}({R_{3}}))\rangle_{\mathrm{CFT}^{\bigotimes mn}}}{\langle\sigma_{g{}_m}(u_{\star}({R_{3}}))\sigma_{g_m^{-1}}(-u_{\star}({R_{3}})) \rangle^n_{\mathrm{CFT}^{\bigotimes m}}}.
\end{align}
%\begin{align}
%	S_R(A:B)=&\lim_{{m,n} \to 1}\frac{1}{1-n}\log\frac{ \langle\sigma_{g^{}_B g_A^{-1}}(u^{}_{L_{2}})\sigma_{g_B^{-1}}(u^{}_{L_{3}})\sigma_{g_B}(\bar{u}^{}_{L_{3}})\rangle_{\mathrm{CFT}^{\bigotimes mn}}}{\langle\sigma_{g_m^{-1}}(u^{}_{L_{3}})\sigma_{g_m}(\bar{u}^{}_{L_{3}}) \rangle^n_{\mathrm{CFT}^{\bigotimes m}}}\notag\\
%	&+\lim_{{m,n} \to 1} \frac{1}{1-n}\log\frac{ \langle\sigma_{g^{}_A g_B^{-1}}(u^{}_{R_{2}})\sigma_{g_B^{-1}}(u^{}_{R_{3}})\sigma_{g^{}_B}(\bar{u}^{}_{R_{3}})\rangle_{\mathrm{CFT}^{\bigotimes mn}}}{\langle\sigma_{g{}_m}(u^{}_{R_{3}})\sigma_{g_m^{-1}}(\bar{u}^{}_{R_{3}}) \rangle^n_{\mathrm{CFT}^{\bigotimes m}}}.
%\end{align}
Utilizing \cref{w-z map} and form of three-point function, the final expression for the reflected entropy may be obtained as 
\begin{align}\label{SR adj.Phase4(ii)}
	S_R(A:B)=& \frac{c}{3}\log\left[\frac{4 u_h(\sqrt{\Delta_{L_2}}-\sqrt{\Delta_{L_3}})(-u_h+\Delta_{L_2} \Delta_{L_3})}{\epsilon \, u^{}_{L_{2}}u^{}_{L_{3}}} \right]\nonumber\\
	&+\frac{c}{3}\log\left[\frac{4(\sqrt{\Delta_{R_2}}-\sqrt{\Delta_{R_3}})(-u_h+\Delta_{R_2} \Delta_{R_3})}{\epsilon \, u^{}_{R_{2}}u^{}_{R_{3}}} \right].
\end{align}
\paragraph{EWCS:} The bulk EWCS for this phase is given by the sum of the lengths of two geodesics in which one connects the point  $L_2$ to an arbitrary point $(u^{}_{L_{3}},\rho^{}_Y,t)$ on the island surface and another joins the point $R_2$ to an arbitrary point $(u^{}_{R_{3}},\rho^{}_Z,\tilde{t})$ on the island surface for the right TFD copy. These geodesics are shown as orange dashed curves in \cref{fig_adj.EE phase4}. Now by utilizing the embedding coordinates of points $L_2$ and $(u^{}_{L_{3}},\rho^{}_Y,t)$ in \cref{Geodesic length}, the length of the first geodesic may be written as 
\begin{align}\label{length QY}
	L=\log \left[\left(\frac{2u_h(\sqrt{\Delta_{L_{2}}}-\sqrt{\Delta_{L_{3}}})^2 +u^{}_{L_{2}} u^{}_{L_{3}}}{u^{}_{L_{2}} u^{}_{L_{3}}}\right)\cosh\rho^{}_Y-\sinh\rho^{}_Y\right]+\rho_\epsilon.
\end{align}
To obtain the EWCS, we need to extremize the above expression with respect to $\rho^{}_Y$. The extremum value of $\rho^{}_Y$ is given as
\begin{align}
	\rho^{}_Y= \frac{1}{2}\log \left(\frac{2u_h(\sqrt{\Delta_{L_{2}}}-\sqrt{\Delta_{L_{3}}})^2 +u^{}_{L_{2}} u^{}_{L_{3}}}{u_h (\sqrt{\Delta_{L_{2}}}-\sqrt{\Delta_{L_{3}}})^2}\right),
\end{align}
Substituting the above in \cref{length QY} and adding the contribution for the right TFD copy, the corresponding bulk EWCS may be obtained as
\begin{align}
	E_{W}(A:B)=& \frac{1}{4 G_N}\log\left[\frac{2 e^{\rho_\epsilon} u_h(\sqrt{\Delta_{L_2}}-\sqrt{\Delta_{L_3}})(-u_h+\Delta_{L_2} \Delta_{L_3})}{u^{}_{L_{2}}u^{}_{L_{3}}} \right]\nonumber\\
	&+\frac{1}{4 G_N}\log\left[\frac{2 e^{\rho_\epsilon}(\sqrt{\Delta_{R_2}}-\sqrt{\Delta_{R_3}})(-u_h+\Delta_{R_2} \Delta_{R_3})}{u^{}_{R_{2}}u^{}_{R_{3}}} \right].
\end{align}
Here also the reflected entropy obtained in \cref{SR adj.Phase4(ii)} matches exactly with twice of the EWCS upon utilizing \cref{bulk-bdy relation} and Brown-Henneaux relation.
% which again verify the holographic duality mentioned earlier.

\subsection*{Phase-III}
The reflected entropy or the bulk EWCS in this phase is similar to the third case of the \cref{adj.EE phase2}, shown as dark green dashed curves in \cref{fig_adj.EE phase4}. Therefore, the bulk EWCS is given by \cref{island ewcs}.

\subsection{Page curve} In this subsection we illustrate the analogue of the Page curves for the reflected entropy for two adjacent subsystems in a BCFT$_2$ on an AdS$_2$ black hole background.
% Here distinct entropy phase transitions arise from different combinations of coordinate parametrization.

\subsubsection{Case-I}\label{sec:adj 1HM-island}
\begin{figure}[H]
	\centering
	\begin{subfigure}[b]{0.45\textwidth}
		\centering
		\includegraphics[scale=.37]{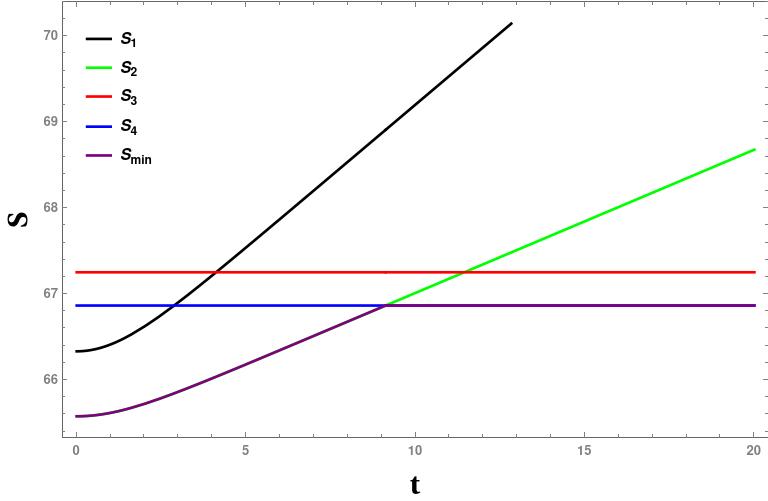}
		\caption{} 
		\label{adj plot(1HM-island)}
	\end{subfigure}
	\hfill
	\begin{subfigure}[b]{0.45\textwidth}
		\centering
		\includegraphics[scale=.35]{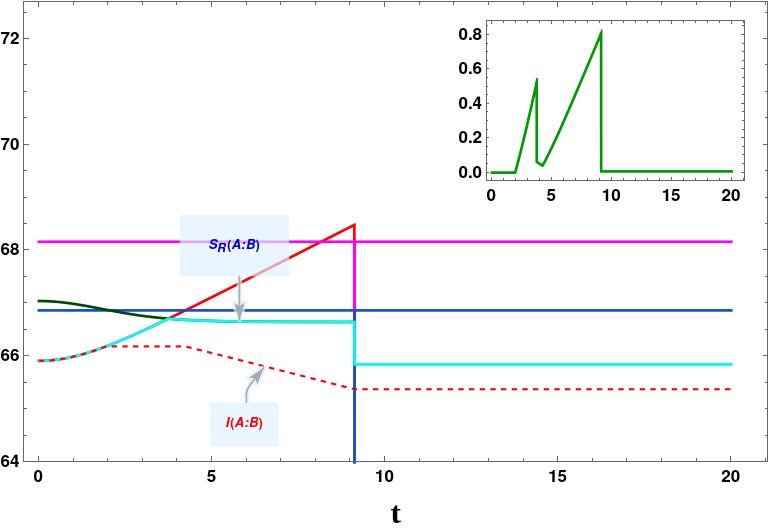}
		\caption{}
		\label{adj SR plot(1HM-island)}
	\end{subfigure}
	\caption{(a) Page curve of the EE for subsystems $A \cup B$. Here purple colour shows the minimum EE among all the phases. (b) Page curve of the reflected entropy between subsystems $A$ and $B$. Here cyan colour represents minimum $S_R$ and red dashed line is mutual information (All graphs are in units of $c$). The inset plot shows the deviation from saturation of the Markov gap \cref{Markov gap}. The Page curves for the EE and $S_R$ is obtained with $u_h=1,\Delta_{L_1}=0.9,\Delta_{L_2}=3 \times10^{-2},\Delta_{L_3}=2 \times10^{-4},\Delta_{R_1}=0.8,\Delta_{R_2}=2\times 10^{-2},\Delta_{R_3}=2 \times10^{-4},\rho^{}_B=-0.3,\rho_\epsilon=100$. }
\end{figure}

The EE phase transition between \hyperref[adj.EE phase1]{phase-2} and \hyperref[adj.EE phase1]{phase-4} may be obtained by utilizing a small brane angle (corresponding to a small boundary entropy) and taking subsystem $B$ away from the boundary, as shown in \cref{adj plot(1HM-island)}. The Page time $T^{\text{adj}}_{{2\rightarrow4}}$ for this transition is given as
\begin{align}
	T^{\text{adj}}_{{2\rightarrow4}}= u_h \cosh^{-1}\left(\frac{u^{2}_h-u_h \left(e^{2 \rho_B}+1\right)  (\Delta_{L_3}+\Delta_{R_3})+\Delta_{L_3} \Delta_{R_3}}{2 u_h e^{2 \rho_B} \sqrt{\Delta_{L_3}\Delta_{R_3}}}\right).
\end{align}
We now investigate the Page curve for the reflected entropy in these EE phases, depicted in \cref{adj SR plot(1HM-island)}. Initially in the EE \hyperref[adj.EE phase2]{phase-2} the reflected entropy increases with time as the bulk EWCS is the HM surface, then remains constant until Page time as the growth rate of the bulk EWCS which lands on the HM surface is almost similar to that of the HM surface. After the Page time, in the EE \hyperref[adj.EE phase4]{phase-4} it saturates to another constant value. From the Page curve of the reflected entropy we observe that initially the Markov gap is zero as the bulk EWCS phase has no non-trivial boundaries but after some time for the same bulk EWCS phase this gap becomes non-zero as the mutual information undergoes a phase transition. After that this gap increases to a value greater than $\frac{2c}{3}\log{2}$ as the bulk EWCS phase has two non-trivial boundaries. Furthermore, after the Page time, the Markov gap saturates to the lower bound mentioned in \cref{Markov gap}.

\subsubsection{Case-II}\label{sec:adj 1HM-dome}
\begin{figure}[H]
	\centering
	\begin{subfigure}[b]{0.45\textwidth}
		\centering
		\includegraphics[scale=.37]{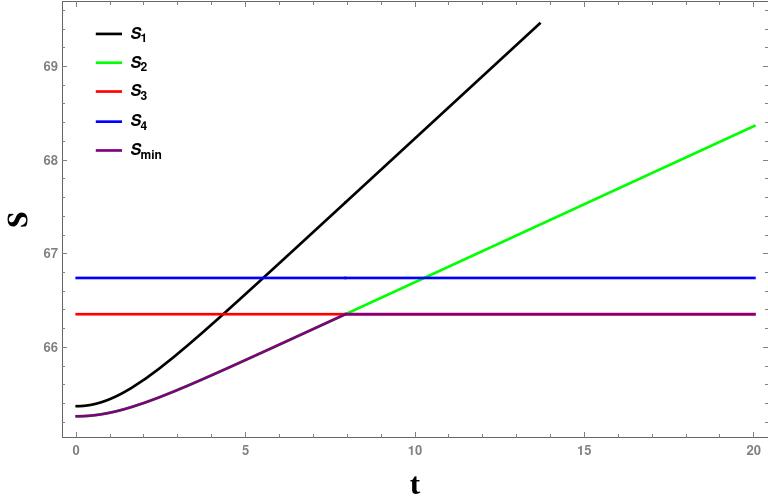}
		\caption{}
		\label{adj plot(1HM-dome)}
	\end{subfigure}
	\hfill
	\begin{subfigure}[b]{0.45\textwidth}
		\centering
		\includegraphics[scale=.35]{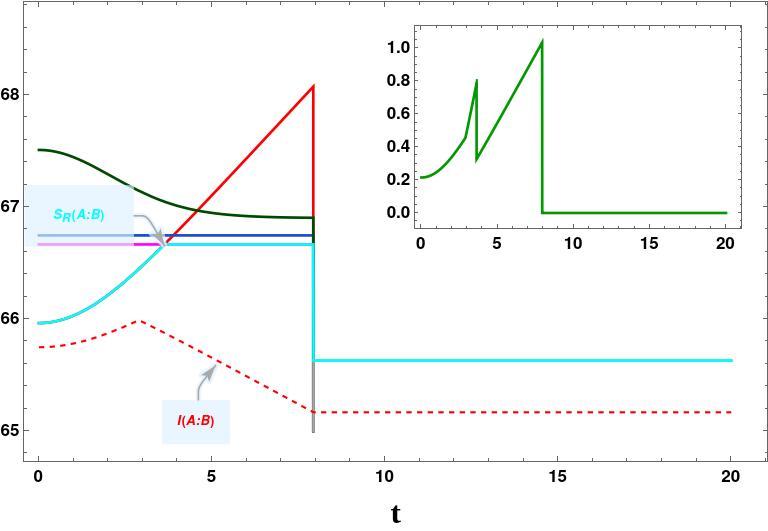}
		\caption{}
		\label{adj SR plot(1HM-dome)}
	\end{subfigure}
	\caption{(a) Page curve of the EE for subsystems $A \cup B$. Here purple colour shows the minimum EE among all the phases. (b) Page curve of the reflected entropy between subsystems $A$ and $B$. Here cyan colour represents minimum $S_R$ and red dashed line is mutual information (All graphs are in units of $c$). The inset plot shows the deviation from saturation of the Markov gap \cref{Markov gap}. The Page curves for the EE and $S_R$ is obtained with $u_h=1,\Delta_{L_1}=0.4,\Delta_{L_2}=5.8 \times 10^{-2}, \Delta_{L_3}=4 \times 10^{-5}, \Delta_{R_1}=0.2, \Delta_{R_2}=.915 \times 10^{-2}, \Delta_{R_3}=5 \times 10^{-5},\rho^{}_B=-0.12, \rho_\epsilon=100$.}
\end{figure}

The EE transition between \hyperref[adj.EE phase2]{phase-2} and \hyperref[adj.EE phase3]{phase-3} may be obtained by taking a small brane angle and subsystem $B$ is relatively away form the boundary than the previous case, as depicted in \cref{adj plot(1HM-dome)}. The Page time for this transition is given as 
\begin{align}\label{Page time 1HM-dome(adj)}
	T^{\text{adj}}_{{2\rightarrow3}}=	u_h \cosh ^{-1}\left(\frac{u_h e^{2 \rho_B}  \left(\sqrt{\Delta_{L_1}}-\sqrt{\Delta_{L_3}}\right)^2 \left(\sqrt{\Delta_{R_1}}-\sqrt{\Delta_{R_3}}\right)^2}{2 u^{}_{L_1}u^{}_{R_1} \sqrt{\Delta_{L_3}\Delta_{R_3}} }-\frac{(\Delta_{L_3}+\Delta_{R_3}) }{2 \sqrt{\Delta_{L_3}\Delta_{R_3}}}\right).
\end{align}
The Page curve of the reflected entropy for these EE phases is shown in \cref{adj SR plot(1HM-dome)}. The reflected entropy in \hyperref[adj.EE phase2]{phase-2} first increases as the bulk EWCS is the HM surface, then remains constant until Page time as the bulk EWCS lands on the EOW brane. Finally after the Page time, it saturates to another constant value in \hyperref[adj.EE phase3]{phase-3}. Initially the Markov gap should be zero as the bulk EWCS has no non-trivial boundary however it is observed to be non-zero which contradicts the geometric interpretation of the Markov gap described in \cref{Markov gap} suggesting a critical re-examination of this issue in the context of the AdS/BCFT scenario. Subsequently with time this gap increases to a value greater than $\frac{2c}{3}\log{2}$ as two non-trivial boundaries of the bulk EWCS appear in the corresponding  phase. Furthermore after the Page time, this gap saturates to the lower bound mentioned in \cref{Markov gap} which was expected as the computation of the reflected entropy reduces to the usual CFT which involves the contribution from the OPE channel of the corresponding BCFT correlators.

\subsubsection{Case-III}\label{sec:adj 2HM-1HM-dome}
\begin{figure}[H]
	\centering
	\begin{subfigure}[b]{0.45\textwidth}
		\centering
		\includegraphics[scale=.37]{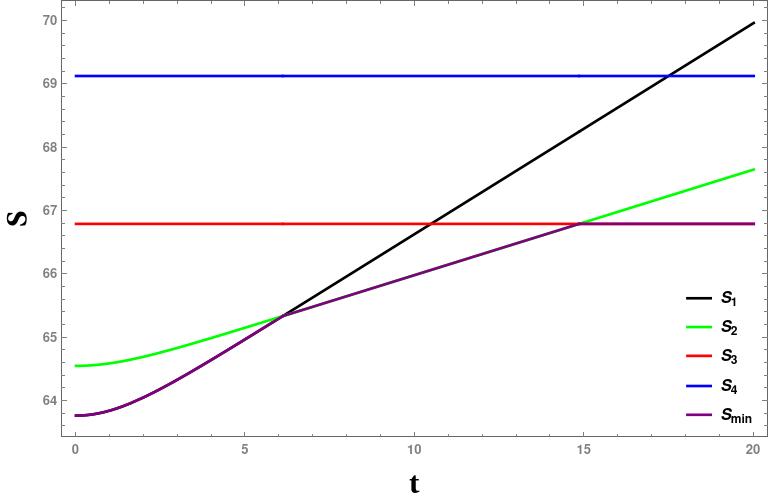}
		\caption{}
		\label{adj plot(2HM-1HM-dome)}
	\end{subfigure}
	\hfill
	\begin{subfigure}[b]{0.45\textwidth}
		\centering
		\includegraphics[scale=.35]{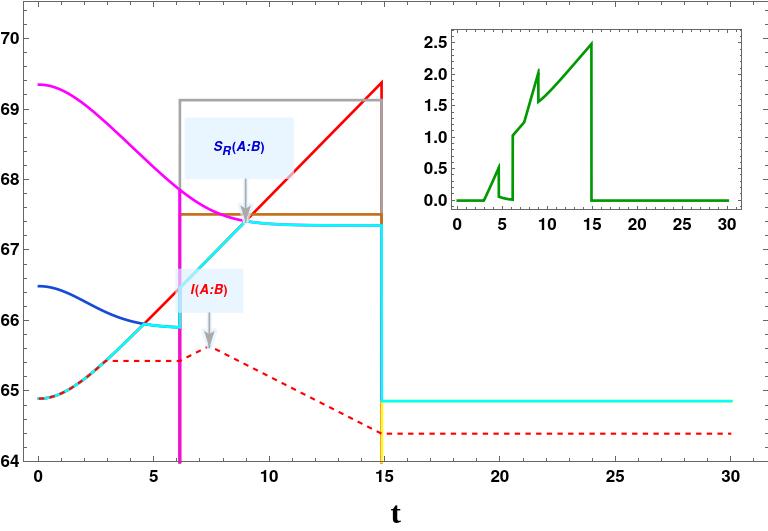}
		\caption{}
		\label{adj SR plot(2HM-1HM-dome)}
	\end{subfigure}
	\caption{(a) Page curve of the EE for subsystems $A \cup B$. Here purple colour shows the minimum EE among all the phases. (b) Page curve of the reflected entropy between subsystems $A$ and $B$. Here cyan colour represents minimum $S_R$ and red dashed line is mutual information (All graphs are in units of $c$). The inset plot shows the deviation from saturation of the Markov gap \cref{Markov gap}. The Page curves for the EE and $S_R$ is obtained with $u_h=1,\Delta_{L_1}=0.6, \Delta_{L_2}=15 \times 10^{-4}, \Delta_{L_3}=5 \times 10^{-10}, \Delta_{R_1}=0.592223, \Delta_{R_2}=9.9 \times 10^{-4}, \Delta_{R_3}=4.5 \times 10^{-10}, \rho^{}_B=-3.7, \rho_\epsilon=100$.}
\end{figure}

The EE transition between \hyperref[adj.EE phase1]{phase-1} and \hyperref[adj.EE phase2]{phase-2} at time $T^{\text{adj}}_{{1\rightarrow2}}$ which is same as $T^{\text{disj}}_{{1\rightarrow2}}$, and \hyperref[adj.EE phase2]{phase-2} and \hyperref[adj.EE phase3]{phase-3} at time $T^{\text{adj}}_{{2\rightarrow3}}$ may be obtained by taking a relatively large brane angle and subsystem $B$ is far away from the boundary than the previous two cases as depicted in \cref{adj plot(2HM-1HM-dome)}. The Page times for these EE transition $T^{\text{adj}}_{{1\rightarrow2}}$ and $T^{\text{adj}}_{{2\rightarrow3}}$ are given in \cref{Page time 2HM-1HM} and \cref{Page time 1HM-dome(adj)} respectively. In the EE \hyperref[adj.EE phase1]{phase-1}, the reflected entropy increases initially as the bulk EWCS is the HM surface and then slowly decreases until Page time $T^{\text{adj}}_{{1\rightarrow2}}$ as the growth rate of the bulk EWCS which lands on the HM surface is lower than that of the HM surface. After the Page time in the EE \hyperref[adj.EE phase2]{phase-2}, it again increases as the bulk EWCS is again the HM surface and then remains constant until Page time $T^{\text{adj}}_{{2\rightarrow3}}$. Finally in the EE \hyperref[adj.EE phase3]{phase-3}, it saturates to a constant value. The reflected entropy Page curve is shown in \cref{adj SR plot(2HM-1HM-dome)}. The Page curve of the reflected entropy indicates that in the EE \hyperref[adj.EE phase1]{phase-1}, initially the Markov gap is zero as there are no non-trivial boundaries of the bulk EWCS, however as earlier this gap becomes non zero with time due to a phase transition in the mutual information. Subsequently as earlier this gap increases to a value greater than $\frac{2c}{3}\log{2}$ as two non-trivial boundaries of the bulk EWCS appear in this phase. Furthermore after the Page time in the EE \hyperref[adj.EE phase2]{phase-2}, this gap is always greater than $\frac{2c}{3}\log{2}$. Finally, in the EE \hyperref[adj.EE phase3]{phase-3}, this gap saturates to the lower bound given in \cref{Markov gap} which was also expected (c.f. \cref{sec:adj 1HM-dome}).

\subsubsection{Case-IV}\label{sec:adj 2HM-dome}
\begin{figure}[H]
	\centering
	\begin{subfigure}[b]{0.45\textwidth}
		\centering
		\includegraphics[scale=.37]{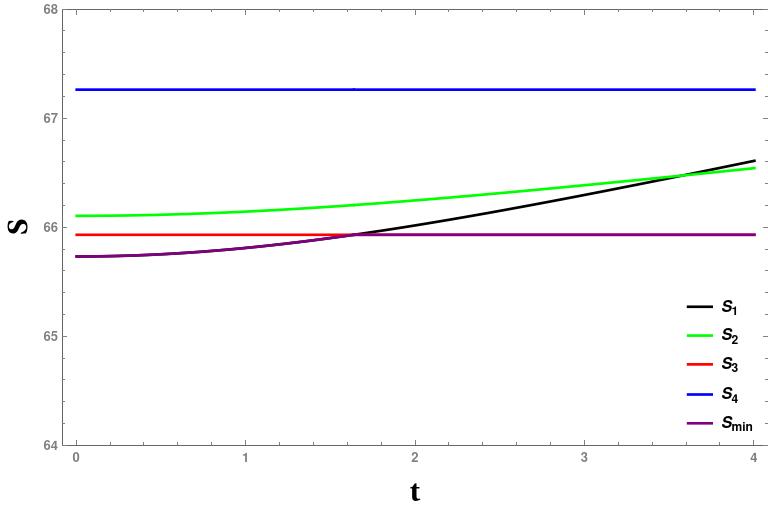}
		\caption{}
		\label{adj plot(2HM-dome)}
	\end{subfigure}
	\hfill
	\begin{subfigure}[b]{0.45\textwidth}
		\centering
		\includegraphics[scale=.355]{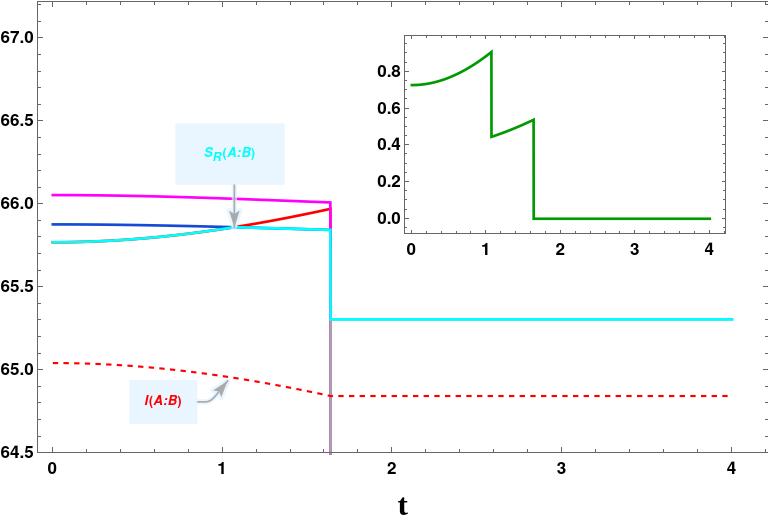}
		\caption{}
		\label{adj SR plot(2HM-dome)}
	\end{subfigure}
	\caption{(a) Page curve of the EE for subsystems $A \cup B$. Here purple colour shows the minimum EE among all the phases. (b) Page curve of the reflected entropy between subsystems $A$ and $B$. Here cyan colour represents minimum $S_R$ and red dashed line is mutual information (All graphs are in units of $c$). The inset plot shows the deviation from saturation of the Markov gap \cref{Markov gap}. The Page curves for the EE and $S_R$ is obtained with $u_h=1,\Delta_{L_1}=0.15, \Delta_{L_2}=1.8 \times 10^{-2}, \Delta_{L_3}=2 \times 10^{-3}, \Delta_{R_1}=0.1, \Delta_{R_2}=1.5 \times 10^{-2}, \Delta_{R_3}=1 \times 10^{-3}, \rho^{}_B=-0.9, \rho_\epsilon=100$.}
\end{figure}

The EE transition between the EE \hyperref[adj.EE phase1]{phase-1} and \hyperref[adj.EE phase3]{phase-3} may be obtained by using a small brane angle and taking the subsystem $B$ is close to the boundary than first two cases as shown in \cref{adj plot(2HM-dome)}. The Page time for this transition is given as 
\begin{align}\label{Page time 2HM-dome(adj)}
	T^{\text{Adj.}}_{{1\rightarrow3}}=\cosh ^{-1}\left(\frac{\Delta_{L_1}+\Delta_{R_1}}{\sqrt{\Delta_{L_1} \Delta_{R_1}}}+\frac{\Delta_{L_3}+\Delta_{R_3}}{\sqrt{\Delta_{L_3} \Delta_{R_3}}}-A^2\right),
\end{align}
where 
\begin{align}
	A^2=& \left(\frac{\Delta_{L_1}}{\Delta_{R_1}}+\frac{\Delta_{R_1}}{\Delta_{L_1}}\right)+\left(\frac{\Delta_{L_3}}{\Delta_{R_3}}+\frac{\Delta_{R_3}}{\Delta_{L_3}}\right)-8 \Delta_{L_1} \Delta_{L_3} \Delta_{R_1} \Delta_{R_3} \left(\frac{\Delta_{L_1}+\Delta_{L_3}}{\sqrt{\Delta_{L_1} \Delta_{L_3}}}+\frac{\Delta_{R_1}+\Delta_{R_3}}{\sqrt{\Delta_{R_1} \Delta_{R_3}}}\right)\nonumber\\
	&+20+\frac{2 \Delta_{L_1} (-\Delta_{L_3}+2 \Delta_{R_1}+\Delta_{R_3})+2 \Delta_{L_3} (\Delta_{R_1}+2 \Delta_{R_3})-2 \Delta_{R_1} \Delta_{R_3}}{\sqrt{\Delta_{L_1} \Delta_{L_3} \Delta_{R_1} \Delta_{R_3}}}.
\end{align}
From the Page curve of the reflected entropy, we observe that in the EE \hyperref[adj.EE phase1]{phase-1}, the reflected entropy increases initially as the bulk EWCS is the HM surface and then slowly decreases until Page time as the growth rate of the bulk EWCS which lands on the HM surface is lower than that of the HM surface. After the Page time, in the EE \hyperref[adj.EE phase3]{phase-3}, it saturates to a constant value. Here also initially the Markov gap should be zero as the bulk EWCS has no non-trivial boundaries however it is observed to be non zero which again suggests further critical analysis of the geometric interpretation. Subsequently till the Page time $T^{\text{adj}}_{{1\rightarrow3}}$, it is always greater than $\frac{2c}{3}\log{2}$ as the bulk EWCS has two non-trivial boundaries. Furthermore after the Page time, it saturates to the lower bound given in \cref{Markov gap} which was also expected (c.f. \cref{sec:adj 1HM-dome}).

\section{Summary and discussion}\label{Summary}
In this article, we have investigated the mixed state entanglement structure through the reflected entropy, in the KR braneworld model with the radiation bath located in a gravitational background. In particular, we considered an AdS$_3$ black string geometry truncated by a EOW brane for which the lower dimensional effective perspective consists of a gravitating radiation bath. In this connection, the dual BCFT$_2$ is defined on an eternal AdS$_2$ black hole background. We have obtained the reflected entropy for various bipartite mixed state configurations involving two disjoint and adjacent subsystems in the BCFT$_2$. Furthermore, we have elucidated the phase structure for the reflected entropy arising from various factorizations of the twist field correlators in the large central charge limit after identifying the EE phases. Subsequently, we have computed the corresponding EWCS in the dual bulk AdS$_3$ black string geometry for the bipartite mixed states under consideration. It is further demonstrated that our holographic computations
match identically with the field theory replica technique results in the large central charge limit for all the phases.

Following the above, we have also obtained the Page curves for the EE and within each phase of the EE, we observed rich phase structure for the holographic reflected entropy depending on the boundary entropy, the location of the subsystems and their relative separation from the boundary. Furthermore, we have also compared the reflected entropy with the holographic mutual information in order to investigate the Markov gap. For all the phases of the reflected entropy, we have shown the deviation of the Markov gap from its saturation value. Interestingly, we observed that for the phases of the bulk EWCS involving the HM surface in the bulk geometry, the Markov gap is non-zero even when there are no non-trivial boundaries of the EWCS, which is not obvious from the standard geometric interpretation provided in \cite{Hayden:2021gno}. As a result, it will be interesting to investigate this interpretation in the braneworld geometry.

From the Page curve of the reflected entropy we observed that there are sudden jumps or drops at the Page time whose origin may be attributed to the specific choice of the R\'enyi entanglement entropy saddle i.e., fully connected and fully disconnected replica wormhole saddles. These discontinuities may be smoothened out by considering all possible R\'enyi entanglement entropy saddles \cite{Penington:2019kki}. However we may still miss some R\'enyi reflected entropy saddles. Thus to obtain a proper (continuous and smooth) Page curve for the reflected entropy as in \cite{Akers:2022max}, one should include contributions from all R\'enyi reflected entropy saddles in each R\'enyi entanglement entropy saddle in the corresponding gravitational path integral which is computationally challenging and beyond the scope of the present work.

There are several future directions to explore. It will be interesting to investigate other mixed state entanglement and correlation measures such as entanglement negativity, entanglement of purification, balance partial entanglement in this braneworld model. One may also generalize our study to higher dimensions and to multipartite entanglement and correlations. This model may also be generalized for two boundary BCFT with black holes induced on the each of the two corresponding EOW branes. It would be interesting to investigate the interaction between these two black holes in this scenario.
Another significant generalization of this braneworld model could be the study of mixed state entanglement 
for two CFTs defined on a black hole background communicating through a shared interface. We leave these interesting open issue for future consideration.

%One may also generalize this braneworld model to the case where the bath CFT is deformed by a $T\bar{T}$ deformation.

\section*{Acknowledgement}
The work of GS is partially supported by the Dr Jagmohan Garg Chair Professor position at the Indian Institute of Technology, Kanpur.

\appendix
\section{Geodesics between two minimal surfaces}\label{Appendix}
The EWCS for some configurations may be obtained by computing the geodesic distance between two geodesics. In the embedding coordinates, the length of the geodesics ending on the bulk points $X_i$ and $X_j$ is given by
\begin{align}\label{Geodesic Length}
	\sigma(X_i,X_j)&=\cosh ^{-1}\xi_{ij}, ~~~~~~~~~~ \xi_{ij}=-X_i\cdot X_j.
\end{align}

\subsection{Geodesic between a fixed boundary point and a bulk geodesic} 
A spacelike geodesic anchored on two boundary points $X_1$ and $X_3$ may be parametrized by an affine parameter $\lambda$ as follows
\begin{align}
	X_{13}^A(\lambda)=\frac{X^A_1 e^{-\lambda}+X^A_3 e^{\lambda}}{\sqrt{-2X_1\cdot X_3^{}}}.
\end{align}
By utilizing \cref{Geodesic Length}, the geodesic length between a fixed boundary point $X_2$ and a geodesic $X_{13}(\lambda)$ may be written as
\begin{align}\label{g213}
\sigma(X_{13},X_2)= \cosh^{-1}\left[\frac{\xi_{12} e^{-\lambda}+\xi_{23} e^{-\lambda}}{\sqrt{2 \xi_{13}}}\right].
\end{align}
Now by extremizing this length over $\lambda$ and substituting back the extremum value in \cref{g213} we may obtained the EWCS between a fixed boundary point and a bulk geodesic as
\begin{align}
	E_W= \frac{1}{4 G_N}\cosh ^{-1}\left(\sqrt{\frac{2 \xi_{12}\xi_{23}}{\xi_{13}}}\right).
\end{align}

%
%\begin{align}
%\lambda_\star= \frac{1}{2}\log \left(\frac{\xi_{12}}{\xi_{23}}\right),
%\end{align}

\subsection{Geodesic between one bulk point and three boundary point}

The spacelike geodesic $X_{14}(\lambda)$ anchored on two bulk points $X_1$ and $X_4$ is given as 
\begin{align}
X_{14}(\lambda)=e^{\lambda } \left(\frac{X_4}{\sqrt{2 \xi_{14}}}-\frac{X_1}{(2 \xi_{14})^{3/2}}\right)+e^{-\lambda } \left(\frac{X_1}{\sqrt{2 \xi_{14}}}-\frac{X_4}{(2 \xi_{14})^{3/2}}\right),
\end{align}
and a spacelike geodesic $X_{23}(\lambda_p)$ anchored between two boundary points $X_2$ and $X_3$ is given as
\begin{align}
X_{23}(\lambda_p)=\frac{e^{-\lambda_p} X_2+e^{\lambda_p} X_3}{\sqrt{2 \xi_{23}}}.
\end{align}
Now by utilizing the \cref{Geodesic Length}, the length of the geodesic between $X_{14}(\lambda)$ and $X_{23}(\lambda_p)$ may be written as
\begin{align}\label{L}
\sigma({X_{14},X_{23}})= \cosh^{-1} \left(-X_{14}\cdot X_{23}\right).
\end{align}
By extremizing this length over $\lambda$ and $\lambda_p$ and putting back their extremum value in \cref{L},the EWCS may be obtained as 
\begin{align}\label{bulk point EWCS formula}
E_W= \cosh ^{-1}\left(\frac{2 \xi_{14} \xi_{34}-\sqrt{\frac{(\xi_{12}-2 \xi_{14} \xi_{24}) (2 \xi_{13} \xi_{14}-\xi_{34}) (\xi_{13}-2 \xi_{14} \xi_{34})}{2 \xi_{12} \xi_{14}-\xi_{24}}}-\xi_{13}}{2 \xi_{14}^{3/2} \sqrt{{\xi_{23}}} \sqrt{\frac{2 \xi_{14} \xi_{34}-\xi_{13}}{2 \xi_{12} \xi_{14}-\xi_{24}}}}\right).
\end{align}

\bibliographystyle{utphys}

\bibliography{ref}

\providecommand{\href}[2]{#2}\begingroup\raggedright\begin{thebibliography}{10}

\bibitem{Hawking:1975vcx}
S.~W. Hawking, ``{Particle Creation by Black Holes},''
  \href{http://dx.doi.org/10.1007/BF02345020}{{\em Commun. Math. Phys.}
  {\bfseries 43} (1975) 199--220}. [Erratum: Commun.Math.Phys. 46, 206 (1976)].

\bibitem{Hawking:1976ra}
S.~W. Hawking, ``{Breakdown of Predictability in Gravitational Collapse},''
  \href{http://dx.doi.org/10.1103/PhysRevD.14.2460}{{\em Phys. Rev. D}
  {\bfseries 14} (1976) 2460--2473}.

\bibitem{Penington:2019npb}
G.~Penington, ``{Entanglement Wedge Reconstruction and the Information
  Paradox},'' \href{http://dx.doi.org/10.1007/JHEP09(2020)002}{{\em JHEP}
  {\bfseries 09} (2020) 002}, \href{http://arxiv.org/abs/1905.08255}{{\ttfamily
  arXiv:1905.08255 [hep-th]}}.

\bibitem{Almheiri:2019psf}
A.~Almheiri, N.~Engelhardt, D.~Marolf, and H.~Maxfield, ``{The entropy of bulk
  quantum fields and the entanglement wedge of an evaporating black hole},''
  \href{http://dx.doi.org/10.1007/JHEP12(2019)063}{{\em JHEP} {\bfseries 12}
  (2019) 063}, \href{http://arxiv.org/abs/1905.08762}{{\ttfamily
  arXiv:1905.08762 [hep-th]}}.

\bibitem{Almheiri:2019hni}
A.~Almheiri, R.~Mahajan, J.~Maldacena, and Y.~Zhao, ``{The Page curve of
  Hawking radiation from semiclassical geometry},''
  \href{http://dx.doi.org/10.1007/JHEP03(2020)149}{{\em JHEP} {\bfseries 03}
  (2020) 149}, \href{http://arxiv.org/abs/1908.10996}{{\ttfamily
  arXiv:1908.10996 [hep-th]}}.

\bibitem{Almheiri:2019yqk}
A.~Almheiri, R.~Mahajan, and J.~Maldacena, ``{Islands outside the horizon},''
  \href{http://arxiv.org/abs/1910.11077}{{\ttfamily arXiv:1910.11077
  [hep-th]}}.

\bibitem{Penington:2019kki}
G.~Penington, S.~H. Shenker, D.~Stanford, and Z.~Yang, ``{Replica wormholes and
  the black hole interior},''
  \href{http://dx.doi.org/10.1007/JHEP03(2022)205}{{\em JHEP} {\bfseries 03}
  (2022) 205}, \href{http://arxiv.org/abs/1911.11977}{{\ttfamily
  arXiv:1911.11977 [hep-th]}}.

\bibitem{Almheiri:2020cfm}
A.~Almheiri, T.~Hartman, J.~Maldacena, E.~Shaghoulian, and A.~Tajdini, ``{The
  entropy of Hawking radiation},''
  \href{http://dx.doi.org/10.1103/RevModPhys.93.035002}{{\em Rev. Mod. Phys.}
  {\bfseries 93} no.~3, (2021) 035002},
  \href{http://arxiv.org/abs/2006.06872}{{\ttfamily arXiv:2006.06872
  [hep-th]}}.

\bibitem{Ryu:2006bv}
S.~Ryu and T.~Takayanagi, ``{Holographic derivation of entanglement entropy
  from AdS/CFT},'' \href{http://dx.doi.org/10.1103/PhysRevLett.96.181602}{{\em
  Phys. Rev. Lett.} {\bfseries 96} (2006) 181602},
  \href{http://arxiv.org/abs/hep-th/0603001}{{\ttfamily arXiv:hep-th/0603001}}.

\bibitem{Ryu:2006ef}
S.~Ryu and T.~Takayanagi, ``{Aspects of Holographic Entanglement Entropy},''
  \href{http://dx.doi.org/10.1088/1126-6708/2006/08/045}{{\em JHEP} {\bfseries
  08} (2006) 045}, \href{http://arxiv.org/abs/hep-th/0605073}{{\ttfamily
  arXiv:hep-th/0605073}}.

\bibitem{Hubeny:2007xt}
V.~E. Hubeny, M.~Rangamani, and T.~Takayanagi, ``{A Covariant holographic
  entanglement entropy proposal},''
  \href{http://dx.doi.org/10.1088/1126-6708/2007/07/062}{{\em JHEP} {\bfseries
  07} (2007) 062}, \href{http://arxiv.org/abs/0705.0016}{{\ttfamily
  arXiv:0705.0016 [hep-th]}}.

\bibitem{Faulkner:2013ana}
T.~Faulkner, A.~Lewkowycz, and J.~Maldacena, ``{Quantum corrections to
  holographic entanglement entropy},''
  \href{http://dx.doi.org/10.1007/JHEP11(2013)074}{{\em JHEP} {\bfseries 11}
  (2013) 074}, \href{http://arxiv.org/abs/1307.2892}{{\ttfamily arXiv:1307.2892
  [hep-th]}}.

\bibitem{Engelhardt:2014gca}
N.~Engelhardt and A.~C. Wall, ``{Quantum Extremal Surfaces: Holographic
  Entanglement Entropy beyond the Classical Regime},''
  \href{http://dx.doi.org/10.1007/JHEP01(2015)073}{{\em JHEP} {\bfseries 01}
  (2015) 073}, \href{http://arxiv.org/abs/1408.3203}{{\ttfamily arXiv:1408.3203
  [hep-th]}}.

\bibitem{Page:1993wv}
D.~N. Page, ``{Information in black hole radiation},''
  \href{http://dx.doi.org/10.1103/PhysRevLett.71.3743}{{\em Phys. Rev. Lett.}
  {\bfseries 71} (1993) 3743--3746},
  \href{http://arxiv.org/abs/hep-th/9306083}{{\ttfamily arXiv:hep-th/9306083}}.

\bibitem{Page:1993df}
D.~N. Page, ``{Average entropy of a subsystem},''
  \href{http://dx.doi.org/10.1103/PhysRevLett.71.1291}{{\em Phys. Rev. Lett.}
  {\bfseries 71} (1993) 1291--1294},
  \href{http://arxiv.org/abs/gr-qc/9305007}{{\ttfamily arXiv:gr-qc/9305007}}.

\bibitem{Page:2013dx}
D.~N. Page, ``{Time Dependence of Hawking Radiation Entropy},''
  \href{http://dx.doi.org/10.1088/1475-7516/2013/09/028}{{\em JCAP} {\bfseries
  09} (2013) 028}, \href{http://arxiv.org/abs/1301.4995}{{\ttfamily
  arXiv:1301.4995 [hep-th]}}.

\bibitem{Almheiri:2019qdq}
A.~Almheiri, T.~Hartman, J.~Maldacena, E.~Shaghoulian, and A.~Tajdini,
  ``{Replica Wormholes and the Entropy of Hawking Radiation},''
  \href{http://dx.doi.org/10.1007/JHEP05(2020)013}{{\em JHEP} {\bfseries 05}
  (2020) 013}, \href{http://arxiv.org/abs/1911.12333}{{\ttfamily
  arXiv:1911.12333 [hep-th]}}.

\bibitem{Dong:2020uxp}
X.~Dong, X.-L. Qi, Z.~Shangnan, and Z.~Yang, ``{Effective entropy of quantum
  fields coupled with gravity},''
  \href{http://dx.doi.org/10.1007/JHEP10(2020)052}{{\em JHEP} {\bfseries 10}
  (2020) 052}, \href{http://arxiv.org/abs/2007.02987}{{\ttfamily
  arXiv:2007.02987 [hep-th]}}.

\bibitem{Kawabata:2021vyo}
K.~Kawabata, T.~Nishioka, Y.~Okuyama, and K.~Watanabe, ``{Replica wormholes and
  capacity of entanglement},''
  \href{http://dx.doi.org/10.1007/JHEP10(2021)227}{{\em JHEP} {\bfseries 10}
  (2021) 227}, \href{http://arxiv.org/abs/2105.08396}{{\ttfamily
  arXiv:2105.08396 [hep-th]}}.

\bibitem{Rozali:2019day}
M.~Rozali, J.~Sully, M.~Van~Raamsdonk, C.~Waddell, and D.~Wakeham,
  ``{Information radiation in BCFT models of black holes},''
  \href{http://dx.doi.org/10.1007/JHEP05(2020)004}{{\em JHEP} {\bfseries 05}
  (2020) 004}, \href{http://arxiv.org/abs/1910.12836}{{\ttfamily
  arXiv:1910.12836 [hep-th]}}.

\bibitem{Chen:2020uac}
H.~Z. Chen, R.~C. Myers, D.~Neuenfeld, I.~A. Reyes, and J.~Sandor, ``{Quantum
  Extremal Islands Made Easy, Part I: Entanglement on the Brane},''
  \href{http://dx.doi.org/10.1007/JHEP10(2020)166}{{\em JHEP} {\bfseries 10}
  (2020) 166}, \href{http://arxiv.org/abs/2006.04851}{{\ttfamily
  arXiv:2006.04851 [hep-th]}}.

\bibitem{Chen:2020hmv}
H.~Z. Chen, R.~C. Myers, D.~Neuenfeld, I.~A. Reyes, and J.~Sandor, ``{Quantum
  Extremal Islands Made Easy, Part II: Black Holes on the Brane},''
  \href{http://dx.doi.org/10.1007/JHEP12(2020)025}{{\em JHEP} {\bfseries 12}
  (2020) 025}, \href{http://arxiv.org/abs/2010.00018}{{\ttfamily
  arXiv:2010.00018 [hep-th]}}.

\bibitem{Deng:2020ent}
F.~Deng, J.~Chu, and Y.~Zhou, ``{Defect extremal surface as the holographic
  counterpart of Island formula},''
  \href{http://dx.doi.org/10.1007/JHEP03(2021)008}{{\em JHEP} {\bfseries 03}
  (2021) 008}, \href{http://arxiv.org/abs/2012.07612}{{\ttfamily
  arXiv:2012.07612 [hep-th]}}.

\bibitem{Suzuki:2022xwv}
K.~Suzuki and T.~Takayanagi, ``{BCFT and Islands in two dimensions},''
  \href{http://dx.doi.org/10.1007/JHEP06(2022)095}{{\em JHEP} {\bfseries 06}
  (2022) 095}, \href{http://arxiv.org/abs/2202.08462}{{\ttfamily
  arXiv:2202.08462 [hep-th]}}.

\bibitem{Grimaldi:2022suv}
G.~Grimaldi, J.~Hernandez, and R.~C. Myers, ``{Quantum extremal islands made
  easy. Part IV. Massive black holes on the brane},''
  \href{http://dx.doi.org/10.1007/JHEP03(2022)136}{{\em JHEP} {\bfseries 03}
  (2022) 136}, \href{http://arxiv.org/abs/2202.00679}{{\ttfamily
  arXiv:2202.00679 [hep-th]}}.

\bibitem{Geng:2020qvw}
H.~Geng and A.~Karch, ``{Massive islands},''
  \href{http://dx.doi.org/10.1007/JHEP09(2020)121}{{\em JHEP} {\bfseries 09}
  (2020) 121}, \href{http://arxiv.org/abs/2006.02438}{{\ttfamily
  arXiv:2006.02438 [hep-th]}}.

\bibitem{Geng:2020fxl}
H.~Geng, A.~Karch, C.~Perez-Pardavila, S.~Raju, L.~Randall, M.~Riojas, and
  S.~Shashi, ``{Information Transfer with a Gravitating Bath},''
  \href{http://dx.doi.org/10.21468/SciPostPhys.10.5.103}{{\em SciPost Phys.}
  {\bfseries 10} no.~5, (2021) 103},
  \href{http://arxiv.org/abs/2012.04671}{{\ttfamily arXiv:2012.04671
  [hep-th]}}.

\bibitem{Geng:2021iyq}
H.~Geng, S.~L\"ust, R.~K. Mishra, and D.~Wakeham, ``{Holographic BCFTs and
  Communicating Black Holes},''
  \href{http://dx.doi.org/10.1007/JHEP08(2021)003}{{\em JHEP} {\bfseries 08}
  (2021) 003}, \href{http://arxiv.org/abs/2104.07039}{{\ttfamily
  arXiv:2104.07039 [hep-th]}}.

\bibitem{Geng:2021mic}
H.~Geng, A.~Karch, C.~Perez-Pardavila, S.~Raju, L.~Randall, M.~Riojas, and
  S.~Shashi, ``{Entanglement phase structure of a holographic BCFT in a black
  hole background},'' \href{http://dx.doi.org/10.1007/JHEP05(2022)153}{{\em
  JHEP} {\bfseries 05} (2022) 153},
  \href{http://arxiv.org/abs/2112.09132}{{\ttfamily arXiv:2112.09132
  [hep-th]}}.

\bibitem{Geng:2021hlu}
H.~Geng, A.~Karch, C.~Perez-Pardavila, S.~Raju, L.~Randall, M.~Riojas, and
  S.~Shashi, ``{Inconsistency of islands in theories with long-range
  gravity},'' \href{http://dx.doi.org/10.1007/JHEP01(2022)182}{{\em JHEP}
  {\bfseries 01} (2022) 182}, \href{http://arxiv.org/abs/2107.03390}{{\ttfamily
  arXiv:2107.03390 [hep-th]}}.

\bibitem{Takayanagi:2011zk}
T.~Takayanagi, ``{Holographic Dual of BCFT},''
  \href{http://dx.doi.org/10.1103/PhysRevLett.107.101602}{{\em Phys. Rev.
  Lett.} {\bfseries 107} (2011) 101602},
  \href{http://arxiv.org/abs/1105.5165}{{\ttfamily arXiv:1105.5165 [hep-th]}}.

\bibitem{Fujita:2011fp}
M.~Fujita, T.~Takayanagi, and E.~Tonni, ``{Aspects of AdS/BCFT},''
  \href{http://dx.doi.org/10.1007/JHEP11(2011)043}{{\em JHEP} {\bfseries 11}
  (2011) 043}, \href{http://arxiv.org/abs/1108.5152}{{\ttfamily arXiv:1108.5152
  [hep-th]}}.

\bibitem{Karch:2000ct}
A.~Karch and L.~Randall, ``{Locally localized gravity},''
  \href{http://dx.doi.org/10.1088/1126-6708/2001/05/008}{{\em JHEP} {\bfseries
  05} (2001) 008}, \href{http://arxiv.org/abs/hep-th/0011156}{{\ttfamily
  arXiv:hep-th/0011156}}.

\bibitem{Karch:2000gx}
A.~Karch and L.~Randall, ``{Open and closed string interpretation of SUSY CFT's
  on branes with boundaries},''
  \href{http://dx.doi.org/10.1088/1126-6708/2001/06/063}{{\em JHEP} {\bfseries
  06} (2001) 063}, \href{http://arxiv.org/abs/hep-th/0105132}{{\ttfamily
  arXiv:hep-th/0105132}}.

\bibitem{Raju:2020smc}
S.~Raju, ``{Lessons from the information paradox},''
  \href{http://dx.doi.org/10.1016/j.physrep.2021.10.001}{{\em Phys. Rept.}
  {\bfseries 943} (2022) 1--80},
  \href{http://arxiv.org/abs/2012.05770}{{\ttfamily arXiv:2012.05770
  [hep-th]}}.

\bibitem{Geng:2022dua}
H.~Geng, L.~Randall, and E.~Swanson, ``{BCFT in a black hole background: an
  analytical holographic model},''
  \href{http://dx.doi.org/10.1007/JHEP12(2022)056}{{\em JHEP} {\bfseries 12}
  (2022) 056}, \href{http://arxiv.org/abs/2209.02074}{{\ttfamily
  arXiv:2209.02074 [hep-th]}}.

\bibitem{Vidal:2002zz}
G.~Vidal and R.~F. Werner, ``{Computable measure of entanglement},''
  \href{http://dx.doi.org/10.1103/PhysRevA.65.032314}{{\em Phys. Rev. A}
  {\bfseries 65} (2002) 032314},
  \href{http://arxiv.org/abs/quant-ph/0102117}{{\ttfamily
  arXiv:quant-ph/0102117}}.

\bibitem{Plenio:2005cwa}
M.~B. Plenio, ``{Logarithmic Negativity: A Full Entanglement Monotone That is
  not Convex},'' \href{http://dx.doi.org/10.1103/PhysRevLett.95.090503}{{\em
  Phys. Rev. Lett.} {\bfseries 95} no.~9, (2005) 090503},
  \href{http://arxiv.org/abs/quant-ph/0505071}{{\ttfamily
  arXiv:quant-ph/0505071}}.

\bibitem{Calabrese:2012ew}
P.~Calabrese, J.~Cardy, and E.~Tonni, ``{Entanglement negativity in quantum
  field theory},'' \href{http://dx.doi.org/10.1103/PhysRevLett.109.130502}{{\em
  Phys. Rev. Lett.} {\bfseries 109} (2012) 130502},
  \href{http://arxiv.org/abs/1206.3092}{{\ttfamily arXiv:1206.3092
  [cond-mat.stat-mech]}}.

\bibitem{Calabrese:2012nk}
P.~Calabrese, J.~Cardy, and E.~Tonni, ``{Entanglement negativity in extended
  systems: A field theoretical approach},''
  \href{http://dx.doi.org/10.1088/1742-5468/2013/02/P02008}{{\em J. Stat.
  Mech.} {\bfseries 1302} (2013) P02008},
  \href{http://arxiv.org/abs/1210.5359}{{\ttfamily arXiv:1210.5359
  [cond-mat.stat-mech]}}.

\bibitem{Calabrese:2014yza}
P.~Calabrese, J.~Cardy, and E.~Tonni, ``{Finite temperature entanglement
  negativity in conformal field theory},''
  \href{http://dx.doi.org/10.1088/1751-8113/48/1/015006}{{\em J. Phys. A}
  {\bfseries 48} no.~1, (2015) 015006},
  \href{http://arxiv.org/abs/1408.3043}{{\ttfamily arXiv:1408.3043
  [cond-mat.stat-mech]}}.

\bibitem{Dutta:2019gen}
S.~Dutta and T.~Faulkner, ``{A canonical purification for the entanglement
  wedge cross-section},'' \href{http://dx.doi.org/10.1007/JHEP03(2021)178}{{\em
  JHEP} {\bfseries 03} (2021) 178},
  \href{http://arxiv.org/abs/1905.00577}{{\ttfamily arXiv:1905.00577
  [hep-th]}}.

\bibitem{Jeong:2019xdr}
H.-S. Jeong, K.-Y. Kim, and M.~Nishida, ``{Reflected Entropy and Entanglement
  Wedge Cross Section with the First Order Correction},''
  \href{http://dx.doi.org/10.1007/JHEP12(2019)170}{{\em JHEP} {\bfseries 12}
  (2019) 170}, \href{http://arxiv.org/abs/1909.02806}{{\ttfamily
  arXiv:1909.02806 [hep-th]}}.

\bibitem{Horodecki:EoP}
B.~M. Terhal, M.~Horodecki, D.~W. Leung, and D.~P. DiVincenzo, ``The
  entanglement of purification,''
  \href{http://dx.doi.org/10.1063/1.1498001}{{\em Journal of Mathematical
  Physics} {\bfseries 43} no.~9, (2002) 4286--4298},
  \href{http://arxiv.org/abs/https://doi.org/10.1063/1.1498001}{{\ttfamily
  https://doi.org/10.1063/1.1498001}}. \url{https://doi.org/10.1063/1.1498001}.

\bibitem{Takayanagi:2017knl}
T.~Takayanagi and K.~Umemoto, ``{Entanglement of purification through
  holographic duality},''
  \href{http://dx.doi.org/10.1038/s41567-018-0075-2}{{\em Nature Phys.}
  {\bfseries 14} no.~6, (2018) 573--577},
  \href{http://arxiv.org/abs/1708.09393}{{\ttfamily arXiv:1708.09393
  [hep-th]}}.

\bibitem{Tamaoka:2018ned}
K.~Tamaoka, ``{Entanglement Wedge Cross Section from the Dual Density
  Matrix},'' \href{http://dx.doi.org/10.1103/PhysRevLett.122.141601}{{\em Phys.
  Rev. Lett.} {\bfseries 122} no.~14, (2019) 141601},
  \href{http://arxiv.org/abs/1809.09109}{{\ttfamily arXiv:1809.09109
  [hep-th]}}.

\bibitem{Wen:2021qgx}
Q.~Wen, ``{Balanced Partial Entanglement and the Entanglement Wedge Cross
  Section},'' \href{http://dx.doi.org/10.1007/JHEP04(2021)301}{{\em JHEP}
  {\bfseries 04} (2021) 301}, \href{http://arxiv.org/abs/2103.00415}{{\ttfamily
  arXiv:2103.00415 [hep-th]}}.

\bibitem{Li:2020ceg}
T.~Li, J.~Chu, and Y.~Zhou, ``{Reflected Entropy for an Evaporating Black
  Hole},'' \href{http://dx.doi.org/10.1007/JHEP11(2020)155}{{\em JHEP}
  {\bfseries 11} (2020) 155}, \href{http://arxiv.org/abs/2006.10846}{{\ttfamily
  arXiv:2006.10846 [hep-th]}}.

\bibitem{Li:2021dmf}
T.~Li, M.-K. Yuan, and Y.~Zhou, ``{Defect extremal surface for reflected
  entropy},'' \href{http://dx.doi.org/10.1007/JHEP01(2022)018}{{\em JHEP}
  {\bfseries 01} (2022) 018}, \href{http://arxiv.org/abs/2108.08544}{{\ttfamily
  arXiv:2108.08544 [hep-th]}}.

\bibitem{Shao:2022gpg}
Y.~Shao, M.-K. Yuan, and Y.~Zhou, ``{Entanglement Negativity and Defect
  Extremal Surface},'' \href{http://arxiv.org/abs/2206.05951}{{\ttfamily
  arXiv:2206.05951 [hep-th]}}.

\bibitem{BasakKumar:2022stg}
J.~Basak~Kumar, D.~Basu, V.~Malvimat, H.~Parihar, and G.~Sengupta, ``{Reflected
  entropy and entanglement negativity for holographic moving mirrors},''
  \href{http://dx.doi.org/10.1007/JHEP09(2022)089}{{\em JHEP} {\bfseries 09}
  (2022) 089}, \href{http://arxiv.org/abs/2204.06015}{{\ttfamily
  arXiv:2204.06015 [hep-th]}}.

\bibitem{Basu:2022reu}
D.~Basu, H.~Parihar, V.~Raj, and G.~Sengupta, ``{Defect extremal surfaces for
  entanglement negativity},''
  \href{http://dx.doi.org/10.1103/PhysRevD.108.106005}{{\em Phys. Rev. D}
  {\bfseries 108} no.~10, (2023) 106005},
  \href{http://arxiv.org/abs/2205.07905}{{\ttfamily arXiv:2205.07905
  [hep-th]}}.

\bibitem{Afrasiar:2022ebi}
M.~Afrasiar, J.~Kumar~Basak, A.~Chandra, and G.~Sengupta, ``{Islands for
  entanglement negativity in communicating black holes},''
  \href{http://dx.doi.org/10.1103/PhysRevD.108.066013}{{\em Phys. Rev. D}
  {\bfseries 108} no.~6, (2023) 066013},
  \href{http://arxiv.org/abs/2205.07903}{{\ttfamily arXiv:2205.07903
  [hep-th]}}.

\bibitem{Afrasiar:2022fid}
M.~Afrasiar, J.~K. Basak, A.~Chandra, and G.~Sengupta, ``{Reflected entropy for
  communicating black holes. Part I. Karch-Randall braneworlds},''
  \href{http://dx.doi.org/10.1007/JHEP02(2023)203}{{\em JHEP} {\bfseries 02}
  (2023) 203}, \href{http://arxiv.org/abs/2211.13246}{{\ttfamily
  arXiv:2211.13246 [hep-th]}}.

\bibitem{Afrasiar:2023jrj}
M.~Afrasiar, J.~K. Basak, A.~Chandra, and G.~Sengupta, ``{Reflected Entropy for
  Communicating Black Holes II: Planck Braneworlds},''
  \href{http://arxiv.org/abs/2302.12810}{{\ttfamily arXiv:2302.12810
  [hep-th]}}.

\bibitem{Basu:2023wmv}
D.~Basu, J.~Lin, Y.~Lu, and Q.~Wen, ``{Ownerless island and partial
  entanglement entropy in island phases},''
  \href{http://arxiv.org/abs/2305.04259}{{\ttfamily arXiv:2305.04259
  [hep-th]}}.

\bibitem{Kumari:2023ops}
A.~Kumari, V.~Raj, and G.~Sengupta, ``{Odd entanglement entropy in boundary
  conformal field theories and holographic moving mirrors},''
  \href{http://arxiv.org/abs/2310.11242}{{\ttfamily arXiv:2310.11242
  [hep-th]}}.

\bibitem{Hayden:2021gno}
P.~Hayden, O.~Parrikar, and J.~Sorce, ``{The Markov gap for geometric reflected
  entropy},'' \href{http://dx.doi.org/10.1007/JHEP10(2021)047}{{\em JHEP}
  {\bfseries 10} (2021) 047}, \href{http://arxiv.org/abs/2107.00009}{{\ttfamily
  arXiv:2107.00009 [hep-th]}}.

\bibitem{Chandrasekaran:2020qtn}
V.~Chandrasekaran, M.~Miyaji, and P.~Rath, ``{Including contributions from
  entanglement islands to the reflected entropy},''
  \href{http://dx.doi.org/10.1103/PhysRevD.102.086009}{{\em Phys. Rev. D}
  {\bfseries 102} no.~8, (2020) 086009},
  \href{http://arxiv.org/abs/2006.10754}{{\ttfamily arXiv:2006.10754
  [hep-th]}}.

\bibitem{Vardhan:2021mdy}
S.~Vardhan, J.~Kudler-Flam, H.~Shapourian, and H.~Liu, ``{Mixed-state
  entanglement and information recovery in thermalized states and evaporating
  black holes},'' \href{http://dx.doi.org/10.1007/JHEP01(2023)064}{{\em JHEP}
  {\bfseries 01} (2023) 064}, \href{http://arxiv.org/abs/2112.00020}{{\ttfamily
  arXiv:2112.00020 [hep-th]}}.

\bibitem{Akers:2022max}
C.~Akers, T.~Faulkner, S.~Lin, and P.~Rath, ``{The Page curve for reflected
  entropy},'' \href{http://dx.doi.org/10.1007/JHEP06(2022)089}{{\em JHEP}
  {\bfseries 06} (2022) 089}, \href{http://arxiv.org/abs/2201.11730}{{\ttfamily
  arXiv:2201.11730 [hep-th]}}.

\bibitem{Ling:2021vxe}
Y.~Ling, P.~Liu, Y.~Liu, C.~Niu, Z.-Y. Xian, and C.-Y. Zhang, ``{Reflected
  entropy in double holography},''
  \href{http://dx.doi.org/10.1007/JHEP02(2022)037}{{\em JHEP} {\bfseries 02}
  (2022) 037}, \href{http://arxiv.org/abs/2109.09243}{{\ttfamily
  arXiv:2109.09243 [hep-th]}}.

\bibitem{KumarBasak:2020ams}
J.~Kumar~Basak, D.~Basu, V.~Malvimat, H.~Parihar, and G.~Sengupta, ``{Islands
  for entanglement negativity},''
  \href{http://dx.doi.org/10.21468/SciPostPhys.12.1.003}{{\em SciPost Phys.}
  {\bfseries 12} no.~1, (2022) 003},
  \href{http://arxiv.org/abs/2012.03983}{{\ttfamily arXiv:2012.03983
  [hep-th]}}.

\bibitem{Lu:2022cgq}
Y.~Lu and J.~Lin, ``{The Markov gap in the presence of islands},''
  \href{http://dx.doi.org/10.1007/JHEP03(2023)043}{{\em JHEP} {\bfseries 03}
  (2023) 043}, \href{http://arxiv.org/abs/2211.06886}{{\ttfamily
  arXiv:2211.06886 [hep-th]}}.

\bibitem{Brown:1986nw}
J.~D. Brown and M.~Henneaux, ``{Central Charges in the Canonical Realization of
  Asymptotic Symmetries: An Example from Three-Dimensional Gravity},''
  \href{http://dx.doi.org/10.1007/BF01211590}{{\em Commun. Math. Phys.}
  {\bfseries 104} (1986) 207--226}.

\bibitem{Kusuki:2019evw}
Y.~Kusuki and K.~Tamaoka, ``{Entanglement Wedge Cross Section from CFT:
  Dynamics of Local Operator Quench},''
  \href{http://dx.doi.org/10.1007/JHEP02(2020)017}{{\em JHEP} {\bfseries 02}
  (2020) 017}, \href{http://arxiv.org/abs/1909.06790}{{\ttfamily
  arXiv:1909.06790 [hep-th]}}.

\bibitem{Akers:2021pvd}
C.~Akers, T.~Faulkner, S.~Lin, and P.~Rath, ``{Reflected entropy in random
  tensor networks},'' \href{http://arxiv.org/abs/2112.09122}{{\ttfamily
  arXiv:2112.09122 [hep-th]}}.

\bibitem{Czech:2012bh}
B.~Czech, J.~L. Karczmarek, F.~Nogueira, and M.~Van~Raamsdonk, ``{The Gravity
  Dual of a Density Matrix},''
  \href{http://dx.doi.org/10.1088/0264-9381/29/15/155009}{{\em Class. Quant.
  Grav.} {\bfseries 29} (2012) 155009},
  \href{http://arxiv.org/abs/1204.1330}{{\ttfamily arXiv:1204.1330 [hep-th]}}.

\bibitem{Fitzpatrick:2014vua}
A.~L. Fitzpatrick, J.~Kaplan, and M.~T. Walters, ``{Universality of
  Long-Distance AdS Physics from the CFT Bootstrap},''
  \href{http://dx.doi.org/10.1007/JHEP08(2014)145}{{\em JHEP} {\bfseries 08}
  (2014) 145}, \href{http://arxiv.org/abs/1403.6829}{{\ttfamily arXiv:1403.6829
  [hep-th]}}.

\bibitem{Banerjee:2016qca}
P.~Banerjee, S.~Datta, and R.~Sinha, ``{Higher-point conformal blocks and
  entanglement entropy in heavy states},''
  \href{http://dx.doi.org/10.1007/JHEP05(2016)127}{{\em JHEP} {\bfseries 05}
  (2016) 127}, \href{http://arxiv.org/abs/1601.06794}{{\ttfamily
  arXiv:1601.06794 [hep-th]}}.

\bibitem{Cardy:2004hm}
J.~L. Cardy, ``{Boundary conformal field theory},''
  \href{http://arxiv.org/abs/hep-th/0411189}{{\ttfamily arXiv:hep-th/0411189}}.

\end{thebibliography}\endgroup

\end{document}